\documentclass[conference, 12pt, onecolumn]{ieeeconf}

\IEEEoverridecommandlockouts
\usepackage[usenames]{color}
\usepackage{enumerate}
\usepackage{url}
\usepackage{subfigure}
\usepackage{amsfonts,mathrsfs}
\usepackage{amssymb,amsmath}
\usepackage{verbatim}
\usepackage{acronym}
\usepackage{mathtools}
\usepackage{cite}
\usepackage{graphicx}

\usepackage[colorlinks=true,breaklinks=true,bookmarks=true,urlcolor=blue,
     citecolor=blue,linkcolor=blue,bookmarksopen=false,draft=false]{hyperref}
     \usepackage{algorithm}
\usepackage[noend]{algpseudocode}

\def\fskip#1{}

\newtheorem{theorem}{Theorem}

\newtheorem{corollary}{Corollary}

\newtheorem{definition}{Definition}
\newtheorem{example}{Example}

\newtheorem{lemma}{Lemma}

\newtheorem{proposition}[theorem]{Proposition}
\newtheorem{remark}{Remark}

\renewcommand{\theenumi}{\arabic{enumi}}

\def\1{{\bf 1}}

\newcommand{\remove}[1]{}

\def\argmin{\mathop{\rm argmin}}

\allowdisplaybreaks

\begin{document}
\title{An Optimal Control Framework for Online Job Scheduling with General Cost Functions}
\author{\authorblockN{S. Rasoul Etesami*}
\thanks{*Department of Industrial and Systems Engineering and Coordinated Science Lab, University of Illinois at Urbana-Champaign,  Urbana, IL 61801, Email: (etesami1@illinois.edu). This work is supported by the National Science Foundation under Grant No. EPCN-1944403.}
}
\maketitle
\begin{abstract}
We consider the problem of online job scheduling on a single machine or multiple unrelated machines with general job and machine-dependent cost functions. In this model, each job $j$ has a processing requirement (length) $v_{ij}$ and arrives with a nonnegative nondecreasing cost function $g_{ij}(t)$ if it has been dispatched to machine $i$, and this information is revealed to the system upon arrival of job $j$ at time $r_j$. The goal is to dispatch the jobs to the machines in an online fashion and process them preemptively on the machines so as to minimize the generalized integral completion time $\sum_{j}g_{i(j)j}(C_j)$. Here $i(j)$ refers to the machine to which job $j$ is dispatched, and $C_j$ is the completion time of job $j$ on that machine. It is assumed that jobs cannot migrate between machines and that each machine has a fixed unit processing speed that can work on a single job at any time instance. In particular, we are interested in finding an online scheduling policy whose objective cost is competitive with respect to a slower optimal offline benchmark, i.e., the one that knows all the job specifications a priori and is slower than the online algorithm. We first show that for the case of a single machine and special cost functions $g_j(t)=w_jg(t)$, with nonnegative nondecreasing $g(t)$, the highest-density-first rule is optimal for the generalized fractional completion time. We then extend this result by giving a speed-augmented competitive algorithm for the general nondecreasing cost functions $g_j(t)$ by utilizing a novel optimal control framework. This approach provides a principled method for identifying dual variables in different settings of online job scheduling with general cost functions. Using this method, we also provide a speed-augmented competitive algorithm for multiple unrelated machines with nondecreasing convex functions $g_{ij}(t)$, where the competitive ratio depends on the curvature of the cost functions $g_{ij}(t)$.
\end{abstract}

\section{Introduction}
Job scheduling is one of the fundamental problems in operations research and computer science. Broadly speaking, its goal is to schedule a collection of jobs with different specifications to a set of machines by minimizing a certain performance metric. Depending on the application domain, various performance metrics have been proposed and analyzed over the past decades, with some of the notable ones being the weighted \emph{completion} time $\sum_jw_jC_j$, where $C_j$ denotes the completion time of job $j$; weighted \emph{flow} time $\sum_jw_j (C_j-r_j)$, where $r_j$ is the release time of job $j$; or a generalization of both, such as weighted $\ell_k$-flow time $(\sum_jw_j(C_j-r_j)^k)^{\frac{1}{k}}$ \cite{leung2004handbook,graham1979optimization,angelopoulos2019primal,anand2012resource,im2011online}. In fact, one can capture all of these performance metrics using a general form $\sum_jg_j(C_j)$, where $g_j(\cdot)$ is a general nonnegative and nondecreasing cost function. In this paper, we focus on this most general performance metric and develop online algorithms with bounded competitive ratios under certain assumptions on the structure of the cost functions $g_j(t)$.  For instance, by choosing $g_j(t)=w_jt$ and $g_j(t)=w_j(t-r_j)$ one can obtain the weighted completion time and the weighted flow time cost functions, respectively.  Although minimizing flow time is often more difficult and considered more important in the past literature, since our analysis is based on the general cost functions $g_j(t)$, we do not distinguish between these two cases, and our results continue to hold for either case.  Of course, each of the performance metrics that we discussed above was written for only a single machine. However, one can naturally extend them to multiple unrelated machines by setting $\sum_jg_{i(j)j}(C_j)$, where $i(j)$ denotes the machine to which job $j$ is dispatched, and $g_{i(j)j}$ is the cost function associated with job $j$ if it is dispatched to machine $i(j)$.

The job-scheduling problems have been extensively studied in the past literature under both offline and online settings. In the offline setting, it is assumed that all the job specifications (i.e., processing lengths, release times, and cost functions) are known and given to the scheduler a priori. In the online setting that we consider in this paper, the scheduler only learns a job specification upon the job's arrival at the system, at which point the scheduler must make an irrevocable decision.  Here, an irrevocable decision means that at any current time $t$, an online scheduler can not change its decisions on how the jobs were assigned and processed on the machines before time $t$. Therefore, an immediate question here is whether an online scheduler can still achieve a performance ``close" to that of the offline one despite its lack of information ahead of time. The question has been addressed using the notion of \emph{competitive ratio}, which has frequently been used as a standard metric for evaluating the performance of online algorithms. In this paper, we shall also use the competitive ratio to evaluate the performance guarantees of our devised online algorithms.

In this paper, we allow \emph{preemptive} schedules, meaning that the processing of a job on a machine can be interrupted because of the existence or arrival of the other jobs. This is much-needed in a deterministic setting \cite{bansal2009weighted,leonardi2007approximating,lucarelli2016online}, because even for a single machine, there are strong lower bounds for the competitive ratio of any online algorithm.  For instance, it was shown in \cite{lucarelli2016online} that no deterministic algorithm has bounded competitive ratio when preemptions are not allowed even for a single machine with arbitrary large speed augmentation.  Moreover, we consider \emph{nonmigratory} schedules in which a dispatched job must stay on the same machine until its completion and is not allowed to migrate to other machines. In fact, for various reasons, such as to increase the lifetimes of machines or reduce failures in job completions, nonmigratory schedules are quite desirable in practical applications \cite{anand2012resource}.  Furthermore, in this paper, we assume that all the machines have unit speeds, and at any time, a machine can work on one job only.  This setting is more restrictive than the \emph{rate allocation} setting in which a machine can work on multiple jobs at the same time by fractionally distributing its unit processing speed among the jobs \cite{im2014selfishmigrate,im2018competitive}.  In fact, in the rate allocation setting, an online scheduler has extra freedom to split its computational power fractionally among the pending jobs to achieve a better competitive ratio.  



%

Unfortunately, even for a simple weighted flow time problem on three unrelated machines, it is known that no online algorithm can achieve a bounded competitive ratio \cite{garg2007minimizing}. To overcome that obstacle, in this paper we adopt the \emph{speed augmentation} framework, which was first proposed by \cite{kalyanasundaram2000speed} and subsequently used for various online job scheduling problems \cite{chadha2009competitive,devanur2018primal,anand2012resource,angelopoulos2019primal,im2015competitive}. More precisely, in the speed augmentation framework, one compares the performance of the online scheduler with that of a weaker optimal offline benchmark, i.e., the one in which each machine has a fixed slower speed of $\frac{1}{1+\epsilon}, \epsilon>0$. In other words, an online scheduler can achieve a bounded competitive ratio if the machines run $1+\epsilon$ times faster than those in the optimal offline benchmark.

In general, there are two different approaches to devising competitive algorithms for online job scheduling. The first method is based on the potential function technique, in which one constructs a clever potential function and shows that the proposed algorithm behaves well compared to the offline benchmark in an amortized sense. Unfortunately, constructing potential functions can be very nontrivial and often requires a good ``guess." Even if one can come up with the good potential function, such analysis provides little insight about the problem, and the choice of the potential function is very specific to a particular problem setup \cite{im2011online,im2018competitive,im2015competitive,im2011tutorial}. An alternative and perhaps more powerful technique, which we shall use in this paper, is the one based on linear/convex programming duality and dual-fitting \cite{anand2012resource,devanur2018primal,angelopoulos2019primal}. In this approach, one first models the offline job scheduling problem as a mathematical program and then utilizes this program to develop an online algorithm that preserves KKT optimality conditions as much as possible over the course of the algorithm. Following this approach, one can construct an online feasible primal solution (i.e., the solution generated by the algorithm) together with a properly ``fitted" dual solution, and then show that the cost increments in the primal objective (i.e., the increase in the cost of the algorithm due to its decisions) and those of the dual objective are within a certain factor from each other. As a result, the cost increments of the primal and dual feasible solutions due to the arrival of a new job remain within a certain factor from each other, which establishes a competitive ratio for the devised algorithm because of the weak duality. However, one major difficulty here is that of carefully selecting the dual variables, which, in general, could be highly nontrivial. As one of the contributions of this paper, we provide a principled way of setting dual variables by using results from optimal control and the minimum principle. As a by-product, we show how one can recover some of the earlier dual-fitting results and extend them to more complicated nonlinear settings. 

\subsection{Related Work}
It is known that without speed augmentation, there is no competitive online algorithm for minimizing weighted flow time \cite{chekuri2001algorithms}. The first online competitive algorithm with speed augmentation for minimizing flow time on a single machine was given by \cite{kalyanasundaram2000speed}. In \cite{im2011online}, a potential function was constructed to show that a natural online greedy algorithm is $(1+\epsilon)$-speed $O(k\epsilon^{-(2+\frac{2}{k})})$-competitive for minimizing the $\ell_k$-norm of weighted flow time on unrelated machines. This result was improved by \cite{anand2012resource} to a $(1+\epsilon)$-speed $O(k\epsilon^{-(2+\frac{1}{k})})$-competitive algorithm, which was the first analysis of online job scheduling that uses the dual-fitting technique. In that algorithm, each machine works based on the highest residual density first (HRDF) rule, such that the residual density of a job $j$ on machine $i$ at time $t$ is given by the ratio of its weight $w_{ij}$ over its remaining length ${v}_{ij}(t)$, i.e., $\rho_{ij}(t):=\frac{w_{ij}}{v_{ij(t)}}$. In particular, a newly released job is dispatched to a machine that gives the least increase in the objective of the offline linear program. Our algorithm for online job scheduling with generalized cost functions was partly inspired by the primal-dual algorithm in \cite{devanur2018primal}, which was developed for a different objective of minimizing the sum of the energy and weighted flow time on unrelated machines. However, unlike the work in \cite{devanur2018primal}, for which the optimal dual variables can be precisely determined using natural KKT optimality conditions, the dual variables in our setting do not admit a simple closed-form characterization. Therefore, we follow a different path to infer certain desired properties by using a dynamic construction of dual variables that requires new ideas. 

Online job scheduling on a \emph{single} machine with general cost functions of the form $g_j(t)=w_jg(t-r_j)$, where $g(t)$ is a general nonnegative nondecreasing function, has been studied in \cite{angelopoulos2019primal}. In particular, it has been shown in \cite{angelopoulos2019primal} that the highest density first (HDF) rule is optimal for minimizing the fractional completion time on a single machine, and it was left open for multiple machines. Here, the \emph{fractional objective} means that the contribution of a job to the objective cost is proportional to its remaining length. The analysis in \cite{angelopoulos2019primal} is based on a primal-dual technique that updates the optimal dual variables upon arrival of a new job by using a fairly complex two-phase process. We obtain the same result here using a much simpler process that was inspired by dynamic programming and motivated our optimal control formulation, wherein we extended this result to arbitrary nondecreasing cost functions $g_j(t)$. The problem of minimizing the generalized flow time $\sum_{i,j}w_{ij}g(t-r_j)$ on unrelated machines for a convex and nondecreasing cost function $g(\cdot)$ has recently been studied in \cite{nguyen2013lagrangian}, where it is shown that a greedy dispatching rule similar to that in \cite{anand2012resource}, together with the HRDF scheduling rule, provides a competitive online algorithm under a speed-augmented setting. The analysis in \cite{nguyen2013lagrangian} is based on nonlinear Lagrangian relaxation and dual-fitting as in \cite{anand2012resource}. However, the competitive ratio in \cite{nguyen2013lagrangian} depends on additional assumptions on the cost function $g(t)$ and is a special case of our generalized completion time problem on unrelated machines. In particular, our algorithm is different in nature from the one in \cite{nguyen2013lagrangian} and is based on a simple primal-dual dispatching scheme. The competitive ratios that we obtain in this work follow organically from our analysis and require less  assumptions on the cost functions.

The generalized flow time problem on a single machine with special cost functions $g_j(t)=g(t-r_j)$ has been studied in \cite{im2014online}. It was shown that for nondecreasing nonnegative function $g(\cdot)$, the HDF rule is $(2+\epsilon)$-speed $O(1)$-competitive; the HDF rule is, in essence, the best online algorithm one can hope for under the speed-augmented setting.  \cite{nguyen2013lagrangian} uses Lagrangian duality for online scheduling problems beyond linear and convex programming. The problem of rate allocation on a single machine with the objective of minimizing weighted flow/completion time when jobs of unknown size arrive online (i.e., the \emph{nonclairvoyant} setting) has been studied in \cite{im2015competitive,im2014selfishmigrate,im2018competitive}.  Moreover, \cite{im2016fair} give an $(1+\epsilon)$-speed $O(\frac{1}{\epsilon^2})$-competitive algorithm for fair rate allocation over unrelated machines.

The \emph{offline} version of job scheduling on a single or multiple machines has also received much attention in the past few years \cite{leung2004handbook,williamson2011design}.  \cite{azar2005convex} use a convex program to give a $2$-approximation algorithm for minimizing the $\ell_k$-norm of the loads on unrelated machines.  \cite{megow2018dual} study offline scheduling on a machine with varying speed and provide a polynomial-time approximation scheme for minimizing the total weighted completion time $\sum_jw_jg(C_j)$, even without preemption. Moreover, a tight analysis of HDF for the very same class of problems was given by \cite{hohn2015performance}.  \cite{bansal2014geometry} studied the offline version of a very general scheduling problem on a single machine; the online version of that problem is considered in this paper. More precisely, \cite{bansal2014geometry} provide a preemptive $O(\log\log nP)$-approximation algorithm for minimizing the generalized completion time $\sum_jg_j(C_j)$, where $n$ is the number of jobs and $P$ is the maximum job length. This result has recently been extended in \cite{moseley2019scheduling} to the case of multiple identical machines, where it was shown that using preemption and migration, there exists an $O(\log\log nP)$-approximation algorithm for minimizing the offline generalized completion time, assuming that all jobs are available at the same time.  \cite{angelopoulos2019primal} considered the online generalized completion time $\sum_jg_j(C_j)$ problem on a single machine and provided a rate allocation algorithm that is $(1+\epsilon)$-speed $\frac{4(1+\epsilon)^2}{\epsilon^2}$-competitive, assuming differentiable and monotone concave cost functions $g_j(t)$. We note that the rate allocation problem is a relaxation of the problem we consider here. This paper is the first to study the generalized completion time $\sum_jg_j(C_j)$ problem under the online and speed augmented setting. In particular, for both single and multiple unrelated machines, we provide online preemptive nonmigratory algorithms whose competitive ratios depend on the curvature of the cost functions.

\subsection{Organization and Contributions}

In Section \ref{sec:problem-formulation}, we provide a formulation of the \emph{\underline{g}eneralized \underline{i}ntegral \underline{c}ompletion} time on a \emph{\underline{s}ingle} machine (GICS),  and its fractional relaxation, namely,  \emph{\underline{g}eneralized \underline{f}rational \underline{c}ompletion} time on a \emph{\underline{s}ingle} machine (GFCS).  We also show how an online algorithm for GFCS can be converted to an online algorithm for GIC-S with only a small loss in the speed/competitive ratio, hence allowing us to only focus on designing competitive algorithms for the fractional problem.  In particular, we consider a special \emph{\underline{h}omogeneous} case of GFCS that we refer to as HGFCS, for which the cost functions are of the form $g_j(t)=w_jg(t)$, where $w_j\ge 0$ is a constant weight and $g(t)$ is an arbitrary nonnegative nondecreasing function.  We show that the HDF is an optimal online schedule for HGFCS by determining the optimal dual variables using a simple backward process. Using the insights obtained from this special homogeneous case and in order to handle the general GFCS problem,  in Section \ref{sec:optimal-control}, we provide an optimal control formulation for the offline version of GFCS with identical release times. This formulation allows us to use tools from optimal control,  such as the minimum principle and Hamilton-Jacobi-Bellman (HJB) equation, to set the dual variables in GFCS as close as possible to the optimal dual variables.  In Section \ref{sec:heterogeneous-cost}, we consider the GFCS problem and use our choice of dual variables to design an online algorithm as an iterative application of the offline GFCS with identical release times. In that regard, we deduce our desired properties on the choice of dual variables by making a connection to a network flow problem. These results together will allow us to obtain an $2K(1+\epsilon)$-speed $\frac{2(1+\epsilon)}{\epsilon}$-competitive algorithm for GICS, assuming monotonicity of the cost functions $g_j(t)$, where $K=\sup_{j,t\ge r_j}\frac{(t-r_j)g_j''(t)}{g_j'(t)}$.  In Section \ref{sec:unrelated-competitive}, we extend this result to online scheduling for \emph{\underline{g}eneralized \underline{i}ntegral \underline{c}ompletion} time on \emph{\underline{u}nrelated} machines (GICU), and its fractional relaxation, namely, \emph{\underline{g}eneralized \underline{f}rational \underline{c}ompletion} time on \emph{\underline{u}nrelated} machines (GFCU).  In particular, we obtain an $2K\theta(1+\epsilon)$-speed $\frac{2\theta(1+\epsilon)}{\epsilon}$-competitive algorithm for GICU by assuming monotonicity and convexity of the cost functions $g_{ij}(t)$, where $K=\sup_{i,j,t\ge r_j}\frac{(t-r_j)g_{ij}''(t)}{g_{ij}'(t)}$ and $\theta=\sup_{i,j,t\ge v}\frac{g_{ij}(t+v)-g_{ij}(t)}{vg'_{ij}(t)}$ are curvature parameters. To the best of our knowledge, our devised online algorithms are the first speed-augmented competitive algorithms for such general cost functions on a single or multiple unrelated machines.  We conclude the paper by identifying some future directions of research in Section \ref{sec:conclusions}.  Finally,  we present another application of the optimal control framework to generalization of some of the existing dual-fitting results in Appendix I. We relegate omitted proofs to Appendix II.  Table 1 summarizes the results of this paper in comparison to previous work.

\begin{table}[t]
\caption {{\footnotesize Summary of Results for Generalized Integral Completion Time for a Single or Multiple Nonmigratory Machines with Preemption}} \label{tab:title}\medskip
{\tiny \hspace{0.3cm}\begin{tabular}{| c || c | c | c || c | c |}
\hline
\textbf{$g(\cdot), g_j(\cdot)$} &\multicolumn{3}{c||}{\textbf{Single Machine}} &\multicolumn{2}{c|}{\textbf{Multiple Machines}}\\
\cline{2-6}
& \textbf{$\sum_jw_jg(C_j)$} & \textbf{$\sum_jw_jg(C_j-r_j)$} & \textbf{$\sum_jg_j(C_j)$} & \textbf{$\sum_jw_{ij}g(C_{j}-r_j)$} & \textbf{$\sum_jg_{ij}(C_{j})$}\\
\hline
\hline
\textbf{Convex} &  & $\substack{\big(1+\epsilon, \frac{1+\epsilon}{\epsilon}\big)-\mbox{FIFO}\\ \mbox{identical job density}\\ \mbox{\cite{angelopoulos2019primal}}}$ & & $\substack{\big(\frac{1}{1-3\epsilon}, \frac{2K(g,\epsilon)}{\epsilon}\big)-\mbox{HRDF}\\ K(g,\epsilon): \mbox{ a function of}\ g, \epsilon\\\mbox{unrelated machines}\\\mbox{\cite{nguyen2013lagrangian}}}$ & $\boldsymbol{\substack{\big(2K\theta(1+\epsilon), \frac{2\theta(1+\epsilon)}{\epsilon}\big)\\\mbox{{\bf unrelated machines}}}}$\\
\cline{3-3}\cline{5-5}
\textbf{} &  &  $\substack{\big(2+\epsilon, O(1)\big)-\mbox{WSETF}\\ \mbox{nonclairvoyant}\\\mbox{\cite{fox2013online}}}$ &  & $\substack{\big(3+\epsilon, O(1)\big)-\mbox{WSETF}\\ \mbox{nonclairvoyant}\\\mbox{identical machines}\\\mbox{\cite{fox2013online}}}$ & $\boldsymbol{\substack{K=1+\sup\limits_{j,t\ge r_j}\frac{(t-r_j)g_j''(t)}{g'_j(t)}\\ \theta=\!\!\!\sup\limits_{i,j,t\ge r_j}\!\!\!\!\!\frac{g_{ij}(t+v)-g_{ij}(t)}{vg'_{ij}(t)}}}$\\
\hline
\hline
\textbf{Concave} &  &  $\substack{\big(1+\epsilon, \frac{1+\epsilon}{\epsilon}\big)-\mbox{HDF}\\ \mbox{\cite{angelopoulos2019primal}}}$  & $\substack{\boldsymbol{\big((1+\epsilon)^2, \frac{(1+\epsilon)^2}{\epsilon^2}\big)}\\\mbox{{\bf twice differentiale costs}}\\ \smallskip}$ & &\\
\cline{3-4}
\textbf{} &  & $\substack{\big(1+\epsilon, O(\frac{1}{\epsilon^2})\big)-\mbox{WLAPS}\\ \mbox{nonclairvoyant}\\\mbox{\cite{im2014online}}}$ & $\substack{\big(1+\epsilon, \frac{4(1+\epsilon)^2}{\epsilon^2}\big)\\ \mbox{rate allocation setting}\\ \mbox{\cite{angelopoulos2019primal}}}$ & &\\
\hline
\hline
\textbf{General} & $\substack{\big(1+\epsilon, \frac{1+\epsilon}{\epsilon}\big)-\mbox{HDF}\\ \mbox{\cite{angelopoulos2019primal}}}$ &  $\substack{\big(1+\epsilon, \frac{(1+\epsilon)}{\epsilon}\big)-\mbox{FIFO}\\ \mbox{identical job density}\\ \mbox{\cite{angelopoulos2019primal}}}$ & $\substack{ \boldsymbol{\big(2K(1+\epsilon), \frac{2(1+\epsilon)}{\epsilon}\big)}}$ & &\\
\cline{2-3}
\textbf{} & {\bf $\substack{\boldsymbol{\big(1+\epsilon, \frac{1+\epsilon}{\epsilon}\big)}-\mbox{HDF}\\ \mbox{backward dual-fitting}}$} & $\substack{\big(2+\epsilon, O(1)\big)-\mbox{HDF}\\\mbox{\cite{im2014online}}}$ & $\substack{ \boldsymbol{K=1+\sup\limits_{j,t\ge r_j}\frac{(t-r_j)g_j''(t)}{g'_j(t)}}}$ & &\\
\hline
\end{tabular}}\par\medskip
{\tiny The $(s,c)$ notation describes an algorithm which is $c$-speed $c$-competitive.  In the above table, we use the following abbreviates: Weighted Late Arrival Processor Sharing (WLAPS),  Highest Density First (HDF),  Highest Residual Density First (HRDF), First-In-First-Out (FIFO), Weighted Shortest Elapsed Time First (WSETF).  The results of this paper are presented in bold notation.}
\end{table}

\section{Problem Formulation and Preliminary Results}\label{sec:problem-formulation}

Consider a single machine that can work on at most one unfinished job $j$ at any time instance $t\in \mathbb{R}_+$. Moreover, assume that the machine has a fixed unit processing speed, meaning that it can process only a unit length of a job $j$ per unit of time. We consider a clairvoyant setting in which each job $j$ has a known length $v_j$ and a job-dependent cost function $g_j(t)$, which is revealed to the machine only at its arrival time $r_j\ge 0$.  Here, $g_j(t), t\ge r_j$ is a nonnegative nondecreasing differentiable function with $g_j(r_j)=0$, and $t$ refers to the elapsed time on the machine regardless of whether job $j$ is being processed or is waiting to be processed at time $t$.  In other words, as long as job $j$ is not fully processed, its presence on the machine contributes to the delay cost that includes both the processing duration of job $j$ and the waiting time due to execution of other jobs.  Note that in the online setting, the machine does not know a priori the job specifications $v_j, r_j, g_j(t)$, and learns them only upon release of job $j$ at time $r_j$. Given a time instance $t\ge r_j$, let us use $v_j(t)$ to denote the remaining length of job $j$ at time $t$, such that $v_j(r_j)=v_j$. We say that the \emph{completion time} of the job $j$ is the first time $C_j> r_j$ at which the job is fully processed, i.e., $v_j(C_j)=0$. Of course, $C_j$ depends on the type of schedule that the machine is using to process the jobs, and we have not specified the schedule type here. The generalized integral completion time problem on a single machine (GICS) is to find an online schedule that minimizes the objective cost $\sum_jg_j(C_j)$.

In this paper, we shall focus only on devising competitive algorithms for \emph{fractional} objective functions. This is a common approach for obtaining a competitive, speed-augmented scheduling algorithm for various \emph{integral} objective functions \cite{becchetti2006online,angelopoulos2019primal,im2018competitive,fox2013online}.  In the fractional problem, only the remaining fraction $\frac{v_j(t)}{v_j}\leq 1$ of job $j$ at time $t$ contributes $g'_j(t)$ amount to the delay cost of job $j$.  Thus, the objective cost of the generalized fractional completion time on a single machine (GFCS) is given by
\begin{align}\nonumber
\sum_j\int_{r_j}^{\infty}\frac{v_j(t)}{v_j}g'_j(t)dt.
\end{align}   
Note that the fractional cost is a lower bound for the integral cost in which the entire unit fraction $\frac{v_j}{v_j}=1$ of a job $j$ receives a delay cost of $g'_j(t)$ such that $\int_{r_j}^{C_j}1\times g'_j(t)dt=g_j(C_j)$. The following lemma, whose proof is given in Appendix II, is a slight generalization of the result in \cite{chadha2009competitive}, which establishes a ``black-box" reduction between designing a competitive, speed-augmented algorithm for the GFCS problem and its integral counterpart GICS.

\begin{definition}
An online algorithm is called $s$-speed $c$-competitive if it can achieve a competitive ratio of $c$ given that the machine runs $s$ times faster than that in the optimal offline algorithm.
\end{definition}

\begin{lemma}\label{lemm-s-c}
Given any $\epsilon\in (0, 1]$, an $s$-speed $c$-competitive algorithm for the GFCS problem can be converted to an $(1+\epsilon)s$-speed $\frac{1+\epsilon}{\epsilon}c$-competitive algorithm for the GICS problem.
\end{lemma}

Although our ultimate goal is to devise online algorithms for the GICS, in the remainder of this paper, we only focus on deriving online algorithms that are competitive for the GFCS.  However, using Lemma \ref{lemm-s-c},  the results can be extended to the integral objectives by incurring a small multiplicative loss in speed and competitive ratio. 

\begin{remark}
It was shown in \cite{angelopoulos2019primal} that for special cost functions $g_j(t)=w_jg(t)$, one can bypass the black-box reduction of Lemma \ref{lemm-s-c}, and obtain
an improvement to the competitive ratio and the speed by a factor of $O(\frac{1}{\epsilon})$ and $O(1+\epsilon)$, respectively. Thus, it would be interesting to see whether the same direct approach can be applied to our generalized cost setting. 
\end{remark}

Next, we introduce a natural LP formulation for the GFCS problem and postpone its extension to multiple unrelated machines to Section \ref{sec:unrelated-competitive}.  The derivation of such LP formulation follows similar steps as those in \cite{angelopoulos2019primal,fox2013online,anand2012resource}, and we provide here for the sake of completeness.  Let us use $x_j(t)$ to denote the rate at which job $j$ is processed in an infinitesimal interval $[t, t+dt]$, such that $dv_j(t)=-x_j(t)dt$. Thus, using integration by parts, we can write the above objective as
\begin{align}\nonumber
\sum_j\int_{r_j}^{\infty}\frac{v_j(t)}{v_j}g'_j(t)dt=\sum_j\left(\frac{v_j(t)}{v_j}g_j(t){|_{r_j}^{\infty}}+\int_{r_j}^{\infty}\frac{g_j(t)}{v_j}x_j(t)dt\right)=\sum_j\int_{r_j}^{\infty}\frac{g_j(t)}{v_j}x_j(t)dt,
\end{align}
where the second equality holds because $g_j(r_j)=0$ and $v_j(\infty)=0$.  Now, for simplicity and by some abuse of notation, let us redefine $g_j(t)$ to be its scaled version $\frac{g_j(t)}{v_j}$.

Then, the \emph{offline} GFCS problem is given by the following LP, which is also a fractional relaxation for the offline GICS problem.
\begin{align}\label{eq:H_primal_single machine}
&\min \sum_{j}\int_{r_j}^{\infty}g_j(t)x_j(t)dt\cr 
\mbox{subject to}&\qquad \int_{r_j}^{\infty}\frac{x_j(t)}{v_j}\ge 1, \ \forall j\cr 
&\qquad \sum_{j}x_j(t)\leq 1, \forall t\cr 
&\qquad x_j(t)\ge 0, \ \forall j, t.
\end{align}
Here, the first constraint implies that every job $j$ must be fully processed. The second constraint ensures that the machine has a unit processing speed at each time instance $t$. Finally, the integral constraints $x_j(t)\in \{0,1\}, \forall j,t$, which are necessary to ensure that at each time instance at most one job can be processed, are relaxed to $x_j(t)\ge 0, \forall j, t$. The dual of the LP \eqref{eq:H_primal_single machine} is given by
\begin{align}\label{eq:H_dual_single machine}
&\max \sum_{j}\alpha_j-\int_{0}^{\infty}\beta_t dt\cr 
\mbox{subject to} &\qquad \frac{\alpha_j}{v_j}\leq \beta_t+g_j(t), \ \forall j,t\ge r_j\cr 
&\qquad \alpha_j,\beta_t\ge 0, \ \forall j, t.
\end{align}
Therefore, our goal in solving the GFCS problem is to devise an online algorithm whose objective cost is competitive with respect to the optimal offline LP cost \eqref{eq:H_primal_single machine}.    

\subsection{A Special Homogeneous Case}\label{sec:single-g(t)}

In this section, we consider a homogeneous version of the GFCS problem, namely HGFCS, which is for the specific cost functions $g_j(t)=w_jg(t)$, where $w_j\ge 0$ is a constant weight reflecting the importance of job $j$, and $g(t)$ is a general nonnegative nondecreasing function. Again by some abuse of notation, the scaled cost function is given by $g_j(t)=\rho_jg(t)$, where $\rho_j:=\frac{w_j}{v_j}$ denotes the density of job $j$. The assumption that $g_j(r_j)=0, \forall j$ requires $g(r_j)=0, \forall j$. However, this relation can be assumed without loss of generality by shifting each function to $g_j(t)=\rho_j(g(t)-g(r_j))$. This change only adds a constant term $\sum_{j}w_jg(r_j)$ to the objective cost. If we rewrite \eqref{eq:H_primal_single machine} and \eqref{eq:H_dual_single machine} for this special class of cost functions, we obtain, 
\begin{align}\label{eq:PD_special_single}
&\min \sum_{j}\int_{r_j}^{\infty}\rho_jg(t)x_j(t)dt\ \ \ \ \ \ \ \ \ \ \ \ \ \ \ \ \ \ \ \  \qquad\qquad \max \sum_{j}\alpha_j-\int_{0}^{\infty}\beta_t dt\cr 
\mbox{subject to} &\qquad \int_{r_j}^{\infty}\frac{x_j(t)}{v_j}\ge 1, \ \forall j \ \ \ \ \ \ \ \ \ \ \ \ \ \ \ \ \ \ \mbox{subject to}  \qquad \frac{\alpha_j}{v_j}\leq \beta_t+\rho_jg(t), \ \forall j,t\ge r_j\cr 
&\qquad \sum_{j}x_j(t)\leq 1, \forall t \ \ \ \ \ \ \ \ \ \ \ \ \ \ \ \ \ \ \ \ \ \ \ \ \ \ \ \ \             \qquad\qquad \alpha_j,\beta_t\ge 0, \ \forall j, t \cr 
&\qquad x_j(t)\ge 0, \ \forall j, t.
\end{align}

Next, in order to obtain an \emph{optimal online} schedule for the HGFCS problem, we generate an integral feasible solution (i.e., $x_j(t)\in \{0,1\}$) to the primal LP \eqref{eq:PD_special_single} together with a feasible dual solution of the same objective cost. The integral feasible solution is obtained simply by following the highest density first (HDF) schedule: \emph{among all the alive jobs, process the one that has the highest density}. More precisely, if the set of alive jobs at time $t$ is denoted by 
\begin{align}\nonumber
\mathcal{A}(t):=\{j: v_j(t)>0, t\ge r_j\},
\end{align}
the HDF rule schedules the job $\arg\max_{j\in A(t)}\rho_j$ at time $t$, where ties are broken arbitrarily.  


Next, let us apply the HDF rule on the original instance with $n$ jobs, and let $\cup_{\ell=1}^{k_j}[t_{j_{\ell}}, t_{j_{\ell}}+\bar{v}_{j_{\ell}}]$ be the disjoint time intervals in which job $j$ is being processed.  Here $0\leq t_{j_{k_j}}<\ldots<t_{j_2}<t_{j_1}$ are the time instances at which the HDF schedule preempts execution of other jobs in order to process job $j$,  and $\bar{v}_{j_{k_j}},\ldots,\bar{v}_{j_1}$ are the length portions of job $j$ that are processes between those consecutive preemption times. In particular, we note that $\sum_{\ell=1}^{k_j}\bar{v}_{j_{\ell}}=v_j$. We define the \emph{split} instance to be the one with $N:=\sum_{j=1}^{n}k_{j}$ jobs, where the $k_j$ jobs (subjobs in the original instance) associated with job $j$ have the same density $\rho_j$, lengths $\bar{v}_{j_1},\ldots,\bar{v}_{j_{k_j}}$, and release times $t_{j_1},\ldots,t_{j_{k_j}}$. The motivation for introducing the split instance is that we do not need to worry about time instances at which a job is interrupted/resumed because of arrival/completion of newly released jobs.  Therefore, instead of keeping a record of the time instances at which the HDF schedule preempts a job in the original instance, we can treat each subjob separately as a new job in the split instance.  This allows us to easily generate a dual optimal solution for the split instance and then convert it into an optimal dual solution for the original instance. 


\begin{lemma}\label{lemm:split}
Let $\mathcal{I}_o$ be an original instance of the HGFCS problem with the corresponding split instance $\mathcal{I}_s$.  Then, the fractional completion time of HDF for both $\mathcal{I}_o$ and $\mathcal{I}_s$ is the same. In particular, HDF is an optimal online schedule for $\mathcal{I}_s$ with respect to the fractional completion time, and the optimal dual variables for $\mathcal{I}_s$ can be obtained in a closed-form.
\end{lemma}
\emph{Proof:} As HDF performs identically on both the split and original instances, the cost of HDF on both instances is the same.  In the split instance $\mathcal{I}_s$,  each new job (which would have been a subjob in the original instance $\mathcal{I}_o$) is released right after completion of the previous one. The order in which these jobs are released is exactly the one dictated by the HDF. Therefore, any work-preserving schedule (and, in particular, HDF rule) that uses the full unit processing power of the machine would be optimal for the split instance. Furthermore, we can fully characterize the optimal dual solution for the split instance in a closed form. To see this, let us relabel all the $N:=\sum_{j=1}^{n}k_{j}$ jobs in increasing order of their processing intervals by $1,2,\ldots,N$. Then,
\begin{align}\label{eq:beta-alpha-1}
&\bar{\beta}_t=\rho_k(g(t_k+\bar{v}_k)-g(t))+\sum_{j=k+1}^{N}\rho_{j}(g(t_j+\bar{v}_j)-g(t_j)), \ \ \mbox{if}\ \ t\in[t_k, t_k+\bar{v}_k],\cr 
&\bar{\alpha}_{k}=\bar{v}_{k}(\bar{\beta}_{t_{k}}+\rho_kg(t_{k})), 
\end{align}  
form optimal dual solutions to the split instance 
\begin{align}\label{eq:split-dual}
\max \{\sum_{k=1}^{N}\bar{\alpha}_{k}-\int_{0}^{\infty}\bar{\beta}_t dt: \ \frac{\bar{\alpha}_{k}}{\bar{v}_{k}}\leq \bar{\beta}_t+\rho_kg(t), \ \bar{\alpha}_{k},\bar{\beta}_t\ge 0, \ \forall k,t\ge t_{k}\}.
\end{align}
Intuitively, $\bar{\beta}_t$ in \eqref{eq:beta-alpha-1} is a piecewise decreasing function of time $t$, which captures the time variation of the fractional completion time for $\mathcal{I}_s$ as one follows the HDF schedule. In response,  $\bar{\alpha}_{k}$ is chosen to satisfy the complementary slackness condition with respect to the choice of $\bar{\beta}_t$.  More precisely, by the definition of dual variables in \eqref{eq:beta-alpha-1}, the dual constraint $\frac{\bar{\alpha}_{k}}{\bar{v}_{k}}\leq \bar{\beta}_t+\rho_kg(t)$ is satisfied by equality for the entire time period $[t_{k},t_{k}+\bar{v}_{k}]$, during which job $k$ is scheduled. To explain why, we note that $x_{k}(t)=1$ for $t\in [t_{k},t_{k}+v_{k}]$. Thus, for any such $t$ and using \eqref{eq:beta-alpha-1}, we have,
\begin{align}\nonumber
&\bar{v}_k[\bar{\beta}_{t}+\rho_kg(t)]=\bar{v}_k[\rho_k(g(t_k\!+\!\bar{v}_k)-g(t))+\!\!\!\!\sum_{j=k+1}^{N}\!\!\rho_{j}(g(t_j\!+\!\bar{v}_j)-g(t_j))+\rho_kg(t)]=\bar{v}_k[\bar{\beta}_{t_k}+\rho_kg(t_k)]=\bar{\alpha}_k.
\end{align}  
Therefore, the dual constraint is tight whenever the corresponding primal variable is positive, which shows that the dual variables in \eqref{eq:beta-alpha-1} together with the integral primal solution generated by the HDF produce an optimal pair of primal-dual solutions to the split instance.\hfill $\blacksquare$  

\begin{definition}
We refer to diagrams of the optimal dual solutions \eqref{eq:beta-alpha-1} in the split instance as the $\alpha$-plot and $\beta$-plot. More precisely, in both plots, the $x$-axis represents the time horizon partitioned into the time intervals $\cup_{k=1}^N[t_{k}, t_{k}+\bar{v}_{k}]$ with which HDF processes subjobs. In the $\alpha$-plot, we draw a horizontal line segment at the height $\frac{\bar{\alpha}_{k}}{\bar{v}_{k}}$ for subjob $k$ and within its processing interval $[t_{k}, t_{k}+\bar{v}_{k}]$. In the $\beta$-plot we simply plot $\bar{\beta}_t$ as a function of time. We refer to line segments of the subjobs $k=j_{\ell}$ that are associated with job $j$ as $j$-steps (see Example \ref{ex:alpha-beta}).
\end{definition}

Next, in Algorithm \ref{alg:update}, we describe a simple backward process for converting optimal dual solutions $(\bar{\alpha}, \bar{\beta}_t)$ in the split instance to optimal dual solutions $(\alpha, \beta_t)$ for the original HGFCS instance with the same objective cost. The correctness of Algorithm \ref{alg:update} is shown in Lemma \ref{lemm:convert} whose proof is given in Appendix II.

\begin{algorithm}[H]
\caption{A Dual Update Process for HGFCS}\label{alg:update}
{\bf Input:} The $\alpha,\beta$-plots obtained from the optimal split instance,  given by \eqref{eq:beta-alpha-1}.  
 
\noindent 
{\bf Output:} Optimal dual $\alpha,\beta$-plots for the HGFCS problem.

\noindent
Given $\alpha,\beta$-plots obtained from the optimal split instance, update these plots sequentially by moving backward over the steps (i.e., from right to left) until time $t=0$ as follows:
\begin{itemize}
\item (1) If the current step $k$ is the \emph{first} $j$-step visited from the right, i.e., $k=j_{1}$, reduce its height to $h_j=\min_{t\ge t_{k}}\{\bar{\beta}_t+\rho_kg(t)\}$, and fix it as a \emph{reference} height for job $j$. Otherwise, if $k=j_{\ell}, \ell\ge 2$, reduce the height of step $k$ to its previously set reference height $h_j$.
\item (2) Reduce the height of \emph{all} other unupdated steps on the left side of step $k$ by $\delta_{k}$, where $\delta_{k}$ denotes the height decrease of the current step $k$.  Update the $\beta$-plot accordingly by lowering the value of $\bar{\beta}_t$ by the same amount $\delta_{k}$ for all times prior to the current step (i.e., $\forall t\in[0, t_{k}+\bar{v}_{k}]$).
\end{itemize}
\end{algorithm}

\begin{lemma}\label{lemm:convert}
The reference heights $\{h_j\}_{j=1}^n$ and $\{\beta_t\}_{t\ge 0}$ values obtained from $\alpha,\beta$-plots at the end of Algorithm \ref{alg:update} form feasible dual solutions $(\{\frac{\alpha_j}{v_j}:=h_j\}_{j=1}^{n}, \{\beta_t\}_{t\ge 0})$ to the original online HGFCS instance whose dual cost equals to the optimal cost of the splitted instance OPT.  
\end{lemma}

\begin{figure}[t]
\vspace{-2cm}
\begin{center}
\includegraphics[totalheight=.32\textheight,
width=.45\textwidth,viewport=250 0 850 550]{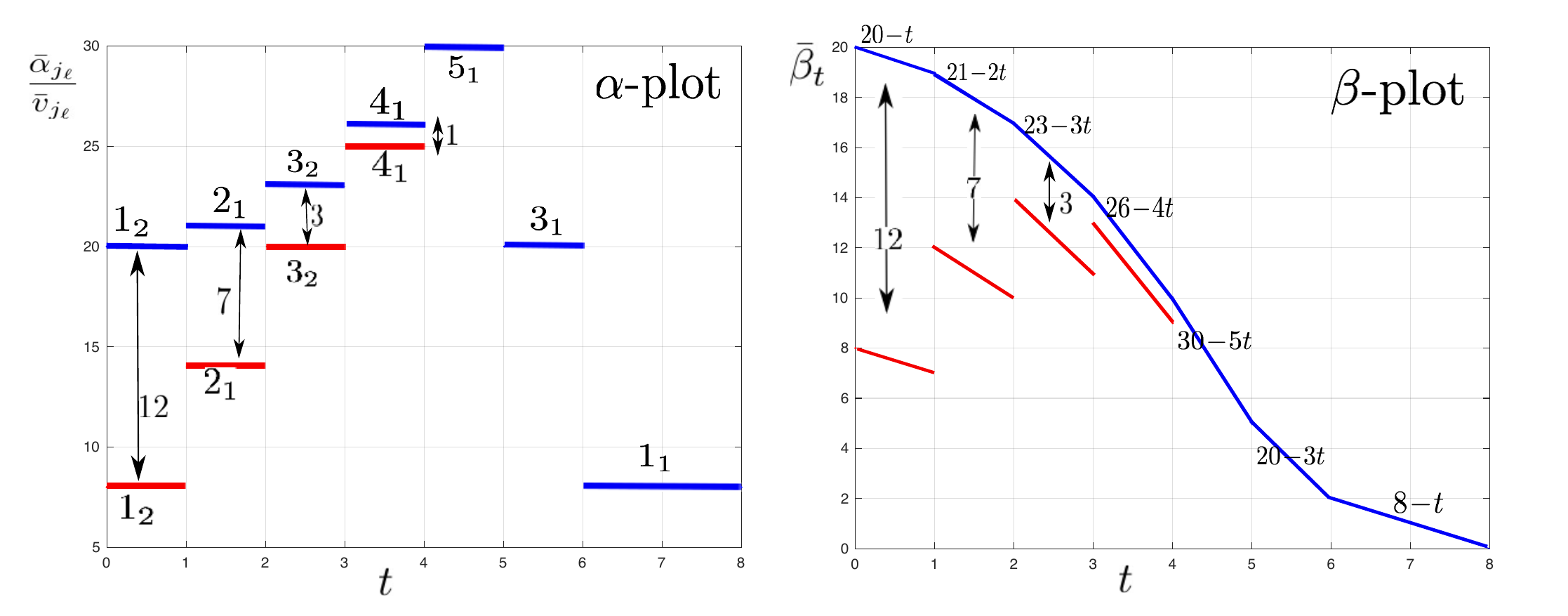} \hspace{0.4in}
\end{center}\vspace{-0.2cm}
\caption{An illustration of the $\alpha,\beta$-plots in Example \ref{ex:alpha-beta}. The blue line segments on the left figure correspond to steps (i.e., subjobs) in the optimal split instance. The red line segments are those associated with the optimal dual variables in the original HGFCS instance, which are obtained at the end of Algorithm \ref{alg:update}. The right figure illustrates the optimal $\beta$-curves for the split and original instances.}\label{fig:alpha-beta}
\end{figure}

\begin{example}\label{ex:alpha-beta}
Consider an (original) instance of the HGFCS with jobs' lengths $\{v_1=3, v_2=1, v_3=2, v_4=1, v_5=1\}$, release times $\{r_1=0, r_2=1, r_3=2, r_4=3, r_5=4\}$, and densities $\{\rho_1=1, \rho_2=2, \rho_3=3, \rho_4=4, \rho_5=5\}$. Moreover, assume that $g(t)=t$ so that the HGFCS problem reduces to the standard fractional completion time problem: 
\begin{align}\nonumber
\min\{\sum_{j}\rho_j\int_{r_j}^{\infty}tx_j(t)dt: \int_{r_j}^{\infty}\frac{x_j(t)}{v_j}\ge 1, \ \sum_{j}x_j(t)\leq 1, \ x_j(t)\ge 0, \ \forall j, t\}.
\end{align}
Now, if we apply HDF on this instance, we get a split instance with 7 subjobs: two $1$-steps of lengths $\bar{v}_{1_1}=2, \bar{v}_{1_2}=1$, which are scheduled over time intervals $[6, 8]$ and $[0,1]$; a $2$-step of length $\bar{v}_{2_1}=1$, which is scheduled over $[1,2]$; two $3$-steps of equal length $\bar{v}_{3_1}=\bar{v}_{3_2}=1$, which are scheduled over intervals $[5,6]$ and $[2,3]$; one $4$-step of length $\bar{v}_{4_1}=1$, which is scheduled over $[3,4]$; and, finally, one $5$-step of length $\bar{v}_{5_1}=1$, which is scheduled over $[4,5]$. These steps for the split instance are illustrated by blue line segments in the $\alpha$-plot in Figure \ref{fig:alpha-beta}. The corresponding optimal $\beta$-plot for the split instance is also given by the continuous blue curve in Figure \ref{fig:alpha-beta}, which was obtained from \eqref{eq:beta-alpha-1}. Now, moving backward in time over the steps, we set $1_1, 3_1,5_1$ as the reference steps for jobs $1,3$, and $5$, respectively. Note that by Algorithm \ref{alg:update}, these steps do not need to be lowered. However, step $4_1$ will be lowered by one unit and set as the reference height for job $4$. Consequently, all the steps before $4_1$ will be lowered by one unit in both the $\alpha$-plot and $\beta$-plot. Continuing in this manner by processing all the remaining steps $3_2,2_1,1_2$, we eventually obtain the red steps in the $\alpha$-plot and the red piecewise curves in the $\beta$-plot which correspond to the optimal dual solutions of the original instance. Note that at the end of this process, all the steps corresponding to a job are set to the same reference height. For instance, the two subjobs $3_1$ and $3_2$ are set to the reference height $20$, i.e., $\frac{\bar{\alpha}_{3_1}}{\bar{v}_{3_1}}=\frac{\bar{\alpha}_{3_2}}{\bar{v}_{3_2}}=20$.\hfill $\blacksquare$       
\end{example}

\begin{theorem}\label{thm:HDF-single}
HDF is an optimal online algorithm for the HGFCS problem with cost functions $g_j(t)=\rho_jg(t)$, where $g(t)$ is an arbitrary nonnegative nondecreasing function.
\end{theorem}

\emph{Proof:} Consider the split instance obtained by applying HDF on the original online HGFCS instance with $n$ jobs. From Lemma \ref{lemm:split}, HDF is an optimal schedule for the split instance whose optimal cost OPT equals the cost of HDF on the original HGFCS instance. Let $(\{\bar{\alpha}_k\}_{k=1}^{N}, \bar{\beta}_t)$ be the optimal dual solution to the split instance. Using Algorithm \ref{alg:update}, one can convert $(\{\bar{\alpha}_k\}_{k=1}^{N}, \bar{\beta}_t)$ to a feasible dual solution $(\{\alpha_j\}_{j=1}^n,\beta_t)$ for the original HGFCS instance with the same objective cost OPT (Lemma \ref{lemm:convert}). Thus, the solution generated by HDF together with $(\{\alpha_j\}_{j=1}^{n},\beta_t)$ forms a feasible primal-dual solution to the original HGFCS instance with the same cost OPT. Therefore, by strong duality, HDF is an optimal online schedule for the original HGFCS instance.\hfill $\blacksquare$
          
Algorithm \ref{alg:update} provides a simple update rule for generating optimal dual variables for HGFCS with special cost functions $g_j(t)=\rho_jg(t)$, that is when all the cost functions share the same basis function $g(t)$. Unfortunately, it quickly becomes intractable when one considers general cost functions $g_j(t)$.  The main reason is that for general cost functions $g_j(t)$, it is known that optimal job scheduling on a single machine is NP-hard even for the offline setting \cite{bansal2014geometry}.  In particular, for general cost functions,  one first needs to construct the entire dual curves and then use them to determine the optimal reference heights. That requires exponential computation in terms of the number of jobs. However, one can still use a backward induction similar to that in Algorithm \ref{alg:update} to set the dual variables as close as possible to their optimal values,  and that is the main idea behind our generalized competitive analysis for GFCS.

More precisely, a closer look at the structure of Algorithm \ref{alg:update} shows that it mimics a dynamic programming update that starts from a dual solution, namely the optimal dual solution of the split instance, and moves backward to fit it into the dual of the original HGFCS instance. This observation suggests that one can formulate the offline GFCS problem as an optimal control problem in which the Hamilton-Jacobi-Bellman (HJB) equation plays the role of Algorithm \ref{alg:update} above and tells us how to fit the dual variables as closely as possible to the dual of the offline GFCS problem. However, a major challenge here is that for the GFCS problem, jobs can arrive  online over time. To overcome that, we use the insights obtained from Algorithm \ref{alg:update} to approximately determine the structure of the optimal $\beta$-curve.  From the red discontinuous curve in the $\beta$-plot of Figure \ref{fig:alpha-beta}, it can be seen that the optimal $\beta$-plot has discontinuous jumps whenever a new job arrives in the system. To mimic that behavior in the general setting,  we proceed as follows.  Upon arrival of a new job $n$ at time $r_n$,  we may need to update the online schedule for future times $t\ge r_n$.  However, since an online scheduler does not know anything about future job arrivals,  we assume that the alive jobs at time $r_n$ are the only ones in the system, which are all available at the same time $r_n$.  In other words, we consider an offline instance of GFCS with identical release times $r_n$, henceforth referred to as GFCS$(r_n)$.  By solving GFCS$(r_n)$ using optimal control,  we update the schedule for $t>r_n$. In this fashion, one only needs to iteratively solve offline optimal control problems with identical release times, as described in the next section.

\section{An Optimal Control Formulation for the Offline GFCS Problem with Identical Release Times}\label{sec:optimal-control}

In this section, we cast the offline GFCS problem with \emph{identical} release times $r_n$ as an optimal control problem.  By abuse of notation, we refer to this problem either in an LP form or an optimal control form as GFCS$(r_n)$.  Consider the time $r_n$ when a new job $n$ is released to the system, and let $\mathcal{A}(r_n)$ be the set of currently alive jobs (excluding job $n$). Now, if we assume that no new jobs are going to be released in the future, then an optimal schedule must solve an offline instance with a set of jobs $\mathcal{A}(r_n)\cup\{n\}$ and identical release times $r_n$, where the length of job $j\in \mathcal{A}(r_n)\cup \{n\}$ is given by its residual length $v_j(r_n)$ at time $r_n$. (Note that for job $n$, we have $v_j(r_n)=v_n$, as this job is released at time $r_n$.) Since the optimal cost of this offline instance depends on the residual lengths of the alive jobs, we shall refer to those residual lengths as \emph{states} of the system at time $r_n$. More precisely, we define the state of job $j$ at time $t$ to be the residual length $v_j(t)$ of that job at time $t$, and the state vector to be $\boldsymbol{v}(t)=(v_1(t),\ldots,v_k(t))^T$, where $k=|\mathcal{A}(r_n)|+1$. Note that since GFCS$(r_n)$ assumes no future arrivals, the dimension of the state vector does not change and equals the number of alive jobs at time $r_n$.

Let us define the \emph{control} input at time $t$ to be $\boldsymbol{x}(t)=(x_1(t),\ldots,x_k(t))^T$, where $x_j(t)$ is the rate at which job $j$ is processed at time $t$. Thus, $v_j(t)=v_j(r_n)-\int_{r_n}^tx_j(\tau)d\tau$, or, equivalently, $\dot{v}_j(t)=-x_j(t)$, with the initial condition $v_j(r_n)$. If we write those equations in a vector form, we obtain
\begin{align}\label{eq:state-dynamics}
\dot{\boldsymbol{v}}(t)=-\boldsymbol{x}(t), \ \boldsymbol{v}(r_n)=(v_1(r_n),\ldots,v_k(r_n))^T.
\end{align}  
Moreover, because of the second primal constraints in \eqref{eq:H_primal_single machine}, we note that at any time, the control vector $\boldsymbol{x}(t)$ must belong to the simplex $\mathcal{X}:=\{u\in[0,1]^k: \sum_{i=1}^ku_j=1\}$. Thus, an equivalent optimal control formulation for \eqref{eq:H_primal_single machine} with identical release times $r_n$ and initial state $\boldsymbol{v}(r_n)$ is given by 
\begin{align}\nonumber
\min &\sum_{j\in \mathcal{A}(r_n)}\int_{r_n}^{\infty}g_j(t)x_j(t)dt, \cr 
 \mbox{subject to}\qquad&\dot{\boldsymbol{v}}(t)=-\boldsymbol{x}(t), \ \boldsymbol{v}(r_n)=(v_1(r_n),\ldots,v_k(r_n))^T, \ \boldsymbol{v}(\infty)=0,\cr 
 &\boldsymbol{x}(t)\in \mathcal{X}, \forall t, 
\end{align}             
where, as before, $g_j(t)$ refers to the original cost function scaled by $\frac{1}{v_j}$. Note that for any $t$, the loss function $\sum_jg_j(t)x_j(t)$ is nonnegative. As $g_j(t)$s can only increase over time, any optimal control $\boldsymbol{x}^o(t)$ must finish the jobs in the finite time interval $[r_n, t_1], \ t_1:=r_n+\sum_jv_j(r_n)$. Thus, without loss of generality, we can replace the upper limit in the integral with $t_1$, which gives us the following optimal control formulation for GFCS$(r_n)$:
\begin{align}\label{eq:optimal_control}
\min &\int_{r_n}^{t_1}\sum_{j\in \mathcal{A}(r_n)}g_j(t)x_j(t)dt, \cr 
 \mbox{subject to}\qquad&\dot{\boldsymbol{v}}(t)=-\boldsymbol{x}(t), \ \boldsymbol{v}(r_n)=(v_1(r_n),\ldots,v_k(r_n))^T, \ \boldsymbol{v}(t_1)=0,\cr 
 &\boldsymbol{x}(t)\in \mathcal{X}, \forall t\in [0, t_1]. 
\end{align}
It is worth noting that we do not need to add nonnegativity constraints $\boldsymbol{v}(t)\ge 0$ to \eqref{eq:optimal_control}, as they implicitly follow from the terminal condition. The reason is that if $v_j(t)< 0$ for some $j$ and $t\leq t_1$, then, as $x_j(t)\ge 0$, the state can only decrease further and remains negative forever, violating the terminal condition $\boldsymbol{v}(t_1)=0$. Therefore, specifying that $\boldsymbol{v}(t_1)=0$ already implies that $\boldsymbol{v}(t)\ge 0, \forall t$. 


\subsection{Solving GFCS$(r_n)$ Using the Minimum Principle}
In this section, we proceed to solve the optimal control problem \eqref{eq:optimal_control} by using the minimum principle.  We first state the following general theorem from optimal control theory that will allow us to characterize the structure of optimal solution to GFCS$(r_n)$ \cite[Theorem 11.3]{basar2020lecture}:

\begin{theorem}[Minimum Principle]\label{thm:min-principle}
Consider the general optimal control problem:
\begin{align}\label{eq:min-princ-thm}
\min_{u(t)\in \mathcal{U}, \forall t}\big\{\int_{t_0}^{t_1}\ell(\boldsymbol{y}(t),\boldsymbol{u}(t),t)dt:\ \ \dot{\boldsymbol{y}}(t)=f(\boldsymbol{y}(t),\boldsymbol{u}(t),t), \ \boldsymbol{y}(t_0)=\boldsymbol{y}_0, \ \boldsymbol{y}(t_1)=\boldsymbol{0}\big\},
\end{align}
where $\boldsymbol{y}_0\in \mathbb{R}^{d_1}$ is the initial state, $\mathcal{U}\subseteq \mathbb{R}^{d_2}$ is the control constraint set,  and $\ell:\mathbb{R}^{d_1+d_2+1}\to\mathbb{R}$ is a scalar-valued cost function of the state vector $\boldsymbol{y}(t)$, control input $\boldsymbol{u}(t)$, and time $t$.  Consider the Hamiltonian function $H(\boldsymbol{y}, \boldsymbol{u},\boldsymbol{p}, t)=\boldsymbol{p}^T
f(\boldsymbol{y}, \boldsymbol{u},t)+\ell(\boldsymbol{y}, \boldsymbol{u},t)$,  and suppose that $\boldsymbol{u}^o(t)$ is the optimal control solution to \eqref{eq:min-princ-thm}. Then, there exists a costate vector $\boldsymbol{p}(t)$ such that $\forall t\in [t_0, t_1]$, $\boldsymbol{u}^o(t)=\argmin_{\boldsymbol{u}\in \mathcal{U}}H(\boldsymbol{y}^o(t), \boldsymbol{u},\boldsymbol{p}(t), t)$ and $\dot{\boldsymbol{p}}(t)=-\nabla_{\boldsymbol{y}}H(\boldsymbol{y}^o(t), \boldsymbol{u}^o(t),\boldsymbol{p}(t), t)$, where $\boldsymbol{y}^o(t)$ is the optimal state trajectory corresponding to $\boldsymbol{u}^o(t)$, i.e.,   $\dot{\boldsymbol{y}^o}(t)=f(\boldsymbol{y}^o(t),\boldsymbol{u}^o(t),t)$ with $\boldsymbol{y}^o(t_0)\!=\!\boldsymbol{y}_0, \boldsymbol{y}^o(t_1)\!=\!\boldsymbol{0}$.
\end{theorem}

As the minimum principle can be viewed as an infinite-dimensional generalization of the Lagrangian duality \cite{liberzon2011calculus}, in the following, the readers who are more familiar with nonlinear programming can think of the \emph{costate} vector as the Lagrangian multipliers, the \emph{Hamiltonian} as the Lagrangian function, and the minimum principle as the saddle point conditions.  By specializing Theorem \ref{thm:min-principle} to our problem setting, one can see that the corresponding Hamiltonian for \eqref{eq:optimal_control} with a costate vector $\boldsymbol{p}(t)$ is given by $H(\boldsymbol{v},\boldsymbol{x},\boldsymbol{p}):=\sum_jg_j(t)x_j(t)-\boldsymbol{p}^T(t)\boldsymbol{x}(t)$.  If we write the minimum principle conditions, we obtain: 
\begin{align}\nonumber
&\dot{p}_j(t)=-\frac{\partial}{\partial v_j}H(\boldsymbol{v},\boldsymbol{x},\boldsymbol{p})=0, \ \forall j \ \ \ \Rightarrow\ \ \  p_j(t)=p_j(t_1), \ \forall j,t\ge r_n\cr 
& \boldsymbol{x}^o(t):=\arg\min_{\boldsymbol{x}\in\mathcal{X}}\{\sum_{j\in \mathcal{A}(r_n)} g_j(t)x_j(t)-\boldsymbol{p}^T(t)\boldsymbol{x}(t)\} \ \ \Rightarrow \ \ x^o_j(t)=
\begin{cases}
1, \ \ \mbox{if} \ j=\arg\max_k (p_k(t_1)-g_k(t))\\
0, \ \ \mbox{else}.
\end{cases} 
\end{align} 
Therefore, for every $j\in \mathcal{A}(r_n)$, the minimum principle optimality conditions for \eqref{eq:optimal_control} with free terminal time $t_1$ and fixed endpoints are given by
\begin{align}\label{eq:costate}
&p_j(t)=p_j(t_1), \ \ \ \forall j, t\ge r_n\cr 
&\dot{v}^o_j(t)=-x^o_j(t), \ \ \forall j, t\ge r_n\cr 
&x^o_j(t)=
\begin{cases}
1, \ \ \mbox{if} \ j=\arg\max_k (p_k(t_1)-g_k(t))\\
0, \ \ \mbox{else},
\end{cases} 
\end{align}
with the boundary conditions $v_j(r_n), v_j(t_1)=0$, and $\sum_j(g_j(t_1)-p_j(t_1))x_j(t_1)=0$. Therefore, we obtain the following corollary about the structure of the optimal offline schedule for GFCS$(r_n)$:
\begin{corollary}\label{corr:curve}
The optimal offline schedule for GFCS$(r_n)$ is obtained by plotting all the job curves $\{p_k(t_1)-g_k(t),\ k\in\mathcal{A}(r_n)\}$, and, at any time $t$, scheduling the job $j$ whose curve $p_j(t_1)-g_j(t)$ determines the upper envelope of all other curves at that time.
\end{corollary}


In order to use Corollary \ref{corr:curve}, one first needs to determine the costate constants $p_k(t_1), k\in \mathcal{A}(r_n)$ (although in some special cases knowing the exact values of $p_k(t_1)$ is irrelevant in determining the structure of the optimal policy, see, e.g., Proposition 1).  These constants can be related to the optimal $\alpha$-dual variables in \eqref{eq:H_dual_single machine} assuming identical release times $r_n$. To see that, let us define $\beta_t:=p_j(t_1)-g_j(t)$ if at time $t$ job $j$ is scheduled, and $\frac{\alpha_k}{v_k(r_n)}:=p_k(t_1), \forall k\in\mathcal{A}(r_n)$. Then, for all the time instances $I_j$ at which job $j$ is scheduled, we have
\begin{align}\nonumber
\beta_t+g_j(t)=p_j(t_1)-g_j(t)+g_j(t)=p_j(t_1)=\frac{\alpha_j}{v_j(r_n)}, \ \ \forall t\in I_j,
\end{align} 
which shows that the dual constraint in \eqref{eq:H_dual_single machine} must be tight. Thus, if we define the dual variables in terms of costates as above, the complementary slackness conditions will be satisfied. Moreover, given an arbitrary time $t\in I_j$ at which job $j$ is scheduled, from the last condition in \eqref{eq:costate}, we have $p_j(t_1)-g_j(t)\ge p_k(t_1)-g_k(t), \forall k$. As we defined $\beta_t=p_j(t_1)-g_j(t)$, we have,
\begin{align}\nonumber
\frac{\alpha_k}{v_k(r_n)}=p_k(t_1)\leq p_j(t_1)-g_j(t)+g_k(t)=\beta_t+g_k(t),
\end{align} 
which shows that the above definitions of dual variables in terms of costates are also dual feasible, and hence must be optimal. As a result, we can recover optimal dual variables from the costate curves and vice versa. In the next section, we will use the HJB equation to determine costate constants $p_j(t_1), \forall j$ in terms of the variations in the optimal value function.  



\subsection{Determining Optimal Dual Variables for GFCS$(r_n)$}
Here, we consider the problem of determining costate constants and hence optimal offline dual variables. To that aim, let us define 
\begin{align}\nonumber
V^o(\boldsymbol{v},t)=\min_{\boldsymbol{x}[t,t_1]\in \mathcal{X}}\Big\{\int_{t}^{t_1}\!\!\!\sum_{j\in \mathcal{A}(t)}g_j(\tau)x_j(\tau)d\tau:\  \dot{\boldsymbol{v}}(\tau)=-\boldsymbol{x}(\tau), \ \boldsymbol{v}(t)=(v_1,\ldots,v_k)^T, \ \boldsymbol{v}(t_1)=0\Big\},
\end{align} 
as the optimal value function for the optimal control problem \eqref{eq:optimal_control}, given initial state $\boldsymbol{v}$ at initial time $t$, where the minimum is taken over all control inputs $\boldsymbol{x}[t,t_1]$ over the time interval $[t,t_1]$ such that $x(\tau)\in\mathcal{X}, \forall \tau\in [t,t_1]$. It is shown in \cite[Section 11.1]{basar2020lecture} and \cite[Section 5.2]{liberzon2011calculus} that at any point of differentiability of the optimal value function, the costate obtained from the minimum principle must be equal to the gradient of the optimal value function with respect to the state variable, i.e., $\boldsymbol{p}(t)=\frac{\partial}{\partial \boldsymbol{v}}V^o(\boldsymbol{v}^o(t),t)$, where $\boldsymbol{v}^o(t)$ denotes the optimal state trajectory obtained by following the optimal control $\boldsymbol{x}^o(t)$. As before, let $I_j$ denote the set of time instances at which the optimal schedule processes job $j$, i.e., $x^o_j(t)=1, \forall t\in I_j$. As we showed that the optimal dual variable $\beta_t$ is given by $\beta_t=p_j(t)-g_j(t), \forall t\in I_j$, we can write,
\begin{align}\label{eq:beta-fit-first}
\beta_t=\frac{\partial}{\partial v_j}V^o(\boldsymbol{v}^o(t),t)-g_j(t), \ \forall t\in I_j.
\end{align}
On the other hand, if we write the HJB equation \cite[Chapter 10]{basar2020lecture} (see also \cite[Section 5.1.3]{liberzon2011calculus}) for the optimal control problem \eqref{eq:optimal_control}, for any initial time $t$ and any initial state $\boldsymbol{v}$, it is known that the optimal value function $V^o(\cdot)$ must satisfy the HJB equation given by
\begin{align}\label{eq:HJB-eq}
-\frac{\partial}{\partial t}V^o(\boldsymbol{v},t)=\min_{\boldsymbol{x}\in \mathcal{X}}\{\sum_{\ell\in \mathcal{A}(t)}\big(g_{\ell}(t)-\frac{\partial}{\partial v_{\ell}}V^o(\boldsymbol{v},t)\big)x_j\},
\end{align} 
where the minimum is achieved for a job $j$ with the smallest $g_{j}(t)-\frac{\partial}{\partial v_j}V^o(\boldsymbol{v},t)$. As a result, the optimal control is given by $x^o_j=1$, and $x^o_{\ell}=0\  \forall \ell\neq j$. Thus, if we write the HJB equation \eqref{eq:HJB-eq} along the optimal trajectory $\boldsymbol{v}^o(t)$ with the associated optimal control $\boldsymbol{x}^o(t)$, we obtain
\begin{align}\label{eq:HJB-simplified}
-\frac{\partial}{\partial t}V^o(\boldsymbol{v}^o(t),t)=g_{j}(t)-\frac{\partial}{\partial v_j}V^o(\boldsymbol{v}^o(t),t) \ \ \forall t\in I_j.
\end{align}
In view of \eqref{eq:beta-fit-first}, \eqref{eq:HJB-simplified} shows that the optimal dual variable $\beta_t$ is given by $\beta_t=\frac{\partial}{\partial t}V^o(\boldsymbol{v}^o(t),t), \forall t\in I_j$. Since the above argument holds for every $I_j$, we have
\begin{align}\label{eq:beta-fit}
\beta_t=\frac{\partial}{\partial t}V^o(\boldsymbol{v}^o(t),t)  \ \ \forall t.
\end{align}
Moreover, from using complementary slackness, we know that the dual constraint $\frac{\alpha_j}{v_j(r_n)}\leq \beta_t+g_j(t)$ is tight for every $t\in I_j$. That,  together with \eqref{eq:HJB-simplified} and \eqref{eq:beta-fit}, implies
\begin{align}\nonumber
\frac{\alpha_j}{v_j(r_n)}=\beta_t+g_j(t)=\frac{\partial}{\partial t}V^o(\boldsymbol{v}^o(t),t)+g_j(t)=\frac{\partial}{\partial v_j}V^o(\boldsymbol{v}^o(t),t), \forall t\in I_j. 
\end{align}
As a result, for every $t\in I_j$ the value of $\frac{\partial}{\partial v_j}V^o(\boldsymbol{v}^o(t),t)$ is a constant that equals the optimal $\alpha$-dual variable for the job $j$ that is currently being processed, i.e., 
\begin{align}\label{eq:alpha_fit-final}
\frac{\alpha_j}{v_j(r_n)}=\frac{\partial}{\partial v_j}V^o(\boldsymbol{v}^o(t),t)  \ \ \forall t\in I_j.
\end{align} 
In other words, the optimal dual variables $\frac{\alpha_j}{v_j(r_n)}$ and $\beta_t$ in the GFCS$(r_n)$ problem are equal to the sensitivity of the optimal value function with respect to the length of job $j$ that is currently being processed and the execution time $t$, respectively.  Here, the sensitivity of the value function with respect to a parameter (e.g., job length or time) refers to the amount of change in the optimal objective value of GFCS$(r_n)$ due to an infinitesimal change in that parameter.

\begin{example}
Consider an instance of GFCS$(r_n)$ with $r_n=0$ and two jobs of lengths $v_1(0)=1$ and $v_2(0)=2$. Moreover, let $g_1(t)=\rho_1t$ and $g_2(t)=\rho_2t$, where $\rho_1>\rho_2$. From the previous section we know that HDF is the optimal schedule for HGFC given those special cost functions. Therefore, the optimal value function is given by 
\begin{align}\label{eq:value-function-example}
V^o(v_1,v_2,t)=\int_{t}^{t+v_1}\!\!\!\!\!\rho_1\tau d\tau+\int_{t+v_1}^{t+v_1+v_2}\!\!\!\!\!\rho_2\tau d\tau=\rho_1\frac{v_1^2+2tv_1}{2}+\rho_2\frac{v_2^2+2v_2(t+v_1)}{2}.
\end{align}
Moreover, the optimal control is $x^o_1(t)=1$ if $t\in[0, 1)$, and $x^o_2(t)=1$ if $t\in[1, 3)$. Thus, the optimal state trajectory is given by
\begin{align}\label{eq:optimal-trejectory-example}
\boldsymbol{v}^o(t)=\begin{cases}
v_1^o(t)=1-t, \ v_2^o(t)=2, \ \ \ \ \ \ \mbox{if} \ t\in [0, 1)\\
v_1^o(t)=0, \ v_2^o(t)=3-t \ \ \ \ \ \ \ \mbox{if} \ t\in [1, 3).
\end{cases} 
\end{align} 
Now, using \eqref{eq:value-function-example} and \eqref{eq:optimal-trejectory-example}, we can write
\begin{align}\nonumber
\beta_t=\frac{\partial}{\partial t}V^o(\boldsymbol{v}^o(t),t)=(\rho_1v_1+\rho_2v_2)_{|_{(\boldsymbol{v}^o(t),t)}}=\begin{cases}
\rho_1(1-t)+2\rho_2, \ \ \ \ \mbox{if} \ t\in [0, 1),\\
\rho_2(3-t) \ \ \ \ \ \ \ \ \ \ \ \ \  \mbox{if} \ t\in [1, 3).
\end{cases} 
\end{align}
On the other hand, we have   
\begin{align}\nonumber
&\frac{\alpha_1}{1}= \frac{\partial}{\partial v_1}V^o(\boldsymbol{v}^o(t),t)=(\rho_1(v_1+t)+\rho_2v_2)_{|_{(\boldsymbol{v}^o(t),t)}}=\rho_1+2\rho_2, \ \ \forall t\in[0, 1),\cr
&\frac{\alpha_2}{2}= \frac{\partial}{\partial v_2}V^o(\boldsymbol{v}^o(t),t)=\rho_2(v_1+v_2+t)_{|_{(\boldsymbol{v}^o(t),t)}}=3\rho_2, \ \ \forall t\in[0, 1).
\end{align}
Now one can see that the above $\alpha_1, \alpha_2, \beta_t$ are optimal dual variables with optimal objective value
\begin{align}\nonumber
\alpha_1+\alpha_2-\int_{0}^{3}\beta_tdt=(\rho_1+2\rho_2)+6\rho_2-(\frac{\rho_1}{2}+4\rho_2)=\frac{\rho_1}{2}+4\rho_2.
\end{align}\hfill $\blacksquare$  
\end{example}

An advantage of using the optimal control framework is its simplicity in deriving good estimates on the optimal dual variables, which have been derived in the past literature using LP duality arguments \cite{anand2012resource}.  Unfortunately, for more complex nonlinear objective functions, adapting LP duality analysis seem difficult, while the optimal control method can still provide useful insights on how to set the dual variables (see, e.g., Appendix I). The idea of deriving such bounds is simple and intuitive. Specifically, by \eqref{eq:beta-fit}, we know that for a single machine $\beta_t=\frac{\partial}{\partial t}V^o(\boldsymbol{v}^o(t),t), \forall t$. Since a larger $\beta_t$ will always be in favor of dual feasibility, instead of finding $V^o(\boldsymbol{v}^o(t),t)$, which might be difficult, we find an upper bound for it. To that end, we can upper-bound $V^o(\boldsymbol{v}^o(t),t)$ by using the cost incurred by any feasible test policy (e.g., HDF) that is typically chosen to be a perturbation of the optimal offline policy. The closer the test policy is to the optimal one, the more accurate the dual bounds that can be obtained. We shall examine this idea in more detail in the subsequent sections.  


\section{A Competitive Online Algorithm for the GFCS Problem}\label{sec:heterogeneous-cost}
In this section, we consider the GFCS problem on a single machine whose offline LP relaxation and its dual are given by \eqref{eq:H_primal_single machine} and \eqref{eq:H_dual_single machine}, respectively. In the online setting, the nonnegative nondecreasing cost functions $g_j(t), j=1,2,\ldots$, are released at time instances $r_1,r_2,\ldots$, and our goal is to provide an online scheduling policy to process the jobs on a single machine and achieve a bounded competitive ratio with respect to the optimal offline LP cost \eqref{eq:H_primal_single machine}. 

As we saw in Section \ref{sec:optimal-control}, the GFCS$(r_n)$ problem can be formulated as the optimal control problem \eqref{eq:optimal_control}. Here, we show how to devise an online algorithm for the GFCS problem by repeated application of the offline problem. The online algorithm that we propose works in a greedy fashion as detailed in Algorithm \ref{alg:network-flow-single}. Intuitively, the online Algorithm \ref{alg:network-flow-single} always schedules jobs based on its most recent optimal offline policy (which assumes no future arrivals) until a new job $n$ is released at time $r_n$. At that time, the algorithm updates its scheduling policy by solving a new GFCS$(r_n)$ to account for the impact of the new job $n$. 


\begin{algorithm}
\caption{An Online Algorithm for the GFCS Problem}\label{alg:network-flow-single}
{\bf Input:} An instance of the GFCS problem $\{v_j,r_j, g_j(t), j=1,2,\ldots\}$, as defined in Section \ref{sec:problem-formulation}.   
 
\noindent 
{\bf Output:} An online schedule that determines what job must be processed at each time $t\ge 0$. 

\noindent
\begin{itemize}
\item Upon arrival of a new job $n$ at time $r_n$, let $\mathcal{A}'(r_n)$ denote the set of alive jobs at time $r_n$, including job $n$, with remaining lengths $\{v_j(r_n), j\in \mathcal{A}'(r_n)\}$. 
\item Solve the offline GFCS$(r_n)$ problem \eqref{eq:optimal_control} for jobs $\mathcal{A}'(r_n)$ and job lengths $\{v_j(r_n), j\in \mathcal{A}'(r_n)\}$.  Let $\boldsymbol{x}^o(t)$ denote the optimal solution to this offline problem, which from \eqref{eq:costate} it must be integral.
\item Schedule the jobs from time $r_n$ onward based on the integral optimal solution $\boldsymbol{x}^o(t), t\ge r_n$. 
\end{itemize}
\end{algorithm}


\noindent
{\bf Time Complexity of Algorithm \ref{alg:network-flow-single}:} The runtime of Algorithm \ref{alg:network-flow-single} is $O(nT_s)$, where $T_s$ is the time complexity of an offline oracle for solving GFCS$(r_n)$ given by \eqref{eq:optimal_control}.  For some special cases such as dominating class of cost functions (Proposition \ref{thm:dominance}), the existence of a polynomial-time oracle for solving GFCS$(r_n)$ is guaranteed.  However, for general nondecreasing cost functions $g_j(t)$ and arbitrary release times $r_j$, even finding the optimal offline schedule on a single machine is strongly NP-hard, with the best known $O(\log\log nP)$-approximation algorithm, where $n$ is the number of jobs, and $P$ is the maximum job length \cite{bansal2014geometry}.  Surprisingly,  if the jobs have identical release times (which is the case for GFCS$(r_n)$ in Algorithm \ref{alg:network-flow-single}),  the GFCS$(r_n)$ problem admits both a polynomial-time $(4+\epsilon)$-approximation algorithm \cite{cheung2017primal}, and a quasi-polynomial time $(1+\epsilon)$-approximation scheme \cite{antoniadis2017qptas}.  Thus,  if obtaining an exact solution to GFCS$(r_n)$ in Algorithm \ref{alg:network-flow-single} is not available, the approximation algorithms in \cite{cheung2017primal} and \cite{antoniadis2017qptas} serve as good proxies for obtaining near-optimal solutions with polynomial/quasi-polynomial complexity.

\smallskip
\noindent
{\bf Setting Dual Variables for Algorithm \ref{alg:network-flow-single}:} In order to analyze performance of Algorithm \ref{alg:network-flow-single},  we first describe a dual-fitting process, henceforth referred to as \emph{dual variables generated by Algorithm \ref{alg:network-flow-single},} which adaptively sets and updates the dual variables in response to the run of Algorithm \ref{alg:network-flow-single}.  More precisely,  upon arrival of a new job $n$ at time $r_n$, we set the $\alpha$-dual variable for the new job $n$ to its \emph{new} optimal value $\alpha'_n$, which is the one obtained by solving the offline dual program \eqref{eq:H_dual_single machine} with identical release times $r_n$ and job lengths $\{v_j(r_n), j\in \mathcal{A}'(r_n)\}$. Moreover,  we replace the \emph{tail} of the \emph{old} $\beta$-variable $\{\beta_t\}_{t\ge r_n}$ (i.e., the portion of $\beta$-dual variables that appear after time $t\ge r_n$) with the \emph{new} optimal dual variable $\{\beta'_t\}_{t\ge r_n}$ obtained by solving \eqref{eq:H_dual_single machine}.  All other dual variables $\{\alpha_j\}_{j\neq n}$ and $\{\beta_t\}_{t<r_n}$ are kept unchanged.  We note that the dual variables generated by Algorithm \ref{alg:network-flow-single} has nothing to do with the algorithm implementation and is given for the sake of analysis.

Next, we will show that setting the dual variables as described above indeed generate a feasible dual solution to the offline dual program \eqref{eq:H_dual_single machine} with \emph{different} release times $r_1,r_2,\ldots$, and initial job lengths $v_1,v_2,\ldots$. To that end, we first consider the following definition:

\begin{definition}\label{def:RNF}
For an arbitrary job $n$ with arrival time $r_n$,  we let $\mbox{RNF}(r_n)$ and $\mbox{RNF}'(r_n)$ denote the optimal values to GFCS$(r_n)$ with job lengths $v_j(r_n)$ in the absence and presence of job $n$, i.e., 
{\small\begin{align}\nonumber
&\mbox{RNF}(r_n):=\min \Big\{\!\!\!\sum_{j\in \mathcal{A}(r_n)}\!\int_{r_n}^{\infty}\!\!\!\!g_j(t)x_j(t)dt:\ \int_{r_n}^{\infty}\!\!\!\!\frac{x_j(t)}{v_j(r_n)}\ge 1, \forall j\in \mathcal{A}(r_n), \sum_{j\in\mathcal{A}(r_n)}\!\!\!x_j(t)\leq 1, \forall t,\ x_j(t)\ge 0\Big\},\cr 
&\mbox{RNF}'(r_n):=\min \Big\{\!\!\!\sum_{j\in \mathcal{A}'(r_n)}\!\int_{r_n}^{\infty}\!\!\!\!g_j(t)x_j(t)dt:\ \int_{r_n}^{\infty}\!\!\!\!\frac{x_j(t)}{v_j(r_n)}\ge 1, \forall j\in \mathcal{A}'(r_n), \sum_{j\in\mathcal{A}'(r_n)}\!\!\!x_j(t)\leq 1, \forall t,\ x_j(t)\ge 0\Big\},
\end{align}}where $\mathcal{A}'(r_n):=\mathcal{A}(r_n)\cup\{n\}$ denotes the set of alive jobs at time $r_n$, including job $n$. More generally, we distinguish the parameters associated with the new instance including job $n$, by adding a prime.
\end{definition}

In the subsequent sections, we find it more convenient to work with the discretized version of the quantities $\mbox{RNF}(r_n)$ and $\mbox{RNF}'(r_n)$, which are cast as a discrete time network flow problem on a bipartite graph. More precisely, let us partition the time horizon into infinitesimal time slots of length $\Delta$ such that for a $\Delta$ that is sufficiently small compared to the job lengths, we can assume, without loss of generality, that the jobs' lengths are integer multiples of $\Delta$ and that jobs arrive and complete only at these integer multiples. Such quantization brings only a negligible error $o(\Delta)$ into our analysis, and it vanishes as $\Delta\to 0$. We note that introducing $\Delta$ is only for the sake of analysis to establish some desired properties of the dual variables generated by the algorithm. In particular, it has no impact on the algorithm implementation or its time complexity. 

Upon arrival of a new job $n$ at time $r_n$, let $\{v_j(r_n), j\in \mathcal{A}(r_n)\}$ be the remaining lengths of alive jobs without the new job $n$. Consider a flow network with a source node $a$ and a terminal node $b$ where the goal is to send $\sum_{j\in\mathcal{A}(r_n)}v_j(r_n)$ units of flow from $a$ to $b$. The source node is connected to $|\mathcal{A}(r_n)|$ nodes, each representing one of the alive jobs. Each directed edge $(a,j)$ has capacity $v_j(r_n)$ and cost $0$. The terminal node $b$ is connected to all the time slots $t=r_n,r_n+\Delta,r_n+2\Delta,\ldots$, where the edge $(t,b)$ has capacity $\Delta$ and zero cost. Finally, we set the capacity of the directed edge $(j,t)$ to $\Delta$ and its cost to $g_j(t)$.  By construction, it should be clear that as $\Delta\to 0$, the optimal flow cost in the residual network flow problem is precisely $\mbox{RNF}(r_n)$,  which also equals to optimal value function $V^o(\boldsymbol{v}(r_n),r_n)$ (all these quantities correspond to the optimal value of the same instance of the GFCS$(r_n)$ problem). Thus, by some abuse of notation, in the remainder of this work, we shall also use $\mbox{RNF}(r_n)$ to refer to the residual network flow representation of the GFCS$(r_n)$ problem. Note that since the capacity of each edge in $\mbox{RNF}(r_n)$ is an integral multiple of $\Delta$, integrality of the min-cost flow implies that in the optimal flow solution, each edge $(j,t)$ either is fully saturated by $\Delta$ units of flow or does not carry any flow. The implication is that the optimal flow assigns at most one job to any time slot $t$ (recall that edge $(j,t)$ has capacity $\Delta$), respecting the constraint that at each time slot, the machine can process at most one job.  

\begin{figure}[t]
\vspace{-1.5cm}
\begin{center}
\includegraphics[totalheight=.25\textheight,
width=.4\textwidth,viewport=0 0 900 900]{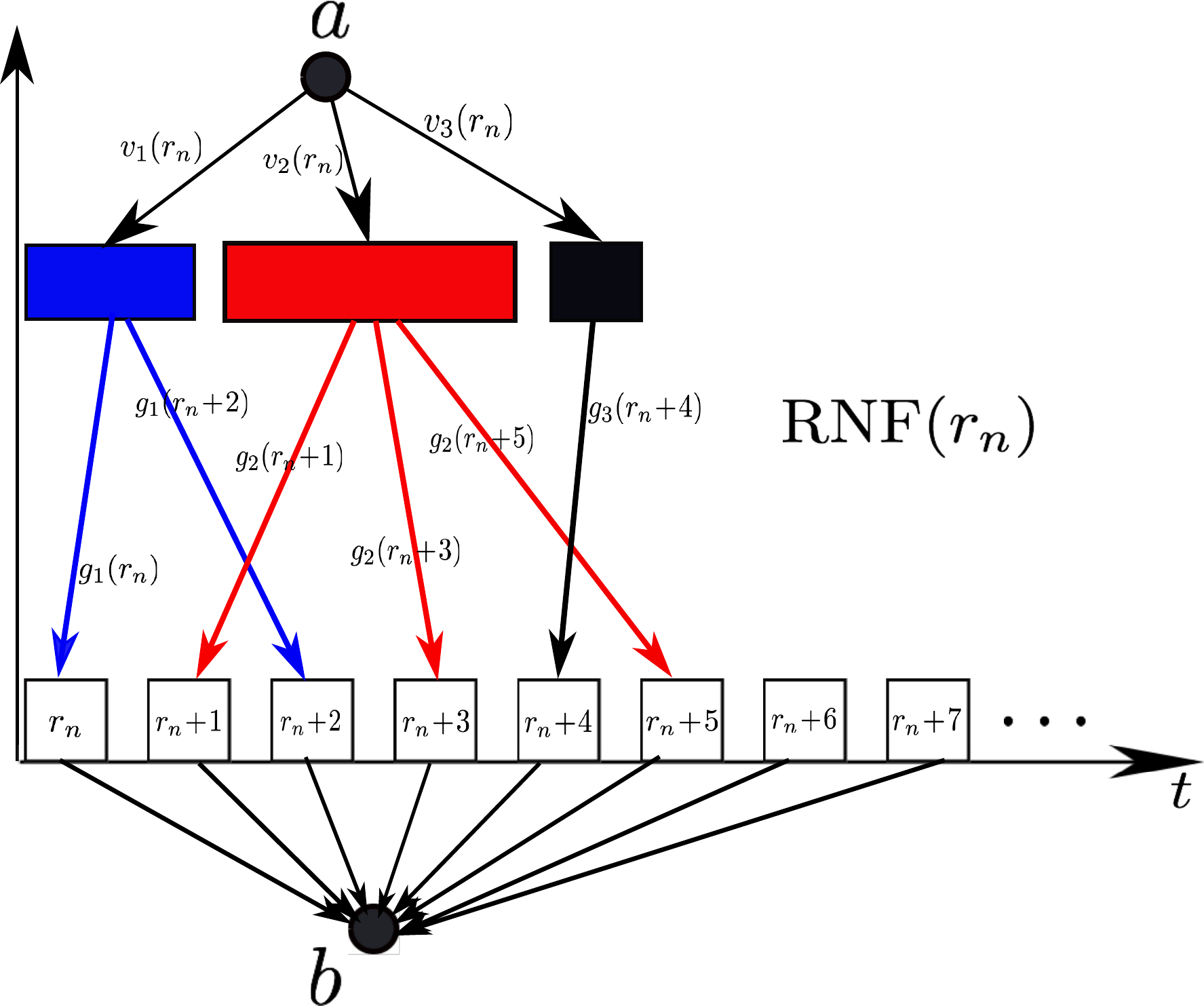} \hspace{0.4in}
\end{center}\vspace{-0.3cm}
\caption{An illustration of the residual network flow  $\mbox{RNF}(r_n)$ with three alive jobs $\mathcal{A}(r_n)=\{1,2,3\}$ and residual job lengths $v_1(r_n)=2$, $v_2(r_n)=3$, and $v_3(r_n)=1$. The optimal flow cost is equal to the sum of the colored-edge costs.}\label{fig:RNF}
\vspace{-0.3cm}
\end{figure}

\begin{remark}
By scaling up all the parameters by a factor of $\frac{1}{\Delta}$, we may assume that the scaled integer time slots are $t=\frac{r_n}{\Delta},\frac{r_n}{\Delta}+1,\ldots$, the edges have capacity $1$, the edge costs are $g_j(t\Delta )\Delta$, and the job lengths are $\frac{v_j(r_n)}{\Delta}$. Henceforth, we will work only with the scaled $\mbox{RNF}(r_n)$; for simplicity, and by some abuse of notations, we will use the same labels to refer to the scaled parameters, i.e., $r_n:=\frac{r_n}{\Delta}, v_j(r_n):=\frac{v_j(r_n)}{\Delta}, g_j(t):=g_j(t\Delta )\Delta$, and $t=r_n,r_n+1,\ldots$ (see Figure \ref{fig:RNF}). Similarly, we again use $\beta_t$ and $\alpha_j$ to denote the dual variables of the scaled system $\Delta\beta_{t\Delta}$ and $\alpha_j$. 
\end{remark}


\begin{lemma}\label{lemm:beta_inflate}
Let $\beta_t$ and $\beta'_t$ be the optimal dual solutions to the linear programs \emph{RNF}$(r_n)$ and $\mbox{\emph{RNF}}'(r_n)$ given in Definition \ref{def:RNF}, respectively. Then $\beta'_t\ge \beta_t, \forall t\ge r_n$. 
\end{lemma}

\emph{Proof:} Consider an instance of $n-1$ jobs with identical release times $r_n$ and lengths $\{v_j(r_n)\}_{j=1}^{n-1}$; $\beta_t$ denotes the optimal dual solution to the $\mbox{RNF}(r_n)$ of this instance. Let $\beta'_t$ denote the optimal dual solution to $\mbox{RNF}'(r_n)$ of the same instance with the additional new job $n$ of length $v_n$ and release time $r_n$. If there are many such optimal dual solutions, we take $\beta'_t$ to be the \emph{maximal} one with the largest value of $\sum_{t\ge r_n}\beta'_t$.  We refer to $\beta_t$ and $\beta'_t$ as the \emph{old} and the \emph{new} solutions, respectively. Note that by monotonicity of edge costs $g_j(\cdot)$, we have $\beta_t=0, \ \forall t\ge T$, and $\beta'_t=0, \ \forall t\ge T'$, where $T:=r_n+\sum_{j=1}^{n-1}v_j(r_n)$, and $T':=T+v_n$. Therefore, if we define $S:=\{t\ge r_n: \beta'_t\ge \beta_t\}$, we have $[T, \infty)\subseteq S$. Moreover, let $\mathcal{N}(S):=\{j\in [n]: x'_j(t)=1\ \mbox{for some}\ t\in S\}$ be the set of all the jobs that, in the new optimal solution, send positive flow to at least one of the time slots in $S$. Furthermore, we define $\bar{S}:=\{t\ge r_n: \beta'_t< \beta_t\}$ and $\bar{\mathcal{N}}(S):=[n]\setminus \mathcal{N}(S)$ to be the complement of $S$ and $\mathcal{N}(S)$, respectively, where we note that $\bar{S}\subseteq [r_n, T)$. To derive a contradiction, let us assume $\bar{S}\neq \emptyset$. We claim that $\alpha'_j\ge \alpha_j, \forall j\in \mathcal{N}(S)\setminus\{n\}$, because if $j\in \mathcal{N}(S)\setminus\{n\}$, there exists $t\in S$ such that $x'_j(t)=1$, and using the complementary slackness condition for the new solution, $\frac{\alpha'_j}{v_j(r_n)}=\beta'_t+g_j(t)$. Thus,
\begin{align}\label{eq:increasing-alpha}
\frac{\alpha'_j}{v_j(r_n)}=\beta'_t+g_j(t)\ge \beta_t+g_j(t)\ge \frac{\alpha_j}{v_j(r_n)}\ \ \ \ \ \ \ \Rightarrow \ \ \ \ \ \ \ \alpha'_j\ge \alpha_j, 
\end{align}
where the first inequality holds because $t\in S$, and the second inequality is due to the dual feasibility of the old solution for the job-slot pair $(j,t)$. On the other hand, $x_j(t)=0, \forall j\in \mathcal{N}(S)\setminus\{n\}, \forall t\in \bar{S}$. Otherwise, if $x_j(t)=1$ for some $j\in \mathcal{N}(S)\setminus\{n\}$ and $t\in \bar{S}$, then 
\begin{align}\nonumber
\beta'_t+g_j(t)\ge \frac{\alpha'_j}{v_j(r_n)}\ge \frac{\alpha_j}{v_j(r_n)}=\beta_t+g_j(t),
\end{align}
contradicting the fact that $t\in\bar{S}$. Here, the first inequality is due to the feasibility of the new solution for the pair $(j,t)$; the second inequality is due to \eqref{eq:increasing-alpha}; and the last equality follows from the complementary slackness condition of the old solution.   

Now, by the monotonicity of $g_j(\cdot)$, we know that the old solution sends exactly one unit of flow to each of the time slots in $[r_n, T)$. As $\bar{S}\subseteq [r_n, T)$, this means that exactly $|\bar{S}|$ units of flow are sent by the old solution to the time slots in $\bar{S}$. Since we just showed that the old solution does not send any flow from $\mathcal{N}(S)\setminus\{n\}$ to $\bar{S}$, the implication is that $|\bar{S}|\leq \sum_{j\in \bar{\mathcal{N}}(S)\setminus\{n\}}v_j(r_n)$ (otherwise, there would not be enough flow to send to $\bar{S}$). On the other hand, by the definition of $\bar{\mathcal{N}}(S)$, we know that the new solution does not send any positive flow from $\bar{\mathcal{N}}(S)$ to $S$. Thus, the flow of all the jobs in $\bar{\mathcal{N}}(S)$ must be sent to $\bar{S}$, and hence $|\bar{S}|\ge \sum_{j\in \bar{\mathcal{N}}(S)}v_j(r_n)$. These two inequalities show that $n\in\mathcal{N}(S)$ and we must have $|\bar{S}|=\sum_{j\in \bar{\mathcal{N}}(S)}v_j(r_n)$. In other words, both the old and new solutions send the entire flow that is going into $\bar{\mathcal{N}}(S)$ toward $\bar{S}$, and thus $x'_j(t)=x_j(t), \forall j\in \bar{\mathcal{N}}(S), \forall t\in \bar{S}$. Therefore, we can decompose the flow network into two parts, $[\bar{\mathcal{N}}(S): \bar{S}]$ and $[\mathcal{N}(S): S]$, with no positive flow from one side to the other in either the old or new solution. However, in that case, $\alpha'':=(\{\alpha_j\}_{j\in \bar{\mathcal{N}}(S)}, \{\alpha'_j\}_{j\in \mathcal{N}(S)})$ and $\beta'':=(\{\beta_t\}_{t\in \bar{S}}, \{\beta'_t\}_{t\in S})$ form another optimal \emph{new} solution with a higher $\beta$-sum, contradicting the maximality of $\{\beta'_t\}$. 

The reason for the optimality of $(\alpha'',\beta'')$ is that if either $j\in \bar{\mathcal{N}}(S), t\in \bar{S}$ or $j\in \mathcal{N}(S), t\in S$, the dual feasibility of $(\alpha'',\beta'')$ follows from the dual feasibility of $(\{\alpha_j\}_{j\in \bar{\mathcal{N}}(S)}, \{\beta_t\}_{t\in \bar{S}})$ or $(\{\alpha'_j\}_{j\in \mathcal{N}(S)}, \{\beta'_t\}_{t\in S})$, respectively. Moreover, for $j\in \bar{\mathcal{N}}(S), t\in S$, the dual feasibility of the old solution implies $\frac{\alpha_j}{v_j(r_n)}\leq \beta_t+g_j(t)\leq \beta'_t+g_j(t)$. Similarly, for $j\in \mathcal{N}(S), t\in \bar{S}$, the dual feasibility of the new solution implies $\frac{\alpha'_j}{v_j(r_n)}\leq \beta'_t+g_j(t)<\beta_t+g_j(t)$. Finally, $(\alpha'',\beta'')$ satisfies the complementary slackness conditions with respect to the optimal new solution $\boldsymbol{x}'(t)$. (Recall that both the old and new solutions coincide over $[\bar{\mathcal{N}}(S): \bar{S}]$ with no positive flow between $[\bar{\mathcal{N}}(S): \bar{S}]$ and $[\mathcal{N}(S): S]$.) \hfill $\blacksquare$

\begin{lemma}\label{lemm:dual feasible-RNF}
The dual solution generated by Algorithm \ref{alg:network-flow-single} is feasible to the dual program \eqref{eq:H_dual_single machine}. 
\end{lemma}

\emph{Proof:} The proof is by induction on the number of jobs. The statement trivially holds when there is only one job in the system, as the optimal solutions to \eqref{eq:H_primal_single machine} coincides with the one generated by $\mbox{RNF}(r_1)$, and hence, the corresponding optimal dual solutions also match. Now suppose the statement is true for the first $n-1$ jobs with release times $r_1\leq \ldots\leq r_{n-1}$, meaning that the dual solution $(\{\beta_t\}_{t\ge 0}, \{\alpha_j\}_{j=1}^{n-1})$ generated by Algorithm \ref{alg:network-flow-single} is a feasible solution to the dual program \eqref{eq:H_dual_single machine} with $n-1$ jobs. Now consider the time $r_n$ when a new job $n$ is released and use $\{v_j(r_n), j\in \mathcal{A}(r_n)\}$ to denote the remaining length of the alive jobs at that time. From the definition of Algorithm \ref{alg:network-flow-single}, we know that $\{\beta_t\}_{t\ge r_{n-1}}$ is an optimal dual solution for $\mbox{RNF}(r_{n-1})$. Thus, by principle of optimality, $\{\beta_t\}_{t\ge r_n}$ must be the optimal dual solution to the instance of jobs $\{v_j(r_n), j\in \mathcal{A}(r_n)\}$ with identical release times $r_n$.

Upon arrival of job $n$, let us use $\{\beta'_t\}_{t\ge r_n}$ to denote the new optimal dual solution to $\mbox{RNF}'(r_n)$. According to the dual solution generated by Algorithm \ref{alg:network-flow-single}, we update the old $\{\beta_t\}_{t\ge 0}$ variables to the concatenation $(\{\beta_t\}_{t<r_n}; \{\beta'_t\}_{t\ge r_n})$. Moreover, from the above argument and the choice of dual variables generated by Algorithm \ref{alg:network-flow-single}, we know that $\{\beta_t\}_{t\ge r_n}$ and $\{\beta'_t\}_{t\ge r_n}$ are optimal dual variables to $\mbox{RNF}(r_n)$ and $\mbox{RNF}'(r_n)$, respectively. Thus, using Lemma \ref{lemm:beta_inflate}, we have $\beta'_t\ge \beta_t , \forall t\ge r_n$. As dual variables $\{\alpha_j\}_{j=1}^{n-1}$ are kept unchanged and are feasible with respect to the old solution $\{\beta_t\}_{t\ge 0}$, $\{\alpha_j\}_{j=1}^{n-1}$ remain feasible with respect to $(\{\beta_t\}_{t<r_n}; \{\beta'_t\}_{t\ge r_n})$. Therefore, we only need to show that the newly set dual variable $\alpha'_n$ also satisfies all the dual constraints for $t\ge r_n$. That conclusion also immediately follows from the dual solution generated by Algorithm \ref{alg:network-flow-single}. The reason is that $\alpha'_n$ is an optimal dual variable for $\mbox{RNF}'(r_n)$ that must be feasible with respect to the optimal $\beta$-variables $\{\beta'_t\}_{t\ge r_n}$. Thus, $\alpha'_n$ is also a feasible solution with respect to $(\{\beta_t\}_{t<r_n}; \{\beta'_t\}_{t\ge r_n})$ for any $t\ge r_n$. \hfill $\blacksquare$

According to the update rule of Algorithm \ref{alg:network-flow-single}, upon arrival of a new job $n$ at time $r_n$, the algorithm updates its schedule for $t\ge r_n$ by resolving the corresponding optimal offline control problem. As a result, the cost increment incurred by Algorithm \ref{alg:network-flow-single} due to such an update is given by $\mbox{RNF}'(r_n)-\mbox{RNF}(r_n)$, which is the difference between the cost of the current schedule and that when the new job $n$ is added to the system. Therefore, we have the following definition:     
\begin{definition}
We define $\Delta_n(\mbox{Alg}):=\mbox{RNF}'(r_n)-\mbox{RNF}(r_n)$ to be the increase in the cost of Algorithm \ref{alg:network-flow-single} due to its schedule update upon the arrival of a new job $n$ at time $r_n$. 
\end{definition}

\begin{lemma}\label{lemm:dual-bound-alg}
Let $\alpha'_n$ be the dual variable generated by Algorithm \ref{alg:network-flow-single} upon arrival of the new job $n$ at time $r_n$. Then, we have $\Delta_n(\mbox{\emph{Alg}})\leq \alpha'_n$.
\end{lemma}

\emph{Proof:} As the algorithm sequentially solves a network flow problem with integral capacities, the feasible primal solution generated by Algorithm \ref{alg:network-flow-single} is also integral. Let $\{x_j(t)\in\{0,1\}: j\in [n-1], t\ge 0\}$ and $(\{\alpha_j\}_{j\in [n-1]},\{\beta_{t}\}_{t\ge 0})$ be the \emph{old} feasible primal and dual solutions generated by the algorithm before the arrival of job $n$, respectively. As $\{g_j(\cdot)\}$ are nondecreasing, we have $\beta_t=0$, $x_j(t)=0, \forall j, t> T:=\sum_{j=1}^{n-1}v_j$. Upon arrival of the job $n$ at time $r_n$, the algorithm updates its primal and dual solutions for $t\ge r_n$ to those obtained from solving $\mbox{RNF}'(r_n)$. We use $\{x'_j(t)\in\{0,1\}: j\in \mathcal{A}(r_n), t\ge r_n\}$ and $(\{\alpha'_j\}_{j\in \mathcal{A}(r_n)}, \{\beta'_t\}_{t\ge r_n})$, respectively, to denote the optimal primal/dual solutions. Again, we note that by the monotonicity of $g_j(\cdot)$, we have $\beta'_t=0$, $x'_j(t)=0, \forall t> T+v_n$. 


\begin{figure}[t]
\vspace{-2.7cm}
\begin{center}
\includegraphics[totalheight=.24\textheight,
width=.39\textwidth,viewport=500 0 1200 700]{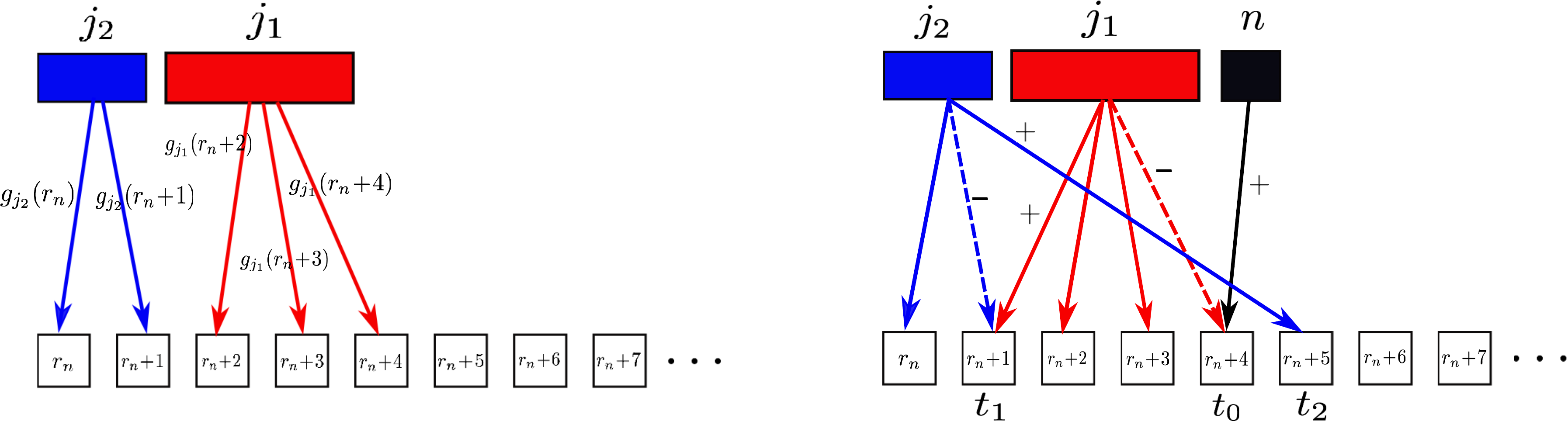} \hspace{0.4in}
\end{center}
\vspace{-0.5cm}
\caption{The left and right figures illustrate the optimal flow before and after the addition of the unit length job $n$, respectively.  $j_1$ and $j_2$ are two alive jobs at time $r_n$ with lengths $v_{j_1}(r_n)=2$ and $v_{j_2}(r_n)=3$, which are represented by red and blue colors, respectively.  Each solid edge $(j,t)$ shows whether a unit of job $j$ is scheduled at time slot $t$, with a cost given by the value on that edge. By adding job $n$,  the new optimal flow in the right figure can be obtained by removing (unscheduling) the dashed edges with negative signs and adding (scheduling) the new solid edges with plus signs. The result is an alternating path $P:=n,t_0,j_1,t_1,j_{2},t_{2}$. The change in the optimal flow cost is the sum of the edge costs along $P$ with respect to plus/minus signs. }\label{fig:flow-lemma}%
\end{figure}

Next, we compute the cost increment of the algorithm due to the introduction of the new job $n$. For simplicity, let us first assume $v_n=1$. Also, assume that solving $\mbox{RNF}' (r_n)$ assigns job $n$ to a time slot $t_0\ge r_n$ (i.e., $x'_n(t_0)=1$). If $t_0>T$, then the new and old solutions are identical, except that now one extra unit of flow is sent over the edge $(n,t_0)$. Therefore, the increase in the flow cost is exactly $\Delta_n(\mbox{Alg})=g_n(t_0)$. Otherwise, if $t_0\in [r_n, T]$, it means that slot $t_0$ was assigned by the old solution to a job $j_1\in \mathcal{A}(r_n)$. Therefore, the new solution must reschedule $j_1$ to a different time slot $t_1\neq t_0$. Note that $t_1\ge r_n$, since in the new solution only the slots that are after $r_n$ are reassigned based on $\mbox{RNF}(r_n)$. Similarly, if $t_1>T$, then the change in the cost of the algorithm is exactly $\Delta_n(\mbox{Alg})=g_n(t_0)-g_{j_1}(t_0)+g_{j_1}(t_1)$. Otherwise, slot $t_1$ was assigned by the old solution to some job $j_2\in \mathcal{A}(r_n)$, and hence, the new solution should reassign job $j_2$ to another slot $t_3\ge r_n$. By repeating this argument, we obtain an alternating path of job-slots $P:=n,t_0,j_1,t_1,\ldots,j_{p},t_{p}$ that starts from job $n$ and ends at some slot $t_{p}>T$. Now, starting from the old solution, we can rematch jobs to slots along the path $P$ to obtain the new solution (see Figure \ref{fig:flow-lemma}). In particular, the increase in the cost of the algorithm is precisely the rematching cost along the path $P$, i.e., 
\begin{align}\label{eq:delta_alg_RNF}
\Delta_n(\mbox{Alg})=g_n(t_0)-g_{j_1}(t_0)+g_{j_1}(t_1)-\ldots+g_{j_{p-1}}(t_{p-1})-g_{j_p}(t_{p-1})+g_{j_p}(t_p).
\end{align}

On the other hand, we know that $(\{\alpha'_j\}_{j\in \mathcal{A}(r_n)}, \{\beta'_t\}_{t\ge r_n})$ is the optimal dual solution to $\mbox{RNF}'(r_n)$. Using complementary slackness and dual feasibility of that solution along path $P$, we have
\begin{align}\label{eq:KKT_optimality_feasibility}
&\beta'_{t_0}=\frac{\alpha'_n}{v_n}-g_n(t_0), \ \ \ \ \ \ \ \ \ \ \ \ \ \ \ \ \ \ \ \ \qquad \frac{\alpha'_{j_1}}{v_{j_1}(r_n)}\leq \beta'_{t_0}+g_{j_1}(t_0),\cr 
&\beta'_{t_1}=\frac{\alpha'_{j_1}}{v_{j_1}(r_n)}-g_{j_1}(t_1), \ \ \ \ \ \ \ \ \ \ \ \ \ \ \ \ \ \ \ \ \frac{\alpha'_{j_2}}{v_{j_2}(r_n)}\leq \beta'_{t_1}+g_{j_2}(t_1),\cr 
&\vdots\cr 
&\beta'_{t_{p-1}}=\frac{\alpha'_{j_{p-1}}}{v_{j_{p-1}}(r_n)}-g_{j_{p-1}}(t_{p-1}), \ \ \ \ \ \ \ \ \ \ \ \frac{\alpha'_{j_p}}{v_{j_p}(r_n)}\leq \beta'_{t_{p-1}}+g_{j_p}(t_{p-1}),\cr 
&\beta'_{t_{p}}=\frac{\alpha'_{j_{p}}}{v_{j_{p}}(r_n)}-g_{j_{p}}(t_{p}),
\end{align} 
where the left side equalities are due to complementary slackness conditions for the optimal solution $(\{\alpha'_j\}_{j\in \mathcal{A}(r_n)}, \{\beta'_t\}_{t\ge r_n})$ over the nonzero flow edges $(n,t_0),(j_1,t_1),\ldots,(j_p,t_p)$ (as $x'_{n}(t_0)=\ldots=x'_{j_p}(t_p)=1$). The right side inequalities in \eqref{eq:KKT_optimality_feasibility} are due to the dual feasibility of $(\{\alpha'_j\}_{j\in \mathcal{A}(r_n)}, \{\beta'_t\}_{t\ge r_n})$ for $\mbox{RNF}(r_n)$, which are written for the job-slot pairs $(j_1,t_0),(j_2,t_1),\ldots,(j_p,t_{p-1})$. (Note that all the jobs in $\mbox{RNF}(r_n)$ have identical release times $r_n$, so $\{\alpha'_j\}_{j\in \mathcal{A}(r_n)}$ must satisfy the dual constraints $\forall t\ge r_n$.) By summing all the relations in \eqref{eq:KKT_optimality_feasibility}, we get
\begin{align}\nonumber
\beta'_{t_{p}}\leq \frac{\alpha'_n}{v_n}-g_n(t_0)+g_{j_1}(t_0)-g_{j_1}(t_1)+g_{j_2}(t_1)-\ldots+g_{j_p}(t_{p-1})-g_{j_{p}}(t_{p})=\frac{\alpha'_n}{v_n}-\Delta_n(\mbox{Alg}),
\end{align}
where the last equality is by \eqref{eq:delta_alg_RNF}. Since by dual feasibility, $\beta'_{t_{p}}\ge 0$ and $v_n=1$, we get $\Delta_n(\mbox{Alg})\leq \alpha'_n$.

Now, if $v_n>1$, then instead of one path $P$ we will have $v_n$ \emph{edge disjoint} paths $P_1,\ldots.P_{v_n}$, meaning that no job-slot pair $(j,t)$ appears more than once in all those paths, simply because each edge $(j,t)$ in $\mbox{RNF}(r_n)$ has a capacity of $1$, so each time slot $t$ is matched to at most one job in either the old or new solution. Thus, all the above analysis can be carried over each path $P_{\ell}, \ell=1,\ldots,v_n$ separately, and we have $\Delta_{P_{\ell}}(\mbox{Alg})\leq \frac{\alpha'_n}{v_n}\ \forall \ell$, where $\Delta_{P_{\ell}}(\mbox{Alg})$ denotes the increment in the algorithm's cost along the path $P_{\ell}$. As all the paths are edge-disjoint, the total cost increment of the algorithm equals $\Delta_n(\mbox{Alg})=\sum_{\ell=1}^{v_n}\Delta_{P_{\ell}}(\mbox{Alg})\leq \sum_{\ell=1}^{v_n}\frac{\alpha'_n}{v_n}=\alpha'_n$.   \hfill $\blacksquare$


\begin{definition}
A class of nondecreasing differentiable cost functions $\mathcal{H}=\{g_{\ell}(t): g_{\ell}(0)=0\}$ is called \emph{monotone substitute} if adding a new job to the offline instance with identical release time can only postpone the optimal scheduling intervals of the existing jobs. 
\end{definition}

For instance, any dominant family of cost functions has the monotone substitute property. Also, any class of nondecreasing cost functions for which the HDF (or HRDF) is the optimal offline policy has monotone substitute property. More generally, any family of nondecreasing cost functions with an insertion type optimal offline policy (i.e., a policy which inserts the new job somewhere between the previously scheduled jobs) possesses monotone substitute property.  Now we are ready to state the main result of this section.


\begin{theorem}\label{thm:single-machine-HGFC}
Let $\mathcal{H}=\{g_{\ell}(t): g_{\ell}(0)=0\}$ be a monotone substitute class of cost functions. Then Algorithm \ref{alg:network-flow-single} is $2K$-speed $2$-competitive for the GFCS problem where $K=1+\sup_{\ell, t\ge r_{\ell}}\frac{(t-r_{\ell})g''_{\ell}(t)}{g_{\ell}'(t)}$.    
\end{theorem}

\emph{Proof:} As before, let $V^o(\boldsymbol{v}^o(t),t)$ denote the optimal value function associated with the offline optimal control problem \eqref{eq:optimal_control} (or, equivalently, $\mbox{RNF}(r_n)$) in the \emph{absence} of job $n$. Then, 
\begin{align}\nonumber
V^o(\boldsymbol{v}^o(t),t)=\sum_{{\ell}\in \mathcal{A}(t)}\int_{t}^{\infty}g_{{\ell}}(\tau)x^o_{\ell}(\tau)d\tau=\sum_{\ell\in \mathcal{A}(t)}\sum_{d=1}^{n_{\ell}}\int_{C_{\ell}^{2d-1}}^{C_{\ell}^{2d}}g_{\ell}(\tau)d\tau,
\end{align}
where $\{\boldsymbol{x}^o(\tau), \tau\ge r_n\}$ is the optimal control for \eqref{eq:optimal_control}, and $I_{\ell}:=[C_{\ell}^{1}, C_{\ell}^{2})\cup\ldots\cup[C_{\ell}^{2n_{\ell}-1}, C_{\ell}^{2n_{\ell}})$ are the subintervals in which job $\ell$ is scheduled by the optimal control, i.e., $x^o_{\ell}(t)=1, \forall t\in I_{\ell}$ and $x^o_{\ell}(t)=0$, otherwise. Note that here, $C_{\ell}:=C_{\ell}^{2n_{\ell}}$ is the optimal completion time of job $\ell$ for the offline instance \eqref{eq:optimal_control}. On the other hand, from \eqref{eq:beta-fit} we know that the optimal $\beta$-variables to \eqref{eq:H_dual_single machine} with identical release times $r_n$ in the \emph{absence} of job $n$ are given by $\beta_t=\frac{\partial}{\partial t}V^o(\boldsymbol{v}^o(t),t), \forall t\ge r_n$. Thus, for any $t\ge r_n$,  
\begin{align}\label{eq:raw-beta}
\beta_t&=\frac{\partial}{\partial t}V^o(\boldsymbol{v}^o(t),t)=\lim_{\delta\to 0^+}\frac{V^o(\boldsymbol{v}^o(t),t+\delta)-V^o(\boldsymbol{v}^o(t),t)}{\delta}\cr 
&\leq \lim_{\delta\to 0^+}\frac{1}{\delta}\Big(\sum_{\ell\in \mathcal{A}(t)}\sum_{d=1}^{n_{\ell}}\int_{C_{\ell}^{2d-1}+\delta}^{C_{\ell}^{2d}+\delta}g_{\ell}(\tau)d\tau-V^o(\boldsymbol{v}^o(t),t)\Big)\cr 
&= \lim_{\delta\to 0^+}\frac{1}{\delta}\sum_{\ell\in \mathcal{A}(t)}\sum_{d=1}^{n_{\ell}}\Big(\int_{C_{\ell}^{2d-1}+\delta}^{C_{\ell}^{2d}+\delta}g_{\ell}(\tau)d\tau-\int_{C_{\ell}^{2d-1}}^{C_{\ell}^{2\ell}}g_{\ell}(\tau)d\tau\Big)\cr 
&=\sum_{\ell\in \mathcal{A}(t)}\sum_{d=1}^{n_{\ell}}[g_{\ell}(C_{\ell}^{2d})-g_{\ell}(C_{\ell}^{2d-1})]:=\hat{\beta}_t,
\end{align}     
where the inequality holds because we can upper-bound the optimal cost $V^o(\boldsymbol{v}^o(t),t+\delta)$ by following a suboptimal schedule that processes the jobs in the same order as the optimal schedule for $V^o(\boldsymbol{v}^o(t),t)$. The only difference here is that since the initial time $t$ is shifted by $\delta$ to the right, all the other scheduling subintervals will also be shifted to the right by $\delta$. By carrying exactly the same analysis in the \emph{presence} of job $n$, we can find an upper bound for the new solution $\beta'_t$ as
\begin{align}\nonumber
\beta'_t\leq \sum_{\ell\in \mathcal{A}'(t)}\sum_{d=1}^{n'_{\ell}}[g_{\ell}(C'^{2d}_{\ell})-g_{\ell}(C'^{2d-1}_{\ell})]:=\hat{\beta}'_t,
\end{align}
where the prime parameters are associated with the $\mbox{RNF}'(r_n)$ instance in the presence of job $n$.   


Next, we note that setting the $\beta$-dual variable higher than those generated by Algorithm \ref{alg:network-flow-single} still preserves dual feasibility. So instead of using optimal $\beta$-variables $\{\beta'_t\}_{t\ge r_n}$ in the dual update process of Algorithm \ref{alg:network-flow-single}, we can use their upper bounds $\{\hat{\beta}'_t\}_{t\ge r_n}$ while keeping the choice of $\alpha$-variables as before. This approach, in view of Lemma \ref{lemm:dual feasible-RNF}, guarantees that the generated dual variables are still feasible solutions to the dual program \eqref{eq:H_dual_single machine}. Note that this additional change in the dual updating process is merely for the sake of analysis and has nothing to do with the algorithm implementation.

Now let us assume that the machine in the optimal offline benchmark has a slower speed of $\frac{1}{1+\epsilon}$, meaning that the optimal benchmark aims to find a schedule for minimizing the \emph{slower} LP: 
\begin{align}\label{eq:dual-slow}
&\min \sum_{j}\int_{r_j}^{\infty}g_j(t)x_j(t)dt  \ \ \ \ \ \ \ \ \ \       \qquad\qquad    \max \sum_{j}\alpha_j-\frac{1}{1+\epsilon}\int_{0}^{\infty}\beta_t dt\cr 
\mbox{subject to}&\qquad \int_{r_j}^{\infty}\frac{x_j(t)}{v_j}\ge 1, \ \forall j\ \ \ \ \ \ \ \ \  \mbox{subject to} \qquad \frac{\alpha_j}{v_j}\leq \beta_t+\rho_jg(t), \ \forall j,t\ge r_j\cr 
&\qquad \sum_{j}x_j(t)\leq \frac{1}{1+\epsilon}, \ \forall t,   \ \ \ \ \ \ \ \   \qquad\qquad\qquad\alpha_j,\beta_t\ge 0, \ \forall j, t.\cr 
&\qquad x_j(t)\ge 0, \ \forall j, t.
\end{align}
Note that any feasible dual solution that is generated by the algorithm for the unit speed LP \eqref{eq:H_dual_single machine} will also be feasible to the dual of the slower system \eqref{eq:dual-slow}, as they both share the same constraints. Upon the arrival of a new job $n$ at time $r_n$, we just showed that the new dual solution generated by the algorithm is feasible, where we recall that only the tail of the dual solution will be updated from $\{\hat{\beta}_t\}_{t\ge r_n}$ to $\{\hat{\beta}'_t\}_{t\ge r_n}$. As we keep $\{\alpha_{j}\}_{j=1}^{n-1}$ unchanged, the cost increment of the updated dual solution with respect to the slower system \eqref{eq:dual-slow} equals $\Delta_s(D)=\alpha'_n-\frac{1}{1+\epsilon}\int_{r_n}^{\infty}(\hat{\beta}'_t-\hat{\beta}_t)dt$. Thus, by using Lemma \ref{lemm:dual-bound-alg} we obtain
\begin{align}\label{eq:delta-alpha-beta}
\Delta_n(\mbox{Alg})\leq \alpha'_n=\Delta_s(D)+\frac{1}{1+\epsilon}\int_{r_n}^{\infty}(\hat{\beta}'_t-\hat{\beta}_t)dt.
\end{align}

Let $I_{j}^r:=[C_{j}^{2r-1}, C_{j}^{2r})$ be the $r$th subinterval in which job $j$ is scheduled by the optimal policy, and define $g_{\ell}(I_{\ell}^{d}):=g_{\ell}(C_{\ell}^{2d})-g_{\ell}(C_{\ell}^{2d-1}), \forall \ell,d$. For any $t\in I_{j}^{r}$, we can write $\hat{\beta}_t$ given in \eqref{eq:raw-beta} as 
\begin{align}\label{eq:beta-structure-hat}
\hat{\beta}_t=g_{j}(C_{j}^{2r})-g_{j}(t)+\sum_{I_{\ell}^d>I_{j}^r}g_{\ell}(I_{\ell}^d)\ \ \ \  \forall t\in I_{j}^{r},
\end{align}
where the summation is taken over all the subintervals $I_{\ell}^d$ on the right side of $I_{j}^{r}$. We have
\begin{equation}\nonumber
\begin{split}
\int_{r_n}^{\infty}\hat{\beta}_tdt&=\sum_{j,r}\int_{I_{j}^r}\hat{\beta}_tdt\\ 
&=\sum_{j,r}\int_{I_{j}^r}\big(g_{j}(C_{j}^{2r})-g_{j}(t)+\sum_{I_{\ell}^d>I_{j}^r}g_{\ell}(I_{\ell}^d)\big)dt\\ 
&=\sum_{j,r}|I_j^r|\big(g_{j}(C_{j}^{2r})+\sum_{I_{\ell}^d>I_{j}^r}g_{\ell}(I_{\ell}^d)\big)-\sum_{j}\int_{I_{j}}g_j(t)dt\\
&=\sum_{j,r}|I_j^r|g_{j}(C_{j}^{2r})+\sum_{j,r}\sum_{I_{\ell}^d>I_{j}^r}|I_j^r|g_{\ell}(I_{\ell}^d)-\sum_{j}\int_{I_{j}}g_j(t)dt\\ 
&=\sum_{j,r}|I_j^r|g_{j}(C_{j}^{2r})+\sum_{\ell,d}(\sum_{I_{j}^r<I_{\ell}^d}|I_j^r|)g_{\ell}(I_{\ell}^d)-\sum_{j}\int_{I_{j}}g_j(t)dt\\ 
&=\sum_{j,r}|I_j^r|g_{j}(C_{j}^{2r})+\sum_{\ell,d}(C_{\ell}^{2d-1}-r_n)g_{\ell}(I_{\ell}^d)-\sum_{j}\int_{I_{j}}g_j(t)dt \\
&=\sum_{\ell,d}(C_{\ell}^{2d}-C_{\ell}^{2d-1})g_{\ell}(C_{\ell}^{2d})+\sum_{\ell,d}(C_{\ell}^{2d-1}-r_n)(g_{\ell}(C_{\ell}^{2d})-g_{\ell}(C_{\ell}^{2d-1}))-\sum_{j}\int_{I_{j}}g_j(t)dt
\end{split}
\end{equation}
\begin{equation}\nonumber
\hspace{-1.2cm}
\begin{split}
&=\sum_{\ell,d}\Big((C_{\ell}^{2d}-r_n)g_{\ell}(C_{\ell}^{2d})-(C_{\ell}^{2d-1}-r_n)g_{\ell}(C_{\ell}^{2d-1})\Big)-\sum_{j}\int_{I_{j}}g_j(t)dt\\
&=\sum_{\ell,d}\int_{I_{\ell}^d}\frac{d}{dt}\{(t-r_n)g_{\ell}(t)\}dt-\sum_{j}\int_{I_{j}}g_j(t)dt\\ 
&=\sum_{\ell}\int_{I_{\ell}}\frac{d}{dt}\{(t-r_n)g_{\ell}(t)\}dt-\sum_{j}\int_{I_{j}}g_j(t)dt\\ 
&=\sum_{\ell}\int_{I_{\ell}}\Big(\frac{d}{dt}\{(t-r_n)g_{\ell}(t)\}-g_{\ell}(t)\Big)dt\\ 
&=\sum_{\ell}\int_{I_{\ell}}(t-r_n)g'_{\ell}(t)dt.
\end{split}
\end{equation}
Similarly, if we use $I'_{\ell}$ to denote the processing interval of job $\ell$ when we are following Algorithm \ref{alg:network-flow-single} in the presence of job $n$, we have $\int_{r_n}^{\infty}\hat{\beta}'_tdt=\sum_{\ell \in \mathcal{A}(r_n)\cup\{n\}}\int_{I'_{\ell}}(t-r_n)g'_{\ell}(t)dt$. Thus, 
\begin{align}\label{eq:beta-difference-single}
\int_{r_n}^{\infty}(\hat{\beta}'_t-\hat{\beta}_t)dt&= \int_{I'_{n}}(t-r_n)g'_{n}(t)dt+\sum_{\ell\in \mathcal{A}(r_n)}\Big(\int_{I'_{\ell}}(t-r_n)g'_{\ell}(t)dt-\int_{I_{\ell}}(t-r_n)g'_{\ell}(t)dt\Big).
\end{align}
On the other hand, we know that 
\begin{align}\label{eq:algo-difference-single}
\Delta_n(\mbox{Alg})=\int_{I'_{n}}g_{n}(t)dt+\sum_{\ell\in \mathcal{A}(r_n)}\Big(\int_{I'_{\ell}}g_{\ell}(t)dt-\int_{I_{\ell}}g_{\ell}(t)dt\Big).
\end{align}
Thus, if we define $h_{\ell}(t):=(t-r_n)g'_{\ell}(t)-Kg_{\ell}(t)$, where $K$ is the constant given in the theorem statement, using \eqref{eq:algo-difference-single} and \eqref{eq:beta-difference-single}, we can write
\begin{align}\label{eq:final-new-old-diiference-single}
\int_{r_n}^{\infty}(\hat{\beta}'_t-\hat{\beta}_t)dt-K\Delta_n(\mbox{Alg})=\int_{I'_{n}}h_{n}(t)dt+\sum_{\ell\in \mathcal{A}(r_n)}(\int_{I'_{\ell}}h_{\ell}(t)dt-\int_{I_{\ell}}h_{\ell}(t)dt).  
\end{align}
From the definition of $K$ and because $r_{\ell}\leq r_n \forall \ell\in \mathcal{A}(r_n)$, we have $(t-r_n)g_{\ell}''(t)-(K-1)g_{\ell}'(t)\leq 0$,  $\forall t\ge r_n, \ell$, which implies that for any $\ell$, the function $h_{\ell}(t)$ is nonincreasing (as the former is the derivative of the latter). Since, by monotone substitute property, adding a new job can only postpone the processing intervals of the alive jobs, the time interval $I'_{\ell}$ can only be shifted further to the right side of $I_{\ell}$. As $h_{\ell}(t)$ is a nonincreasing function, we have $\int_{I'_{\ell}}h_{\ell}(t)dt\leq \int_{I_{\ell}}h_{\ell}(t)dt, \forall \ell\in\mathcal{A}(r_n)$. Moreover, since $h_{n}(r_n)=-Kg_{n}(r_n)= 0$, we have $h_{n}(t)\leq 0, \forall \ell$, and thus $\int_{I'_{n}}h_{n}(t)dt\leq 0$. Those relations together with \eqref{eq:final-new-old-diiference-single} imply that $\int_{r_n}^{\infty}(\hat{\beta}'_t-\hat{\beta}_t)dt\leq K\Delta_n(\mbox{Alg})$. Now, using \eqref{eq:delta-alpha-beta}, we can write
\begin{align}\nonumber
\Delta_n(\mbox{Alg})\leq \Delta_s(D)+\frac{1}{1+\epsilon}\int_{r_n}^{\infty}(\hat{\beta}'_t-\hat{\beta}_t)dt\leq \Delta_s(D)+\frac{K}{1+\epsilon}\Delta_n(\mbox{Alg}),
\end{align}
which implies that $\Delta_n(\mbox{Alg})\leq \frac{1+\epsilon}{1+\epsilon-K}\Delta_s(D)$. As this relation holds at any time that a new job $n$ arrives, by summing over all the jobs, we get $\mbox{Cost}(\mbox{Alg})\leq \frac{1+\epsilon}{1+\epsilon-K}\mbox{Cost}(D_s)$, where $D_s$ is the optimal value of the slower dual program \eqref{eq:dual-slow}. Using weak duality, we obtain that Algorithm \ref{alg:network-flow-single} is $(1+\epsilon)$-speed $\frac{1+\epsilon}{1+\epsilon-K}$-competitive for the GFCS problem. Finally, by selecting $1+\epsilon=2K$, one can see that Algorithm \ref{alg:network-flow-single} is $2K$-speed $2$-competitive for the GFCS problem. \hfill $\blacksquare$

While typically the tradeoff between speed and the competitive ratio of an online scheduling algorithm is characterized by two different functions $f_1(\epsilon)$ and $f_2(\epsilon)$, for simplicity of presentation, in this paper we use a slightly different form by normalizing the competitive ratio to a constant and analyzing the amount of speed that is required to achieve a constant competitive ratio. This approach significantly simplifies the dependence of our bounds on the speeding parameter $\epsilon$ without getting into too many speed-scaling complications. Such a representation is particularly convenient in the case of general cost functions for which the speed-scaling parameters can depend on each of the individual cost functions. We refer to \cite[Theorem 5]{nguyen2013lagrangian} for a direct scaling approach for specific functions $g_j(t)=w_jg(t)$ that uses multicriteria scaling conditions.


\begin{example}
For the family of increasing and differentiable concave cost functions, Algorithm \ref{alg:network-flow-single} is $(1+\epsilon)$-speed $\frac{1+\epsilon}{\epsilon}$-competitive for GFCS. The reason is that for any increasing concave function, we have $g''_{\ell}(t)<0$ and $g'_{\ell}(t)>0, \forall t\ge r_{\ell}$. As a result, $\frac{(t-r_{\ell})g''_{\ell}(t)}{g_{\ell}'(t)}\leq 0, \forall t\ge r_{\ell}$. Thus, by the definition of the curvature ratio, $K=1+\sup_{\ell, t\ge r_{\ell}}\frac{(t-r_{\ell})g''_{\ell}(t)}{g_{\ell}'(t)}=1$.  This in view of Lemma \ref{lemm-s-c} shows that Algorithm \ref{alg:network-flow-single} is $(1+\epsilon)^2$-speed $\frac{(1+\epsilon)^2}{\epsilon^2}$-competitive for the GICS problem. This result is somewhat consistent with the $(1+\epsilon)$-speed $\frac{4(1+\epsilon)^2}{\epsilon^2}$-competitive online algorithm in \cite[Theorem 9]{angelopoulos2019primal} that was given for GICS with concave costs but under a relaxed rate allocation setting.  The class of increasing and differentiable concave functions includes logarithmic functions $g_{\ell}(t)=w_{\ell}\log(1+t)$ that are frequently used for devising proportional fair scheduling algorithms \cite{im2018competitive}. 
\end{example}


It is known that no online algorithm can be $O(1)$-speed $O(1)$-competitive for GFCS \cite[Theorem 3.3]{im2014online}. Thus, the fact that the required speed to achieve a constant competitive ratio in Theorem \ref{thm:single-machine-HGFC} depends on the curvature of cost functions seems unavoidable.  In fact, the curvature ratio $K$ given in Theorem \ref{thm:single-machine-HGFC} is only one way of capturing this dependency.  However, to obtain a better performance guarantee, one needs to obtain tighter bounds on the increase of dual objective function along the solution generated by the online algorithm. Consequently, that requires finding better dual solutions using the HJB value function approximation.  Unfortunately, choosing better dual solutions for general cost functions is very challenging without imposing additional assumptions on the structure of the cost functions.  Having said that, it may be that the curvature ratio given in Theorem \ref{thm:single-machine-HGFC} is close to optimal. If it is the case, it would be interesting to establish a matching lower bound. 


\smallskip
\section{A Competitive Online Algorithm for the GFCU Problem}\label{sec:unrelated-competitive}

The GFCS problem can be naturally extended to multiple unrelated machines. Here we assume that there are $m\ge 2$ unrelated machines and that jobs are released online over time. Upon arrival of a job, a feasible online schedule must dispatch that job to one machine, and a machine can work preemptively on at most one unfinished job that is assigned to it. We only allow nonmigratory schedules in which a job cannot migrate from one machine to another once it has been dispatched. We assume that job $j$ has a processing requirement $v_{ij}$ if it is dispatched to machine $i$ with an associated cost function $\hat{g}_{ij}(t)$.  Moreover, we assume that the specifications of job $j$ are revealed to the system only upon $j$'s arrival at time $r_j$. Given a feasible online schedule, we let $J_i$ be the set of jobs that are dispatched to machine $i$, and $C_j$ be the completion time of job $j$ under that schedule. Therefore, our goal is to find a feasible online schedule that dispatches the jobs to the machines and processes them preemptively on their assigned machines so as to minimize the generalized integral completion time on unrelated machines (GICU) given by $\sum_{i}\sum_{j\in J_i}\hat{g}_{ij}(C_j)$. As before,  and using Lemma \ref{lemm-s-c} adapted for multiple machines,  we only consider the fractional version of that objective cost, where the remaining portion $\frac{v_{ij}(t)}{v_{ij}}$ of a job $j$ that is dispatched to a machine $i$ contributes $\hat{g}'_{ij}(t)$ amount to the delay cost. Therefore, if we use $g_{ij}(t):=\frac{\hat{g}_{ij}(t)}{v_{ij}}$ to denote the scaled cost functions, the objective cost of a feasible schedule for the generalized fractional completion time on unrelated machines (GFCU) is given by 
\begin{align}\label{eq:unrelated-fractional-cost}
\sum_{i}\sum_{j\in J_i}\int_{r_j}^{\infty} \frac{v_{ij}(t)}{v_{ij}}\hat{g}'_{ij}(t)dt=\sum_{i}\sum_{j\in J_i}\int_{r_j}^{\infty} \frac{\hat{g}_{ij}(t)}{v_{ij}}x_{ij}(t)dt=\sum_{i}\sum_{j\in J_i}\int_{r_j}^{\infty} g_{ij}(t)x_{ij}(t)dt,
\end{align}
where  $x_{ij}(t)=-\frac{d}{dt}v_{ij}(t)$ is the rate at which job $j\in J_i$ is processed by the schedule, and the first equality holds by integration by parts and the assumption $g_{ij}(r_j)=0, \forall i,j$.  The following lemma provides a lower bound for the objective cost of the GFCU problem, which is obtained by relaxing the requirement that a job must be processed on only one machine. We will use this LP relaxation as an offline benchmark when we devise a competitive online schedule.  The derivation of such LP resembles that in \cite{anand2012resource}, which is derived here for generalized cost functions.

\begin{lemma}\label{lemm:fractional-relaxation}
The cost of the following LP is at most twice the cost of HGFC on unrelated machines:
\begin{align}\label{eq:H_primal_unrelated}
&\min \sum_{i,j}\int_{r_j}^{\infty}\Big(g_{ij}(t)+d_{ij}\Big)x_{ij}(t)dt\cr 
\mbox{\emph{subject to}}&\qquad \sum_i\int_{r_j}^{\infty}\frac{x_{ij}(t)}{v_{ij}}dt\ge 1, \ \forall j\cr 
&\qquad \sum_{j}x_{ij}(t)\leq 1, \forall i,t\cr 
&\qquad x_{ij}(t)\ge 0, \ \forall i, j, t,
\end{align} 
where $d_{ij}:=\int_{r_j}^{r_j+v_{ij}}\frac{g_{ij}(t)}{v_{ij}}dt$ are constants.
\end{lemma}
\emph{Proof:} Given an arbitrary feasible schedule, let $J_i$ denote the set of all the jobs that are dispatched to machine $i$ at the end of the process, $C_j$ be the completion time of job $j$, and $x^s_{ij}(t)$ be the rate at which the schedule processes job $j\in J_i$. Since in a feasible schedule, a machine $i$ can process at most one job at any time $t$, we must have $\sum_{j\in J_i}x^s_{ij}(t)\leq 1, x^s_{ij}\in\{0,1\}, \forall i,t$. Moreover, as a feasible schedule must process a job entirely, we must have $\int_{r_j}^{\infty}\frac{x^s_{ij}(t)}{v_{ij}}dt\ge 1, \forall i, j\in J_i$. Thus a solution $\boldsymbol{x}^s(t)$ produced by any feasible schedule must satisfy all the constraints in \eqref{eq:H_primal_unrelated} that are relaxations of the length and speed requirement constraints. In fact, the constraints in \eqref{eq:H_primal_unrelated} allow a job to be dispatched to multiple machines or even to be processed simultaneously with other jobs. However, such a relaxation can only reduce the objective cost and gives a stronger benchmark. Finally, the LP cost of the feasible solution $\boldsymbol{x}^s(t)$ equals
\begin{align}\label{eq:bound-on-hgdf}
\sum_{i,j}\int_{r_j}^{\infty}\Big(g_{ij}(t)+d_{ij}\Big)x^s_{ij}(t)dt&=\sum_{i}\sum_{j\in J_i}\int_{r_j}^{\infty}g_{ij}(t)x^s_{ij}(t)dt+\sum_{i}\sum_{j\in J_i}\int_{r_j}^{\infty}d_{ij}x^s_{ij}(t)dt.
\end{align}
Using the definition of $d_{ij}$, and since for any optimal schedule $\int_{r_j}^{\infty}x^s_{ij}(t)dt= v_{ij}, \forall j\in J_i$, we have,
\begin{align}\label{eq:dound-on-d}
\int_{r_j}^{\infty}d_{ij}x^s_{ij}(t)dt=\int_{r_j}^{r_j+v_{ij}}g_{ij}(t)dt\leq \int_{r_j}^{\infty}g_{ij}(t)x^s_{ij}(t)dt,
\end{align}
where the inequality holds by monotonicity of $g_{ij}(t)$, as $\int_{r_j}^{\infty}g_{ij}(t)x^s_{ij}(t)dt$ is minimized when $x^s_{ij}(t)=1$ for $t\in [r_j, r_j+v_{ij}]$ and $x^s_{ij}(t)=0$, otherwise. By substituting \eqref{eq:dound-on-d} into \eqref{eq:bound-on-hgdf}, we find that the LP objective cost of the schedule is at most $2\sum_{i}\sum_{j\in J_i}\int_{r_j}^{\infty}g_{ij}(t)x^s_{ij}(t)dt$, which is twice the fractional cost by the schedule given in \eqref{eq:unrelated-fractional-cost}. Thus, the minimum value of LP \eqref{eq:H_primal_unrelated} is at most twice the minimum fractional cost generated by any feasible schedule. \hfill $\blacksquare$


Finally, we note that the dual program for LP \eqref{eq:H_primal_unrelated} is given by
\begin{align}\label{eq:H_dual_unrelated}
&\max \sum_{j}\alpha_j-\sum_i\int_{0}^{\infty}\beta_{it} dt\cr 
\mbox{subject to}&\qquad \frac{\alpha_j}{v_{ij}}\leq \beta_{it}+g_{ij}(t)+d_{ij}, \ \forall i,j,t\ge r_j\cr 
&\qquad \alpha_j,\beta_{it}\ge 0, \ \forall i,j,t.
\end{align}

In the following section, we first provide a primal-dual online algorithm for the GFCU problem. The algorithm decisions on how to dispatch jobs to machines and process them are guided by a dual solution that is updated frequently upon the arrival of a new job. The process of updating dual variables is mainly motivated by the insights obtained from the case of a single machine with some additional changes to cope with multiple unrelated machines. We then show in Lemma \ref{lemm:dual-feasible-unrelated} that the dual solution generated throughout the algorithm is feasible to the dual program \eqref{eq:H_dual_unrelated}, which by Lemma \ref{lemm:fractional-relaxation} and weak duality gives a lower bound for the optimal value of the offline benchmark. Finally, we bound the gap between the objective cost of the algorithm and the cost of the generated dual solution, which allows us bound the competitive ratio of the algorithm.
\subsection{Algorithm Design and Analysis}

To design an online algorithm for GFCU, we need to introduce an effective dispatching rule. Consider an arbitrary but fixed machine $i$ and assume that currently the alive jobs on machine $i$ are scheduled to be processed over time intervals $\{I_{\ell}, \ell\in \mathcal{A}(r_n)\}$, where we note that each $I_{\ell}$ can itself be the union of disjoint subintervals $I_{\ell}=\cup_d I^{d}_{\ell}$. From the lesson that we learned in the proof of Theorem \ref{thm:single-machine-HGFC}, we shall set the $\beta$-dual variables for machine $i$ to $\hat{\beta}_{it}=\sum_{\ell\in \mathcal{A}(t)}g_{i\ell}(I_{\ell}\cap[t, \infty))$, where $g_{i\ell}(I_{\ell}\cap[t, \infty)):=\sum_d g_{i\ell}(I_{\ell}^d\cap[t, \infty))$ is the total variation of function $g_{i\ell}(t)$ over $I_{\ell}\cap[t, \infty)$. Unfortunately, upon release of a new job $n$ at time $r_n$, we cannot set the new $\alpha$-variable for job $n$, denoted by $\hat{\alpha}'_n$, as high as that in the case of a single machine, i.e., $\alpha'_n$. The reason is that choosing $\hat{\alpha}'_n=\alpha'_n$ may not be feasible with respect to the old $\hat{\beta}_{i't}$-variables of other machines $i'\neq i$. To circumvent that issue, in the case of multiple machines, we slightly sacrifice optimality in favor of generating a feasible dual solution. For that purpose, we do not set $\hat{\alpha}'_n$ as high as before, but rather define it in terms of the \emph{old} $\hat{\beta}_{it}$-variables.  Ideally, we want to set $\hat{\alpha}'_n$ equal to $\min_{t\ge r_n}\{(\hat{\beta}_{it}+g_{in}(t)+d_{in})v_{in}\}$, so that if instead of $\hat{\beta}_{it}$, we had the new optimal $\beta'_{it}$-variables, then we would have obtained the same $\alpha'_n$ as before. Clearly, such an assignment of an $\alpha$-variable to job $n$ is feasible with respect to the $\hat{\beta}_{it}$ of machine $i$. However, to assure that it is also feasible for all other machines, we take another minimum over all the machines by setting 
\begin{align}\label{eq:alpha-hat-prime}
\hat{\alpha}'_n:=\min_{i, t\ge r_n}\{(\hat{\beta}_{it}+g_{in}(t)+d_{in})v_{in}\}.
\end{align}
That guarantees that the new dual variable $\hat{\alpha}'_n$ is also feasible with respect to the $\hat{\beta}_{it}$-variables of all the other machines. Thus, if we let $i^*=\arg\min_{i}\{\min_{t\ge r_n}(\hat{\beta}_{it}+g_{in}(t)+d_{in})v_{in}\}$, we dispatch job $n$ to the machine $i^*$.  On the other hand, to assure that the future dual variables $\hat{\alpha}'_{n+1},\hat{\alpha}'_{n+2},\ldots$ are set as high as possible, we shall update the old $\hat{\beta}_{i^*t}$-variables for machine $i^*$ to their updated versions $\hat{\beta}'_{i^*t}$; doing so also accounts for the newly released job $n$. We do so by inserting the new job $n$ into the old schedule to obtain updated scheduling intervals $I'_{\ell}=\cup_d I'^{d}_{\ell}$ for machine $i^*$, and accordingly define $\hat{\beta}'_{i^*t}=\sum_{\ell\in \mathcal{A}'(t)}g_{i^*\ell}(I'_{\ell}\cap[t, \infty))$ based on this new schedule. As, by complementary slackness conditions, a job is scheduled whenever its dual constraint is tight, we insert job $n$ into the old schedule at time $t^*=\argmin_{t\ge r_n}(\hat{\beta}_{i^*t}+g_{i^*n}(t)+d_{i^*n})v_{i^*n}$ (which is the time at which the dual constraint for job $n$ on machine $i^*$ is tight) and schedule it entirely over $[t^*, t^*+v_{i^*n}]$. Note that this insertion causes all the old scheduling subintervals on machine $i^*$ that were after time $t^*$ to be shifted to the right by $v_{i^*n}$, while the scheduling subintervals that were before time $t^*$ remain unchanged, as in the old schedule. The above procedure in summarized in Algorithm \ref{alg:dispatch-flow}.
    
\begin{algorithm}
\caption{An Online Algorithm for the GFCU Problem}\label{alg:dispatch-flow}
{\bf Input:} An instance of the GFCU problem $\{v_{ij},r_j, g_{ij}(t), i\in [m], j=1,2,\ldots\}$ with nondecreasing convex cost functions $g_{ij}(t)$. 
 
\noindent 
{\bf Output:} An online nonmigratory schedule that assigns each job to a single machine and determines what job must be processed on each machine at any time $t\ge 0$.

\begin{itemize}
\item Upon arrival of a new job $n$ at time $r_n$,  let $\{I_{\ell}:\ell\in \mathcal{A}(r_n)\}$ denote the old scheduling subintervals in the absence of job $n$ for (an arbitrary) machine $i$. Let $\hat{\beta}_{it}=\sum_{\ell\in \mathcal{A}(t)}g_{i\ell}(I_{\ell}\cap[t,\infty))$ be the old $\hat{\beta}_{it}$-variables associated with the old schedule on machine $i$.
\item Dispatch job $n$ to machine $i^*$ for which $i^*=\arg\min_{i}\{\min_{t\ge r_n}(\hat{\beta}_{it}+g_{in}(t)+d_{in})v_{in}\}$,  and let $t^*$ be the minimizing time, that is $t^*=\argmin_{t\ge r_n}(\hat{\beta}_{i^*t}+g_{i^*n}(t)+d_{i^*n})v_{i^*n}$.
\item Form a new schedule by only modifying the scheduling intervals on machine $i^*$ as follows: Over the interval $[r_n, t^*]$, process the jobs on machine $i^*$ based on the old schedule. Schedule the new job $n$ entirely on machine $i^*$ over the interval $I'_{n}=[t^*, t^*+v_{i^*n}]$. Shift all the remaining old scheduling intervals on machine $i^*$ which are after $t^*$ to the right by $v_{i^*n}$. 
\item  Update the tail of the old $\hat{\beta}_{i^*t}$-variables for machine $i^*$ from $\{\hat{\beta}_{i^*t}\}_{t\ge r_n}$ to $\{\hat{\beta}'_{i^*t}\}_{t\ge r_n}$,  where $\hat{\beta}'_{i^*t}=\sum_{\ell\in \mathcal{A}'(t)}g_{i^*\ell}(I'_{\ell}\cap[t, \infty))$, and $\{I'_{\ell}: \ell\in \mathcal{A}'(t)\}$ are the new scheduling intervals on machine $i^*$ that are obtained from the previous stage.  Keep all other dual variables $\{\hat{\beta}_{it}\}_{i\neq i^*}$ unchanged. 
\end{itemize}
\end{algorithm}

\medskip
\noindent
{\bf Time Complexity of Algorithm \ref{alg:dispatch-flow}:} Upon arrival of a new job $n$, Algorithm \ref{alg:dispatch-flow} needs to find the scheduling time $t^*$, which requires solving the minimization problem $\arg\min_{t\ge r_n}\{\hat{\beta}_{it}+d_{in}+g_{in}(t)\}$ for each machine $i$, and then taking the minimum over all the machines in $O(m)$.  Moreover,  using  \eqref{eq:beta-structure-hat}, one can see that $\hat{\beta}_{it}=\sum_{\ell\in \mathcal{A}(t)}g_{i\ell}(I_{\ell}\cap[t, \infty))$ has the form of $\hat{\beta}_{it}=L-g_{ij}(t)$, for some constant $L$ and some job $j$ that is scheduled at time $t$.  We note that unlike Algorithm \ref{alg:network-flow-single}, the scheduling intervals $I_{\ell}$ are not necessarily the optimal offline scheduling intervals. Instead, those intervals are constructed inductively according to the rules of Algorithm \ref{alg:dispatch-flow}, which saves in the optimal offline computational cost. Since Algorithm \ref{alg:dispatch-flow} inserts each new job somewhere between the previously scheduled jobs,  the arrival of a new job can increase the number of scheduling subintervals by at most $2$. As a result,  $\hat{\beta}_{it}$ is a continuous and piecewise monotone concave function with at most $O(n)$ pieces, which can be constructed efficiently. Therefore, solving $\arg\min_{t\ge r_n}\{\hat{\beta}_{it}+d_{in}+g_{in}(t)\}$ requires minimizing the difference of two monotone and smooth convex functions over at most $O(n)$ disjoint subintervals within the support of $\hat{\beta}_{it}$. Thus, Algorithm \ref{alg:dispatch-flow} has an overall runtime $O(nmT_{u})$, where $T_u$ is the time complexity of minimizing the difference of two monotone and smooth univariate convex functions over an interval.

\smallskip
\begin{lemma}\label{lemm:dual-feasible-unrelated}
Assume that $\{g_{i\ell}(t), \forall i,\ell\}$ are nondecreasing convex functions. Then, the dual solution generated by Algorithm \ref{alg:dispatch-flow} is feasible for the dual program \eqref{eq:H_dual_unrelated}.  
\end{lemma}

\emph{Proof:} As we argued above, the new dual variable $\hat{\alpha}'_n$ is feasible with respect to the $\hat{\beta}_{it}$-variables of all the machines and for any $t\ge r_n$. Since we keep all other dual variables $\{\hat{\alpha}_{j}\}_{j=1}^{n-1}$ and $\{\hat{\beta}_{it}\}_{i\neq i^*}$ unchanged, it is enough to show that $\hat{\beta}'_{i^*t}\ge \hat{\beta}_{i^*t}, \forall t\ge r_n$. This inequality also follows from the definition of $\hat{\beta}$-variables and the convexity of cost functions $g_{i\ell}(\cdot)$. More precisely, for any $t\ge r_n$,
\begin{align}\nonumber
\hat{\beta}_{i^*t}=\sum_{\ell\in \mathcal{A}(t)}g_{i^*\ell}(I_{\ell}\cap[t,\infty)), \ \ \ \ \ \ \ \ \ \ \ \ \hat{\beta}'_{i^*t}=\sum_{\ell\in \mathcal{A}'(t)}g_{i^*\ell}(I'_{\ell}\cap[t,\infty)),
\end{align}    
where we note that at any time $t$, $\mathcal{A}(t)\subseteq \mathcal{A}'(t)$ (since by definition $\mathcal{A}'(t)$ contains all the jobs in $\mathcal{A}(t)$ and possibly the new job $n$). As for any $\ell,d$, the subinterval $I'^d_{\ell}$ is either the same as $I^d_{\ell}$, or shifted to the right by $v_{i^*n}$; by the convexity and monotonicity of $g_{i^*\ell}(\cdot)$ we have $g_{i^*\ell}(I'^d_{\ell}\cap[t,\infty))\ge g_{i\ell}(I^d_{\ell}\cap[t,\infty)), \forall d$. Summing this relation for all $d$ and $\ell\in \mathcal{A}'(t)$ shows that $\hat{\beta}'_{it}\ge \hat{\beta}_{i^*t}$. Thus, $\hat{\beta}'_{i^*t}$ can only increase because of the final stage of Algorithm \ref{alg:dispatch-flow}, so dual feasibility is preserved. \hfill $\blacksquare$   

\smallskip
\begin{theorem}\label{thm:unrelated-competitive}
Let $\mathcal{H}=\{g_{i\ell}(t): g_{i\ell}(0)=0\}$ be a family of differentiable nondecreasing convex functions. Then, Algorithm \ref{alg:dispatch-flow} is $2K\theta$-speed $2\theta$-competitive for the GFCU problem, where 
\begin{align}\nonumber
K:=1+\sup_{i,\ell,t\ge r_{\ell}}\frac{(t-r_{\ell})g''_{i\ell}(t)}{g_{i\ell}'(t)}, \qquad\qquad\qquad \theta:=\sup_{i,\ell,t\ge r_{\ell}}\frac{g_{i\ell}(t+v)-g_{i\ell}(t)}{vg'_{i\ell}(t)}.
\end{align}   
\end{theorem}

\emph{Proof:} First, let us assume that each machine in the optimal algorithm has a slower speed of $\frac{1}{1+\epsilon}$. Therefore, using Lemma \ref{lemm:fractional-relaxation}, the following LP and its dual provide a lower bound (up to a factor of 2) for the optimal fractional cost of any schedule with slower unrelated machines.
\begin{align}\label{eq:LP-general_heterogeneous}
&\min \sum_{i,j}\int_{r_j}^{\infty}(g_{ij}(t)+d_{ij})x_{ij}(t)dt \ \ \ \ \ \ \ \ \ \ \ \ \ \ \ \ \ \  \max \sum_{j}\alpha_j-\frac{1}{1+\epsilon}\sum_i\int_{0}^{\infty}\beta_{it} dt\cr
\mbox{subject to}&\qquad \sum_i\int_{r_j}^{\infty}\frac{x_{ij}(t)}{v_{ij}}\ge 1, \ \forall j\ \ \ \ \ \ \ \ \ \ \ \  \mbox{subject to}\qquad\frac{\alpha_j}{v_{ij}}\leq \beta_{it}+g_{ij}(t)+d_{ij}, \ \forall i,j,t\ge r_j\cr 
&\qquad \sum_{j}x_{ij}(t)\leq \frac{1}{1+\epsilon}, \forall i,t\ \ \ \ \ \ \ \ \ \ \ \ \ \ \ \ \ \ \ \qquad\qquad \alpha_j,\beta_{it}\ge 0, \ \forall i,j,t\cr
&\qquad \ x_{ij}(t)\ge 0, \ \forall i,j,t.
\end{align}
By Lemma \ref{lemm:dual-feasible-unrelated}, the dual solution generated by Algorithm \ref{alg:dispatch-flow} is feasible for \eqref{eq:H_dual_unrelated}, and thus, it is also dual feasible for the slower system \eqref{eq:LP-general_heterogeneous}. Now, upon arrival of job $n$ at time $r_n$, let us assume that the algorithm dispatches $n$ to machine $i=i^*$, which from now we fix this machine. Thus, the increment in the (slower) dual objective of the feasible solution generated by the algorithm equals $\Delta_s(D)=\hat{\alpha}'_n-\frac{1}{1+\epsilon}\int_{r_n}^{\infty}(\hat{\beta}'_{it}-\hat{\beta}_{it})dt$. Unfortunately, since in general we may have $\hat{\alpha}'_n\leq \alpha'_n$, we can no longer use Lemma \ref{lemm:dual-bound-alg} to upper-bound the cost increment of the algorithm. Instead, we upper-bound $\Delta_n(\mbox{Alg})$ in terms of $\hat{\alpha}'_n$ directly, and that causes an additional loss factor in the competitive ratio. To that end, let $t^*=\arg\min_{t\ge r_n}\{\hat{\beta}_{it}+d_{in}+g_{in}(t)\}$. We note that before time $t^*$, the old and new schedules are the same, and thus $\int_{t^*}^{t^*+v_{in}}g_{in}(t)dt$ is the new cost due to the scheduling of job $n$. Moreover,  the cost increment between the old and new schedules after time $t^*$, equals
\begin{align}
\sum_{\ell\in \mathcal{A}(t^*)}\int_{(I_{\ell}\cap[t^*,\infty))+v_{in}}g_{i\ell}(t)dt -\sum_{\ell\in \mathcal{A}(t^*)}\int_{I_{\ell}\cap[t^*,\infty]}g_{i\ell}(t)dt= \sum_{\ell\in \mathcal{A}(t^*)}\int_{I_{\ell}\cap[t^*,\infty]}(g_{i\ell}(t+v_{in})-g_{i\ell}(t))dt.
\end{align}
Since we set $\hat{\alpha}'_n=v_{in}(\hat{\beta}_{it^*}+d_{in}+g_{in}(t^*))$, we can write
\begin{align}\nonumber
\frac{\Delta_n(\mbox{Alg})}{\hat{\alpha}'_n}&=\frac{\Delta_n(\mbox{Alg})}{v_{in}(\hat{\beta}_{it^*}+d_{in}+g_{in}(t^*))}= \frac{\int_{t^*}^{t^*+v_{in}}g_{in}(t)dt+\sum_{\ell\in \mathcal{A}(t^*)}\int_{I_{\ell}\cap[t^*,\infty)}(g_{i\ell}(t+v_{in})-g_{i\ell}(t))dt}{v_{in}(d_{in}+g_{in}(t^*))+v_{in}\sum_{\ell\in \mathcal{A}(t^*)}\int_{I_{\ell}\cap[t^*,\infty)}g'_{i\ell}(t)dt}.
\end{align}
Since by the definition of $\theta$ we have $g_{i\ell}(t+v_{in})-g_{i\ell}(t)\leq \theta v_{in}g'_{i\ell}(t), \forall i,\ell,t\ge r_{\ell}$ and $\int_{t^*}^{t^*+v_{in}}g_{in}(t)dt\leq \theta v_{in}(d_{in}+g_{in}(t^*))$, we conclude that $\frac{\Delta_n(\mbox{Alg})}{\hat{\alpha}'_n}\leq \theta$. Therefore,
\begin{align}\label{eq:mutliple-delta-alpha-beta}
\Delta_n(\mbox{Alg})\leq  \theta \hat{\alpha}'_n=\theta \Delta_s(D) +\frac{\theta}{1+\epsilon}\int_{r_n}^{\infty}(\hat{\beta}'_{it}-\hat{\beta}_{it})dt.
\end{align} 
By following the exact same analysis used for the case of the single machine, we know that 
\begin{align}\nonumber
\int_{r_n}^{\infty}\hat{\beta}_{it}&=\sum_{\ell\in\mathcal{A}(r_n)}\int_{I_{\ell}}(t-r_n)g'_{i\ell}(t)dt=\sum_{\ell\in\mathcal{A}(r_n)}\Big(\int_{I_{\ell}\cap[r_n, t^*)}(t-r_n)g'_{i\ell}(t)dt+\int_{I_{\ell}\cap[t^*, \infty)}\!\!\!\!\!(t-r_n)g'_{i\ell}(t)dt\Big),\cr 
\int_{r_n}^{\infty}\hat{\beta}'_{it}&=\sum_{\ell\in\mathcal{A}(r_n)\cup\{n\}}\int_{I'_{\ell}}(t-r_n)g'_{i\ell}(t)dt\cr 
&=\int_{t^*}^{t^*+v_{in}}(t-r_n)g'_{in}(t)dt+\sum_{\ell\in\mathcal{A}(r_n)}\Big(\int_{I_{\ell}\cap[r_n, t^*)}\!\!(t-r_n)g'_{i\ell}(t)dt+\int_{(I_{\ell}\cap[t^*, \infty))+v_{in}}\!\!\!\!\!\!\!\!\!(t-r_n)g'_{i\ell}(t)dt\Big).
\end{align}
Thus, the difference between the above two expressions is given by 
\begin{align}\nonumber
\int_{r_n}^{\infty}(\hat{\beta}'_{it}-\hat{\beta}_{it})dt&=\int_{t^*}^{t^*+v_{in}}\!\!\!\!\!(t-r_n)g'_{in}(t)dt+\sum_{\ell\in\mathcal{A}(t^*)}\Big(\int_{(I_{\ell}\cap[t^*, \infty))+v_{in}}\!\!\!\!\!\!\!\!\!\!\!\!\!(t-r_n)g'_{i \ell}(t)dt-\int_{I_{\ell}\cap[t^*, \infty)}\!\!\!\!\!\!\!\!(t-r_n)g'_{i\ell}(t)dt\Big).
\end{align}
Now let us define $h_{\ell}(t):=(t-r_n)g'_{i\ell}(t)-Kg_{i\ell}(t)$. Then,  
\begin{align}\label{eq:delta-k-beta}
\int_{r_n}^{\infty}(\hat{\beta}'_{it}-\hat{\beta}_{it})dt-K\Delta_n(\mbox{Alg})=\int_{t^*}^{t^*+v_{in}}h_{n}(t)dt+\sum_{\ell\in\mathcal{A}(t^*)}\Big(\int_{(I_{\ell}\cap[t^*, \infty))+v_{in}}h_{\ell}(t)dt-\int_{I_{\ell}\cap[t^*, \infty)}h_{\ell}(t)dt\Big).
\end{align} 
On the other hand, for every $\ell$, $h_{\ell}(t)$ is a nonincreasing function. The reason is that 
\begin{align}\nonumber
h'_{\ell}(t)=(t-r_n)g''_{i\ell}(t)-(K-1)g'_{i\ell}(t)\leq (t-r_{\ell})g''_{i\ell}(t)-(K-1)g'_{i\ell}(t)\leq 0,
\end{align}
where the first inequality is due to the convexity of $g_{i\ell}(t)$ (as $g''_{i\ell}(t)\ge 0$) and the fact that $r_{\ell}\leq r_n$, and the second inequality is by definition of the constant $K$. This, in turn, implies that 
\begin{align}\nonumber
\int_{(I_{\ell}\cap[t^*,\infty))+v_{in}}h_{\ell}(t)dt\leq \int_{I_{\ell}\cap[t^*, \infty)}h_{\ell}(t)dt, \ \ \forall \ell.
\end{align}
Moreover, as $h_{n}(r_n)=-Kg_{in}(r_n)= 0$, we have $h_{n}(t)\leq 0, \forall t\ge r_n$, and thus $\int_{t^*}^{t^*+v_{in}}h_{n}(t)dt\leq 0$. By substituting those relations into \eqref{eq:delta-k-beta}, we can conclude that $\int_{r_n}^{\infty}(\hat{\beta}'_{i^*t}-\hat{\beta}_{i^*t})dt\leq K\Delta_n(\mbox{Alg})$. By substituting this inequality into \eqref{eq:mutliple-delta-alpha-beta} and summing over all the jobs, we obtain
\begin{align}\nonumber
\mbox{Cost}(\mbox{Alg})\leq \theta \mbox{Cost}(D_s)+\frac{\theta K}{1+\epsilon}\mbox{Cost}(\mbox{Alg}),
\end{align}
or, equivalently, $\mbox{Cost}(\mbox{Alg})\leq \frac{(1+\epsilon)\theta}{1+\epsilon-K\theta}\mbox{Cost}(D_s)$, where $D_s$ is the optimal value of the slower dual program \eqref{eq:LP-general_heterogeneous}. Finally, if we choose the speed $1+\epsilon=2K\theta$, the competitive ratio of the algorithm for the GFCU problem is at most $2\theta$. \hfill $\blacksquare$

\begin{example}
As a simple application of Theorem \ref{thm:unrelated-competitive}, if we consider an instance of GFCU with quadratic cost functions of the form $g_{i\ell}(t)=a_{i\ell}t^2+b_{i\ell}t+c_{i\ell}$ with nonnegative coefficients and jobs of at most a unit in length, then we have $K=1+\sup_{i,\ell,t\ge r_{\ell}}\frac{2a_{i\ell}(t-r_{\ell})}{2a_{i\ell}t+b_{i\ell}}\leq 2$ and $\theta=1+\sup_{i,\ell,t\ge v}\frac{a_{i\ell}v}{2a_{i\ell}t+b_{i\ell}}\leq 2$. Thus, for this class of quadratic cost functions Algorithm \eqref{alg:dispatch-flow} is $8$-speed $4$-competitive. 
\end{example}  

\section{Conclusions and Future Directions}\label{sec:conclusions}

In this paper, we considered online scheduling on single or multiple unrelated machines under general job and machine-dependent cost functions. Using results from optimal control and LP duality, we provided a framework for devising competitive online algorithms in a speed-augmented setting and under some mild assumptions on the structure of the cost functions. Here, we focused on a nonmigratory preemptive setting in which the machines have fixed speed and can work on a single job at each time instance.  Moreover,  we assumed that the processing requirement of jobs is a deterministic value $v_{ij}$. However, it is often more realistic to assume that processing requirements are random variables with certain expected values.  One natural idea for such an extension is to leverage our generalized dual-fitting results when the job requirements are replaced by their expected values.  Surprisingly, it has been shown recently in \cite{gupta2020greed} that for the special case of minimizing the expected weighted completion time,  dual-fitting can be very helpful to devise competitive online algorithms with respect to the expected instance. Therefore,  an interesting future direction is to extend this work to stochastic settings with general cost functions using our generalized dual-fitting analysis.


In devising an online algorithm for the GFCS problem, we implicitly assumed access to an oracle for solving the simpler offline subproblem $GFCS(r_n)$. Although solving $GFCS(r_n)$ could be in general NP-hard, as we mentioned earlier, $GFCS(r_n)$ admits polynomial (quasi-polynomial) time approximation algorithm (scheme). Therefore, combining those results from the offline setting to our online analysis can give us competitive online algorithms for GFCS with substantially reduced time complexity. Finally, as we showed above and in Appendix I, optimal control can generalize some of the existing dual-fitting results in the context of job scheduling in a unified manner. Therefore, leveraging this framework to design competitive online algorithms in more complex job scheduling settings, such as multi-queue multi-server systems with nonlinear client-server assignment costs, would be an interesting future research direction.

\section*{Appendix I: A Generalization of the Earlier Dual-Fitting Results}
In this appendix, we apply the results of Section \ref{sec:optimal-control} to a slightly different setting and derive a generalization of the dual bounds given in \cite{anand2012resource}. We use this method to analyze the \emph{highest residual density first} (HRDF) scheduling rule, in which the residual density of a job $j$ at time $t$ is defined to be $\rho_j(t)=\frac{w_j}{v_j(t)}$. More generally, we can define the residual density of a job $j$ at time $t$ with a cost function $g_j(\tau)$ to be $\frac{g_j(\tau)}{v_j(t)}$. (Note that for $g_j(\tau)=w_j$, the two definitions coincide.) It has been shown in \cite{anand2012resource} that for online job scheduling on unrelated machines with $g_{ij}(t)=w_{ij}(t-r_j)^k$, if each machine works based on HRDF and a newly released job is dispatched to the machine that results in the least increase in the $\ell_k$-flow time, then such a scheduling algorithm is competitive in the augmented speed setting with respect to the objective of integral ${\ell}_k$-flow time $\sum_{j}w_{i(j)j}(C_j-r_j)^k$, where $i(j)$ denotes the machine to which job $j$ is dispatched. The analysis in \cite{anand2012resource} is based on dual fitting and a careful choice of dual variables:
\begin{align}\label{eq:dual-fit-HRDF}
&\hat{\beta}_{ir_n}=k\sum_{j\in \mathcal{A}(r_n)}w_{ij}R^{k-1}_j(r_n),\cr
&\hat{\alpha}_n=w_{in}R_n(r_n)^k+\sum_{\substack{j\in\mathcal{A}(r_n)\\ \rho_{ij}(r_n)< \rho_{in}}}w_{ij}\Big((R_{j}(r_n)+v_{ij})^k-R_j(r_n)^k\Big).
\end{align}  
Here, $i$ is the machine to which job $n$ is dispatched, and $R_j(r_n)=C_j-r_n$ is the ``remaining" completion time of job $j$ at time $r_n$, where $C_j$ denotes the completion time of job $j$ following the HRDF rule. However, the choice of the dual variables in \cite{anand2012resource} is not quite immediate and provides little insight about how one can extend those results to more complex cost functions. Here we show how these complicated-looking variables can be obtained systematically by using a simple application of the optimal control framework to a special case of the ${\ell}_k$-flow objective function. It is worth noting that in \cite{anand2012resource}, only the performance of the HRDF schedule for the specific $\ell_k$-norm is considered. The optimal control approach that we adopt here can be used in a more general setting for which we first define an optimal value function according to a scheduling policy that we wish to analyze, and then carry the machinery in Section \ref{sec:optimal-control} to fit the dual variables for that specific policy.


To determine the dual variables for HRDF, let us assume that job $n$ is released at some \emph{arbitrary} time $r_n$ and dispatched to machine $i$ by the algorithm. (From now on, we fix machine $i$ and, for simplicity, drop all the indices that depend on $i$.) By the definition of the HRDF rule, we know that at time $r_n$, the machine must schedule a job $\ell$ with the highest \emph{residual density} $\frac{g_{\ell}(t)}{v_{\ell}(r_n)}$ (rather than the original density $\rho_{\ell}=\frac{g_{\ell}(t)}{v_{\ell}})$. Therefore, the offline optimal control objective function with identical release times $r_{\ell}=r_n, \forall \ell\in \mathcal{A}(r_n)$ is given by
\begin{align}\label{eq:HRDF-formulation}
\min \big\{\int_{r_n}^{\infty}\!\!\!\sum_{\ell\in \mathcal{A}(r_n)}\!\!\!\frac{g_{\ell}(t-r_n)}{v_{\ell}(r_n)}x_{\ell}(t)dt: \  \dot{\boldsymbol{v}}(t)=-\boldsymbol{x}(t),  \boldsymbol{v}(r_n)=(v_1(r_n),\ldots,v_k(r_n))^T,  \boldsymbol{x}(t)\in \mathcal{X}, \forall t\big\}.  
\end{align}
Note that the only difference between the HRDF formulation \eqref{eq:HRDF-formulation} and the earlier formulation \eqref{eq:optimal_control} is that the original cost function $g_{\ell}(t)$ is now scaled by the residual length $\frac{1}{v_{\ell}(r_n)}$ rather than the original length $\frac{1}{v_{\ell}}$, and the argument in $g_{\ell}(t)$ is replaced by $t-r_n$, as we are interested in \emph{flow} time rather than the completion time. To determine a good fit for $\hat{\beta}_t$, let $V^o(\boldsymbol{v}^o(t),t)$ denote the optimal value function for the offline optimal control problem \eqref{eq:HRDF-formulation}, i.e.,
\begin{align}\nonumber
V^o(\boldsymbol{v}^o(t),t)=\sum_{{\ell}\in \mathcal{A}(t)}\int_{t}^{\infty}\frac{g_{\ell}(\tau-r_n)}{v_{\ell}(r_n)}x^o_{\ell}(\tau)d\tau=\sum_{\ell\in \mathcal{A}(t)}\sum_{d=1}^{n_{\ell}}\int_{C_{\ell}^{2d-1}}^{C_{\ell}^{2d}}\frac{g_{\ell}(\tau-r_n)}{v_{\ell}(r_n)}d\tau,
\end{align}
where $\{\boldsymbol{x}^o(\tau), \tau\ge r_n\}$ is the optimal control to \eqref{eq:HRDF-formulation}, and $I_{\ell}:=[C_{\ell}^{1}, C_{\ell}^{2})\cup\ldots\cup[C_{\ell}^{2n_{\ell}-1}, C_{\ell}^{2n_{\ell}})$ are the subintervals in which job $\ell$ is scheduled by the optimal control. Again $C_{\ell}:=C_{\ell}^{2n_{\ell}}$ is the optimal completion time of job $\ell$ for the offline instance \eqref{eq:HRDF-formulation}. On the other hand, from \eqref{eq:beta-fit} we know that the optimal offline dual variable is given by $\beta_{r_n}=\frac{\partial}{\partial r_n}V^o(\boldsymbol{v}^o(r_n),r_n)$. Thus,  
\begin{align}\nonumber
\beta_{r_n}&=\frac{\partial}{\partial r_n}V^o(\boldsymbol{v}^o(r_n),r_n)=\lim_{\delta\to 0^+}\frac{V^o(\boldsymbol{v}^o(r_n),r_n+\delta)-V^o(\boldsymbol{v}^o(r_n),r_n)}{\delta}\cr 
&\leq \lim_{\delta\to 0^+}\frac{1}{\delta}\Big(\sum_{\ell\in \mathcal{A}(r_n)}\sum_{d=1}^{n_{\ell}}\int_{C_{\ell}^{2d-1}+\delta}^{C_{\ell}^{2d}+\delta}\frac{g_{\ell}(\tau-r_n)}{v_{\ell}(r_n)}d\tau-V^o(\boldsymbol{v}^o(r_n),r_n)\Big)\cr 
&= \lim_{\delta\to 0^+}\frac{1}{\delta}\sum_{\ell\in \mathcal{A}(r_n)}\sum_{d=1}^{n_{\ell}}\Big(\int_{C_{\ell}^{2d-1}+\delta}^{C_{\ell}^{2d}+\delta}\frac{g_{\ell}(\tau-r_n)}{v_{\ell}(r_n)}d\tau-\int_{C_{\ell}^{2d-1}}^{C_{\ell}^{2d}}\frac{g_{\ell}(\tau-r_n)}{v_{\ell}(r_n)}d\tau\Big)\cr 
&=\sum_{\ell\in \mathcal{A}(r_n)}\frac{1}{v_{\ell}(r_n)}\sum_{d=1}^{n_{\ell}}[g_{\ell}(C_{\ell}^{2d}-r_n)-g_{\ell}(C_{\ell}^{2d-1}-r_n)],
\end{align}     
where the inequality holds because we can upper-bound the optimal cost $V^o(\boldsymbol{v}^o(r_n),r_n+\delta)$ by following a suboptimal test policy that starts from time $r_n+\delta$ and mimics the same optimal schedule as $V^o(\boldsymbol{v}^o(r_n),r_n)$. If we specialize the above relation to the case of convex functions $g_{\ell}(\tau-r_n)$ and note that $\sum_{d}(C_{\ell}^{2d}-C^{2d-1}_{\ell})=v_{\ell}(r_n)$, we can write
\begin{align}\nonumber
\beta_{r_n}&\leq \sum_{\ell\in \mathcal{A}(r_n)}\frac{1}{v_{\ell}(r_n)}\sum_{d=1}^{n_{\ell}}[g_{\ell}(C_{\ell}^{2d}-r_n)-g_{\ell}(C_{\ell}^{2d-1}-r_n)]\cr 
&\leq \sum_{\ell\in \mathcal{A}(r_n)}\frac{1}{v_{\ell}(r_n)}\sum_{d=1}^{n_{\ell}}g'_{\ell}(C_{\ell}^{2d}-r_n)[C_{\ell}^{2d}-C_{\ell}^{2d-1}]\cr 
&\leq \sum_{\ell\in \mathcal{A}(r_n)}\frac{1}{v_{\ell}(r_n)}\sum_{d=1}^{n_{\ell}}g'_{\ell}(C_{\ell}-r_n)[C_{\ell}^{2d}-C_{\ell}^{2d-1}]\cr 
&=\sum_{\ell\in \mathcal{A}(r_n)}g'_{\ell}(C_{\ell}-r_n),
\end{align}
which suggests to set the $\beta$-dual variable to $\hat{\beta}_{r_n}:=\sum_{\ell\in \mathcal{A}(r_n)}g'_{\ell}(C_{\ell}-r_n)$. To see why that choice of $\hat{\beta}_{r_n}$ coincides with the one given in \eqref{eq:dual-fit-HRDF}, we note that for the special $\ell_k$-flow time, $g_{\ell}(\tau-r_n)$ is given by the convex function $g_{\ell}(\tau-r_n)=w_{i\ell}(\tau-r_n)^k$, and thus $\hat{\beta}_{ir_n}=k\sum_{\ell\in \mathcal{A}(r_n)}w_{i\ell}(C_{\ell}-r_n)^{k-1}$. Finally we note that for the fixed identical release times $r_n$ and homogeneous monotone functions $g_{\ell}(\tau-r_n)=w_{i\ell}(\tau-r_n)^k$, we know from the results of Section \ref{sec:single-g(t)} that the optimal schedule for the offline problem \eqref{eq:HRDF-formulation} is indeed HDF (which also coincides with HRDF in the \emph{offline} setting). Thus the optimal completion times $C_{\ell}$ given above are precisely those obtained by following HRDF.

Next, to fit a dual variable for the optimal $\alpha'_n$, we note that setting the $\alpha$-dual variables lower than the optimal ones always preserves dual feasibility. As we saw in Lemma \ref{lemm:dual-bound-alg}, even for a more general case, $\Delta_n(\mbox{Alg})\leq \alpha'_n$; thus, the lower bound can be chosen to be $\hat{\alpha}_n:=\Delta_n(\mbox{Alg})=V^o(\boldsymbol{v}(r_n)\cup\{n\},r_n)-V^o(\boldsymbol{v}(r_n),r_n)$, i.e., the increase in the cost of the offline optimal control \eqref{eq:HRDF-formulation} due to the addition of a new job $n$. If we simplify the right side by recalling that the optimal offline solution to \eqref{eq:HRDF-formulation} with cost functions $g_{\ell}(\tau-r_n)=w_{i\ell}(\tau-r_n)^k$ is HRDF, we obtain precisely the fractional version of $\hat{\alpha}_n$-variable given in \eqref{eq:dual-fit-HRDF} (the $\hat{\alpha}_n$ in \eqref{eq:dual-fit-HRDF} is written for the integral objective cost, where $V^o(\boldsymbol{v}(r_n),r_n):=\sum_{j\in\mathcal{A}(r_n)}w_{i\ell}(C_{\ell}-r_n)^k$.)

\begin{figure}[t]
\vspace{-3.5cm}
\begin{center}
\includegraphics[totalheight=.25\textheight,
width=.4\textwidth,viewport=0 0 650 650]{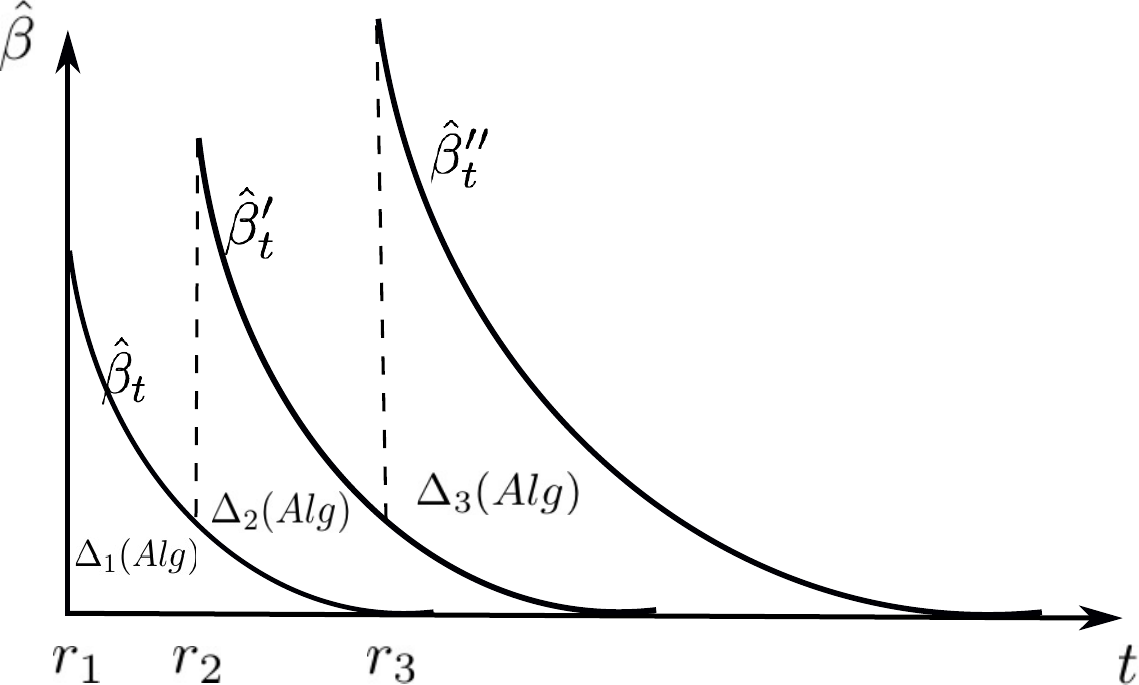} \hspace{0.4in}
\end{center}
\vspace{-0.5cm}
\caption{Upon the arrival of a new job $n$ at time $r_n$, the old tail $\{\hat{\beta}_t\}_{t\ge r_n}$ is updated to the new one $\{\hat{\beta}'_t\}_{t\ge r_n}$. Here the $\hat{\boldsymbol{\beta}}$-curve is given by the upper envelope of the other three curves. The increase in the size of the area under the $\hat{\boldsymbol{\beta}}$-curve due to the arrival of job $n$ is precisely the increase in the integral flow cost of the algorithm $\Delta_n(\mbox{Alg})$, which is also set for the dual variable $\hat{\alpha}_n$. Summing over all jobs gives us the overall area under the $\hat{\beta}$-curve and is equal to the integral flow cost of the algorithm.}\label{fig:beta-increment}
\end{figure}

Finally, to verify why the above choices of dual variables are indeed good, we show that the two terms $\sum_{n}\hat{\alpha}_n$ and $\int_{0}^{\infty}\hat{\beta}$ that appear in the dual objective cost are indeed the same and equal to the integral flow cost of the solution generated by the algorithm. The implication is that the dual variables defined above keep the dual objective cost close to the cost of the primal solution generated by the algorithm. To understand why, we first note that
\begin{align}\nonumber
\int_{r_n}^{\infty}\hat{\beta}_tdt=\int_{r_n}^{\infty}\sum_{\ell\in \mathcal{A}(t)}g'_{\ell}(C_{\ell}-t)dt=\sum_{\ell\in \mathcal{A}(r_n)}\int_{r_n}^{C_{\ell}}g'_{\ell}(C_{\ell}-t)dt=\sum_{\ell\in \mathcal{A}(r_n)}g_{\ell}(C_{\ell}-r_n),
\end{align}
which is the integral objective cost after time $r_n$. Similarly, with job $n$ in the system, we have $\int_{r_n}^{\infty}\hat{\beta}'_tdt=\sum_{\ell\in \mathcal{A}'(r_n)}g_{\ell}(C'_{\ell}-r_n)$. Thus, $\int_{r_n}^{\infty}(\hat{\beta}'_t-\hat{\beta}_t)dt$ is precisely the increment in the integral flow cost due to the arrival of job $n$, i.e., 
\begin{align}\nonumber
\int_{r_n}^{\infty}(\hat{\beta}'_t-\hat{\beta}_t)dt=\Delta_n(\mbox{Alg})=\hat{\alpha}_n, \forall n.
\end{align}
Finally, we note that $\sum_n\int_{r_n}^{\infty}(\hat{\beta}'_{t}-\hat{\beta}_{t})dt$ is precisely the size of the area under the $\hat{\boldsymbol{\beta}}$-curve generated by the algorithm, i.e., $\hat{\boldsymbol{\beta}}=(\{\hat{\beta}_{t}\}_{t=r_1}^{t=r_2}; \{\hat{\beta}'_{t}\}_{t=r_2}^{t=r_3}; \{\hat{\beta}''_{t}\}_{t=r_3}^{t=r_4};\ldots)$ (see Figure \ref{fig:beta-increment}). Thus, if we fit the dual variables as in \eqref{eq:dual-fit-HRDF} (and update the tail of the old $\{\hat{\beta}_t\}_{t\ge r_n}$ to that of the new $\{\hat{\beta}'_t\}_{t\ge r_n}$ upon the arrival of a new job $n$), then
\begin{align}\nonumber
\int_{0}^{\infty}\hat{\boldsymbol{\beta}}dt=\sum_n\hat{\alpha}_n=\mbox{Flow Cost}(\mbox{Alg}).
\end{align}  
The above derivations can be viewed as a generalization of the special $\ell_k$-dual fitting results given in \cite{anand2012resource} (see, e.g., Lemma 5.4).

\section*{Appendix II}

{\bf \emph{Proof of Lemma \ref{lemm-s-c}:}}
Let $\mathcal{A}$ be an online $s$-speed $c$-competitive algorithm for the GFC-S problem. For a fixed $\epsilon\in (0, 1]$, let $t_j$ be the first time that the algorithm $\mathcal{A}$ processes $\frac{1}{1+\epsilon}$ fraction of job $j$.  Then,  by following the schedule $\mathcal{A}$,  we have $v_j(t)\ge \frac{\epsilon}{1+\epsilon}v_j, \forall t\in [r_j, t_j]$.  Therefore,  the contribution of job $j$ to the fractional generalized completion time equals
\begin{align}\nonumber
\int_{r_j}^{\infty}\frac{v_j(t)}{v_j}g'_j(t)dt\ge \int_{r_j}^{t_j}\frac{\epsilon}{1+\epsilon}g'_j(t)dt=\frac{\epsilon}{1+\epsilon}g_j(t_j),
\end{align}
where the inequality holds because $g'_j(t)\ge 0$, and the equality uses $g_j(r_j)=0$.

Now, let us assume that the machine runs faster at a speed of $(1+\epsilon)$, and consider a new online algorithm $\mathcal{B}$ that schedules job $j$ in the first $\frac{v_j}{1+\epsilon}$ time slots that $j$ was scheduled by algorithm $\mathcal{A}$. Then the completion time of job $j$ in schedule $\mathcal{B}$ is $t_j$, and hence the contribution of job $j$ to the integral cost of schedule $\mathcal{B}$ equals $g_j(t_j)$.  Since the same argument holds for any job $j$, by summing over all jobs, we conclude that the integral completion time of $\mathcal{B}$ is at most $\frac{1+\epsilon}{\epsilon}$ times the fractional completion time of $\mathcal{A}$.  Since $\mathcal{A}$ is $s$-speed $c$-competitive for the fractional objective cost and the optimal fractional cost is a lower bound for the optimal integral cost,  the schedule $\mathcal{B}$ is $(1+\epsilon)s$-speed $\frac{1+\epsilon}{\epsilon}c$-competitive for the GICS problem. \hfill $\blacksquare$

\smallskip
{\bf \emph{Proof of Lemma \ref{lemm:convert}:}} First we note that updating $\alpha,\beta$-plots do not change the dual objective value. To see this, assume at the current step $k$ we update both plots by $\delta_{k}$. Then the first term $\sum_{r=1}^N\bar{\alpha}_{r}$ in the dual objective function of the splitted instance \eqref{eq:split-dual} reduces by exactly $\delta_{k}(t_{k}+\bar{v}_{k})$, which is the size of the area shrinked by lowering the height of all the steps prior to the current time $t_{k}+\bar{v}_{k}$. As we also lower the $\beta$-plot by the same amount $\delta_{k}$ for $t\in[0, t_{k}+\bar{v}_{k}]$, the second term $\int_{0}^{\infty}\bar{\beta}_t$ in the dual objective function also decreases by the same amount $\delta_j(t_{k}+\bar{v}_{k})$. Thus the overall effect of updates in the dual objective \eqref{eq:split-dual} at each iteration is zero. This implies that the dual objective value at the end of Algorithm \ref{alg:update} is the same as its initial value, i.e., OPT. 


Next we show that Algorithm \ref{alg:update} terminates properly with a feasible dual solution. Otherwise by contradiction, let $\hat{j}$ be the \emph{first} step whose update at time $\hat{t}:=t_{\hat{j}}+\bar{v}_{\hat{j}}$ violates at least one of the dual constraints, i.e., $\bar{\beta}_{\hat{t}}+\rho_jg(\hat{t})<h_j$ for some $j$. Now let $j_{\ell}$ be the first $j$-step on the right side of $\hat{j}$, and consider the time $t_{j_{\ell}}$ at which $j_{\ell}$ was set to its reference height $h_j$. Defining $\Delta$ to be the height difference between $\hat{j}$ and $h_j$ at time $t_{j_{\ell}}$, we have $\Delta\ge \sum_{k\in I}(\rho_k-\rho_j)(g(t_k+\bar{v}_k)-g(t_k))$, where $I:=[\hat{t}, t_{j_{\ell}}]$, and $k\in I$ refers to all the subjubs (steps) which are scheduled during $I$. This is because first of all the updates prior to $j_{\ell}$ do not change the \emph{relative} height difference between $\hat{j}$ and $j_{\ell}$. Moreover, during the interval $I$ if a subjob of density $\rho_k$ is scheduled over $[t_k, t_k+\bar{v}_k]\subseteq I$, then from \eqref{eq:beta-alpha-1} the first term in $\bar{\beta}_t+\rho_jg(t)$ drops at a negative rate $\rho_kg'(t)$ while the second term increases at a positive rate $\rho_jg'(t)$. As all the intermediate steps $k\in I$ have higher density than $\rho_j$ (otherwise, by HDF rule the subjob $j_{\ell}$ should have been processed earlier), we can write,   
\begin{align}\nonumber
\Delta:&=(\bar{\beta}_{\hat{t}}+\rho_{\hat{j}}g(\hat{t}))-h_j=(\bar{\beta}_{\hat{t}}+\rho_{\hat{j}}g(\hat{t}))-(\bar{\beta}_{t_{j_{\ell}}}+\rho_{j}g(t_{j_{\ell}}))\cr 
&\ge (\bar{\beta}_{\hat{t}}+\rho_{j}g(\hat{t}))-(\bar{\beta}_{t_{j_{\ell}}}+\rho_{j}g(t_{j_{\ell}}))\cr 
&=\sum_{k\in I}(\rho_k-\rho_j)(g(t_k+\bar{v}_k)-g(t_k)). 
\end{align}
In other words, $\Delta$ is larger than the total height decrease that $\bar{\beta}_t+\rho_jg(t)$ incurs over $I$.

To derive a contradiction, it is sufficient to show that $\Delta$ is no less than the total height decrements incurred by the step updates during the interval $I$. Toward this aim let us partition $I$ into subintervals $I=\cup_{j'=1}^{p} I_{j'}$ as we move backward over $I$. Each subinterval $I_{j'}$ starts with the first subjob $j'_1$ outside of the previous one $I_{j'-1}$, and it is just long enough to contain all other $j'$-steps which are inside $I$. By this partitioning and HDF rule, it is easy to see that the first step $j'_1$ of each subinterval $I_{j'}$ must be set as a reference for job $j'$. Now by our choice of $\hat{j}$ we know that all the steps in $I_{j'}$ can be properly set to their reference height at the time of update. Thus using a similar argument as above, the total height decrements due to step updates in the subinterval $I_{j'}$ (except the first step $j'_1$ which is a reference step) is equal to the total height decrease that $\bar{\beta}_t+\rho_{j'}g(t)$ incurs over the interval $I_{j'}$, i.e.,  $\Delta_{j'}=\sum_{k\in I_{j'}}(\rho_k-\rho_{j'})(g(t_k+\bar{v}_k)-g(t_k)).$
 
Finally, we account for total height reduction due to reference updates, denoted by $\sum_{j'}\Delta_{j'_1}$. We do this using a charging argument where we charge height decrements due to reference updates to the subintervals $I_{j'}, j'=1,\ldots,p$. As a result, the total height reduction due to reference updates would be the total charge over all the subintervals. For this purpose, let $L(j'):=\{r_1,r_2,\ldots,r_q\}$ be the longest chain of subintervals such that $r_1$ is index of the first subinterval of lower density on the left side of $I_{j'}$, i.e., $\rho_{r_1}\leq \rho_{j'}$, and inductively, $r_{i+1}$ denotes the index of the first subinterval of lower density on the left side of $I_{r_i}$, i.e., $\rho_{r_{i+1}}\leq \rho_{r_i}$. Then we charge subinterval $I_{j'}$ by $(\rho_{j'}-\rho_{r_q})g(I_{j'})$, where $g(I_{j'}):=\sum_{k\in I_{j'}}(g(t_k+\bar{v}_k)-g(t_k))$ is the total variation of $g(\cdot)$ over $I_{j'}$. Now, as an easy exercise, one can show that the height decrements due to reference updates $\sum_{j'}\Delta_{j'_1}$ is bounded above by the total charge $\sum_{j'}(\rho_{j'}-\rho_{r_q})g(I_{j'})$, i.e.,  $\sum_{j'}\Delta_{j'_1}\leq \sum_{j'}(\rho_{j'}-\rho_{r_q})g(I_{j'})$. Thus the total height reduction over $I$ can be bounded by
\begin{align}\nonumber
\sum_{j'}(\Delta_{j'}+\Delta_{j'_1})&\leq \sum_{j'}\sum_{k\in I_{j'}}(\rho_k-\rho_{j'})(g(t_k+\bar{v}_k)-g(t_k))+\sum_{j'}(\rho_{j'}-\rho_{r_q})g(I_{j'})\cr 
&= \sum_{j'}\sum_{k\in I_{j'}}(\rho_k-\rho_{j'})(g(t_k+\bar{v}_k)-g(t_k))+\sum_{j'}\sum_{k\in I_{j'}}(\rho_{j'}-\rho_{r_q})(g(t_k+\bar{v}_k)-g(t_k))\cr 
&= \sum_{j'}\sum_{k\in I_{j'}}(\rho_{k}-\rho_{r_q})(g(t_k+\bar{v}_k)-g(t_k))\cr
&\leq \sum_{k\in I}(\rho_k-\rho_j)(g(t_k+\bar{v}_k)-g(t_k))\leq \Delta, 
\end{align}
where the second inequality holds because $\rho_{j}\leq \min_{k\in I}\rho_k\leq \rho_{r_q}$. This contradiction establishes the dual feasibility of the generated solution at the end of Algorithm \ref{alg:update}. Finally, let $(\{\bar{\alpha}_{j_{\ell}}\},\beta_t)$ denote the values of $\alpha,\beta$-plots at the end of Algorithm \ref{alg:update}. Since at the end of the algorithm all the $j$-steps are properly set to the same reference height $\frac{\bar{\alpha}_{j_{\ell}}}{\bar{v}_{j_{\ell}}}=h_j, \forall \ell$, we have $h_j=\frac{\sum_{\ell}\bar{\alpha}_{j_{\ell}}}{\sum_{\ell}\bar{v}_{j_{\ell}}}=\frac{\sum_{\ell}\bar{\alpha}_{j_{\ell}}}{v_j}$. This shows that $\alpha_j:=\sum_{\ell}\bar{\alpha}_{j_{\ell}}=h_jv_j$ and $\beta_t$ form feasible dual solutions to the original instance. \hfill $\blacksquare$



\begin{definition}
We say a function $g_1$ \emph{dominates} another function $g_2$ and write $g_1\succ g_2$ if $g'_1(t)> g'_2(t)\ \forall t$. A class of functions $\mathcal{H}=\{g_j\}$ is a \emph{dominating family} if $(\mathcal{H},\succ)$ forms a totally ordered set.  
\end{definition}

\begin{proposition}\label{thm:dominance}
Let $\mathcal{H}=\{g_j(t)\}$ be a dominating family of cost functions. Then the optimal schedule for GFCS$(r_n)$ must process the jobs according to their dominance order, i.e., it first schedules the job with the most dominant cost function, and at the end it schedules the least dominant job.
\end{proposition}      
 
\emph{Proof:} Let us sort the functions in $\mathcal{H}$ as $g_1\succ g_2\succ \ldots\succ g_n$. Note that every curve $p_j(t_1)-g_j(t)$ is simply a $-g_j(t)$ that has been shifted by a constant amount $p_j(t_1)$. Now we argue that every two curves $p_j(t_1)-g_j(t)$ and $p_k(t_1)-g_k(t)$ must intersect exactly once over the interval $[0, t_1]$. The reason is that if $p_j(t_1)-g_j(t)$ and $p_k(t_1)-g_k(t)$ do not intersect, then one of them must lie above the other over the entire horizon $[0, t_1]$. As a result, the job associated with the lower curve will never be executed by the optimal policy $\boldsymbol{x}^o$, which is a contradiction. On the other hand, if two curves intersect in at least two time instances, say $t_a\neq t_b$, then $p_j(t_1)-g_j(t_a)=p_k(t_1)-g_k(t_a)$ and $p_j(t_1)-g_j(t_b)=p_k(t_1)-g_k(t_b)$ imply that $g_j(t_b)-g_j(t_a)=g_k(t_b)-g_k(t_a)$. However, that contradicts the strict dominance of $g_j$ over $g_k$ or vice versa. (Note that $g_j(t_b)-g_j(t_a)=\int_{t_a}^{t_b}g'_j(t)dt$.) Therefore, each curve $p_k(t_1)-g_k(t)$ can participate in determining the upper envelope of the collection of curves $\{p_k(t_1)-g_k(t), k=1,\ldots,n\}$ only if, over the interval $[0,v_1]$, the curve $p_1(t_1)-g_1(t)$ determines the upper envelope; over $[v_1, v_1+v_2]$, the curve $p_2(t_1)-g_2(t)$ determines the upper envelope; and eventually at the last time interval $[t_1-v_n, t_1]$, the curve $p_n(t_1)-g_n(t)$ determines the upper envelope. Thus the optimal policy schedules the jobs according to their dominance order. \hfill $\blacksquare$

\begin{example}
The set of functions $\mathcal{H}=\{\rho_jg(t): \rho_1>\rho_2>\ldots>\rho_n\}$ with nonnegative nondecreasing function $g(t)$ forms a dominating family of cost functions. Therefore, from Proposition \ref{thm:dominance}, the HDF rule is an optimal offline schedule for GFCS$(r_n)$. 
\end{example}

\bibliographystyle{IEEEtran}
\bibliography{thesisrefs}

\begin{thebibliography}{10}
\providecommand{\url}[1]{#1}
\csname url@samestyle\endcsname
\providecommand{\newblock}{\relax}
\providecommand{\bibinfo}[2]{#2}
\providecommand{\BIBentrySTDinterwordspacing}{\spaceskip=0pt\relax}
\providecommand{\BIBentryALTinterwordstretchfactor}{4}
\providecommand{\BIBentryALTinterwordspacing}{\spaceskip=\fontdimen2\font plus
\BIBentryALTinterwordstretchfactor\fontdimen3\font minus
  \fontdimen4\font\relax}
\providecommand{\BIBforeignlanguage}[2]{{%
\expandafter\ifx\csname l@#1\endcsname\relax
\typeout{** WARNING: IEEEtran.bst: No hyphenation pattern has been}%
\typeout{** loaded for the language `#1'. Using the pattern for}%
\typeout{** the default language instead.}%
\else
\language=\csname l@#1\endcsname
\fi
#2}}
\providecommand{\BIBdecl}{\relax}
\BIBdecl

\bibitem{leung2004handbook}
J.~Y. Leung, \emph{Handbook of {S}cheduling: {A}lgorithms, {M}odels, and
  {P}erformance {A}nalysis}.\hskip 1em plus 0.5em minus 0.4em\relax CRC Press,
  2004.

\bibitem{graham1979optimization}
R.~L. Graham, E.~L. Lawler, J.~K. Lenstra, and A.~R. Kan, ``Optimization and
  approximation in deterministic sequencing and scheduling: {A} survey,'' in
  \emph{Annals of Discrete Mathematics}.\hskip 1em plus 0.5em minus 0.4em\relax
  Elsevier, 1979, vol.~5, pp. 287--326.

\bibitem{angelopoulos2019primal}
S.~Angelopoulos, G.~Lucarelli, and T.~N. Kim, ``Primal--dual and dual-fitting
  analysis of online scheduling algorithms for generalized flow-time
  problems,'' \emph{Algorithmica}, vol.~81, no.~9, pp. 3391--3421, 2019.

\bibitem{anand2012resource}
S.~Anand, N.~Garg, and A.~Kumar, ``Resource augmentation for weighted flow-time
  explained by dual fitting,'' in \emph{Proceedings of the Twenty-Third Annual
  ACM-SIAM Symposium on Discrete Algorithms}.\hskip 1em plus 0.5em minus
  0.4em\relax SIAM, 2012, pp. 1228--1241.

\bibitem{im2011online}
S.~Im and B.~Moseley, ``An online scalable algorithm for minimizing
  $\ell_k$-norms of weighted flow time on unrelated machines,'' in
  \emph{Proceedings of the Twenty-Second Annual ACM-SIAM Symposium on Discrete
  Algorithms}.\hskip 1em plus 0.5em minus 0.4em\relax SIAM, 2011, pp. 95--108.

\bibitem{bansal2009weighted}
N.~Bansal and H.-L. Chan, ``Weighted flow time does not admit
  $o(1)$-competitive algorithms,'' in \emph{Proceedings of the Twentieth Annual
  ACM-SIAM Symposium on Discrete Algorithms}.\hskip 1em plus 0.5em minus
  0.4em\relax SIAM, 2009, pp. 1238--1244.

\bibitem{leonardi2007approximating}
S.~Leonardi and D.~Raz, ``Approximating total flow time on parallel machines,''
  \emph{Journal of Computer and System Sciences}, vol.~73, no.~6, pp. 875--891,
  2007.

\bibitem{lucarelli2016online}
G.~Lucarelli, N.~K. Thang, A.~Srivastav, and D.~Trystram, ``Online
  non-preemptive scheduling in a resource augmentation model based on
  duality,'' in \emph{European Symposium on Algorithms (ESA 2016)}, vol.~57,
  no.~63, 2016, pp. 1--17.

\bibitem{im2014selfishmigrate}
S.~Im, J.~Kulkarni, K.~Munagala, and K.~Pruhs, ``Selfishmigrate: {A} scalable
  algorithm for non-clairvoyantly scheduling heterogeneous processors,'' in
  \emph{Proc. 55th Annual Symposium on Foundations of Computer Science}.\hskip
  1em plus 0.5em minus 0.4em\relax IEEE, 2014, pp. 531--540.

\bibitem{im2018competitive}
S.~Im, J.~Kulkarni, and K.~Munagala, ``Competitive algorithms from competitive
  equilibria: {N}on-clairvoyant scheduling under polyhedral constraints,''
  \emph{Journal of the ACM (JACM)}, vol.~65, no.~1, pp. 1--33, 2018.

\bibitem{garg2007minimizing}
N.~Garg and A.~Kumar, ``Minimizing average flow-time: {U}pper and lower
  bounds,'' in \emph{48th Annual IEEE Symposium on Foundations of Computer
  Science (FOCS'07)}.\hskip 1em plus 0.5em minus 0.4em\relax IEEE, 2007, pp.
  603--613.

\bibitem{kalyanasundaram2000speed}
B.~Kalyanasundaram and K.~Pruhs, ``Speed is as powerful as clairvoyance,''
  \emph{Journal of the ACM (JACM)}, vol.~47, no.~4, pp. 617--643, 2000.

\bibitem{chadha2009competitive}
J.~S. Chadha, N.~Garg, A.~Kumar, and V.~Muralidhara, ``A competitive algorithm
  for minimizing weighted flow time on unrelated machines with speed
  augmentation,'' in \emph{Proceedings of the Forty-First Annual ACM Symposium
  on Theory of Computing}.\hskip 1em plus 0.5em minus 0.4em\relax ACM, 2009,
  pp. 679--684.

\bibitem{devanur2018primal}
N.~R. Devanur and Z.~Huang, ``Primal dual gives almost optimal energy-efficient
  online algorithms,'' \emph{ACM Transactions on Algorithms (TALG)}, vol.~14,
  no.~1, pp. 1--30, 2018.

\bibitem{im2015competitive}
S.~Im, J.~Kulkarni, and K.~Munagala, ``Competitive flow time algorithms for
  polyhedral scheduling,'' in \emph{Proc. 56th Annual Symposium on Foundations
  of Computer Science}.\hskip 1em plus 0.5em minus 0.4em\relax IEEE, 2015, pp.
  506--524.

\bibitem{im2011tutorial}
S.~Im, B.~Moseley, and K.~Pruhs, ``A tutorial on amortized local
  competitiveness in online scheduling,'' \emph{SIGACT News}, vol.~42, no.~2,
  pp. 83--97, 2011.

\bibitem{chekuri2001algorithms}
C.~Chekuri, S.~Khanna, and A.~Zhu, ``Algorithms for minimizing weighted flow
  time,'' in \emph{Proceedings of the Thirty-Third Annual ACM Symposium on
  Theory of Computing}.\hskip 1em plus 0.5em minus 0.4em\relax ACM, 2001, pp.
  84--93.

\bibitem{nguyen2013lagrangian}
K.~T. Nguyen, ``Lagrangian duality in online scheduling with resource
  augmentation and speed scaling,'' in \emph{Proc. European Symposium on
  Algorithms}.\hskip 1em plus 0.5em minus 0.4em\relax Springer, 2013, pp.
  755--766.

\bibitem{im2014online}
S.~Im, B.~Moseley, and K.~Pruhs, ``Online scheduling with general cost
  functions,'' \emph{SIAM Journal on Computing}, vol.~43, no.~1, pp. 126--143,
  2014.

\bibitem{im2016fair}
S.~Im and J.~Kulkarni, ``Fair online scheduling for selfish jobs on
  heterogeneous machines,'' in \emph{Proceedings of the 28th ACM Symposium on
  Parallelism in Algorithms and Architectures}.\hskip 1em plus 0.5em minus
  0.4em\relax ACM, 2016, pp. 185--194.

\bibitem{williamson2011design}
D.~P. Williamson and D.~B. Shmoys, \emph{The {D}esign of {A}pproximation
  {A}lgorithms}.\hskip 1em plus 0.5em minus 0.4em\relax Cambridge University
  Press, 2011.

\bibitem{azar2005convex}
Y.~Azar and A.~Epstein, ``Convex programming for scheduling unrelated parallel
  machines,'' in \emph{Proceedings of the Thirty-Seventh Annual ACM Symposium
  on Theory of Computing}.\hskip 1em plus 0.5em minus 0.4em\relax ACM, 2005,
  pp. 331--337.

\bibitem{megow2018dual}
N.~Megow and J.~Verschae, ``Dual techniques for scheduling on a machine with
  varying speed,'' \emph{SIAM Journal on Discrete Mathematics}, vol.~32, no.~3,
  pp. 1541--1571, 2018.

\bibitem{hohn2015performance}
W.~H{\"o}hn and T.~Jacobs, ``On the performance of {S}mith's rule in
  single-machine scheduling with nonlinear cost,'' \emph{ACM Transactions on
  Algorithms (TALG)}, vol.~11, no.~4, pp. 1--30, 2015.

\bibitem{bansal2014geometry}
N.~Bansal and K.~Pruhs, ``The geometry of scheduling,'' \emph{SIAM Journal on
  Computing}, vol.~43, no.~5, pp. 1684--1698, 2014.

\bibitem{moseley2019scheduling}
B.~Moseley, ``Scheduling to approximate minimization objectives on identical
  machines,'' in \emph{Proceedings of the 46th International Colloquium on
  Automata, Languages, and Programming, ICALP}, 2019, pp. 86:1--86:14.

\bibitem{fox2013online}
K.~Fox, S.~Im, J.~Kulkarni, and B.~Moseley, ``Online non-clairvoyant scheduling
  to simultaneously minimize all convex functions,'' in \emph{Approximation,
  Randomization, and Combinatorial Optimization. Algorithms and
  Techniques}.\hskip 1em plus 0.5em minus 0.4em\relax Springer, 2013, pp.
  142--157.

\bibitem{becchetti2006online}
L.~Becchetti, S.~Leonardi, A.~Marchetti-Spaccamela, and K.~Pruhs, ``Online
  weighted flow time and deadline scheduling,'' \emph{Journal of Discrete
  Algorithms}, vol.~4, no.~3, pp. 339--352, 2006.

\bibitem{basar2020lecture}
T.~Ba\c{s}ar, S.~Meyn, and W.~R. Perkins, ``Lecture notes on control system
  theory and design,'' \emph{arXiv preprint arXiv:2007.01367}, 2020.

\bibitem{liberzon2011calculus}
D.~Liberzon, \emph{Calculus of {V}ariations and {O}ptimal {C}ontrol {T}heory:
  {A} {C}oncise {I}ntroduction}.\hskip 1em plus 0.5em minus 0.4em\relax
  Princeton University Press, 2011.

\bibitem{cheung2017primal}
M.~Cheung, J.~Mestre, D.~B. Shmoys, and J.~Verschae, ``A primal-dual
  approximation algorithm for min-sum single-machine scheduling problems,''
  \emph{SIAM Journal on Discrete Mathematics}, vol.~31, no.~2, pp. 825--838,
  2017.

\bibitem{antoniadis2017qptas}
A.~Antoniadis, R.~Hoeksma, J.~Mei{\ss}ner, J.~Verschae, and A.~Wiese, ``A
  {QPTAS} for the general scheduling problem with identical release dates,'' in
  \emph{44th International Colloquium on Automata, Languages, and Programming
  (ICALP 2017)}.\hskip 1em plus 0.5em minus 0.4em\relax Schloss
  Dagstuhl-Leibniz-Zentrum Fuer Informatik, 2017, p. 31:1–31:14.

\bibitem{gupta2020greed}
V.~Gupta, B.~Moseley, M.~Uetz, and Q.~Xie, ``Greed works - {O}nline algorithms
  for unrelated machine stochastic scheduling,'' \emph{Mathematics of
  Operations Research}, vol.~45, no.~2, pp. 497--516, 2020.

\end{thebibliography}

\end{document}